\documentclass[12pt]{article}
\usepackage{amsmath, amssymb}

\usepackage{epsfig}

\newtheorem{theorem}{Theorem}[section]
\newtheorem{proposition}[theorem]{Proposition}
\newtheorem{corollary}[theorem]{Corollary}
\newtheorem{lemma}[theorem]{Lemma}
\newtheorem{definition}[theorem]{Definition}

\newcommand{\rd}{{\rm d}}
\newcommand{\be}{\begin{equation}}
\newcommand{\ee}{\end{equation}}
\newcommand{\bey}{\begin{eqnarray}}
\newcommand{\eey}{\end{eqnarray}}

\newcommand{\bef}{\begin{figure}}
\newcommand{\eef}{\end{figure}}
\newcommand{\bec}{\begin{center}}
\newcommand{\eec}{\end{center}}

\newcommand{\tri}{| \! |\!|}

\newcommand{\eqn}{\begin{eqnarray}}
\newcommand{\eeqn}{\end{eqnarray}}

\newcommand{\bD}{{\bf D}}
\newcommand{\bB}{{\bf B}}
\newcommand{\bA}{{\bf A}}

\newcommand{\bw}{{\bf w}}
\newcommand{\bp}{{\bf p}}

\newcommand{\bu}{{\bf u}}

\newcommand{\Tor}{{\bf T}}
\newcommand{\Fou}{{\cal F}}

\newcommand{\tb}{{\tilde b}}

\newcommand{\tp}{{\tilde p}}

\newcommand{\tu}{{\tilde u}}

\newcommand{\tsi}{{\tilde\sigma}}

\newcommand{\tgamma}{{\tilde\gamma}}
\newcommand{\tGamma}{{\tilde\Gamma}}

\newcommand{\tbp}{{\tilde{\bf p}}}

\newcommand{\tbu}{{\tilde{\bf  u}}}

\renewcommand{\a}{\alpha}

\newcommand{\e}{\varepsilon}
\newcommand{\s}{\sigma}

\newcommand{\om}{{\omega}}

\newcommand{\bP}{{\bf P}}

\newcommand{\bE}{{\bf E}}

\newcommand{\cX}{{\cal X}}
\renewcommand{\iint}{\int \!\! \int}
\newcommand{\bR}{{\mathbb R}}
\newcommand{\bC}{{\bf C}}
\newcommand{\bN}{{\bf N}}
\newcommand{\bZ}{{\mathbb Z}}

\newcommand{\wt}{\widetilde}
\newcommand{\wh}{\widehat}
\newcommand{\ov}{\overline}
\newcommand{\balpha}{\mbox{\boldmath $\alpha$}}

\newcommand{\cI}{{\cal I}}
\newcommand{\cR}{{\cal R}}
\newcommand{\cG}{{\cal G}}

\newcommand{\cC}{{\cal C}}
\newcommand{\cF}{{\cal F}}
\newcommand{\cA}{{\cal A}}
\newcommand{\cB}{{\cal B}}

\newcommand{\cP}{{\cal P}}
\newcommand{\cD}{{\cal D}}
\newcommand{\cV}{{\cal V}}

\newcommand{\cL}{{\cal L}}

\newcommand{\cO}{{\cal O}}
\newcommand{\cT}{{\cal T}}

\input epsf

\def\fb{{\frak b}}

\newcommand{\de}{ \mbox{\rm \scriptsize deg}}
\newcommand{\dee}{ \mbox{\rm \tiny deg}}

\def\fS{{\frak S}}

\date{Mar 23, 2007}

\begin{document}

\title{Quantum diffusion for the Anderson model 
\\
in the scaling limit.}
\author{L\'aszl\'o Erd\H os${}^1$\thanks{Partially
supported by NSF grant DMS-0200235 and EU-IHP Network 
``Analysis and Quantum'' HPRN-CT-2002-0027.}
\\ Manfred  Salmhofer${}^2$\thanks{Partially supported by 
DFG grant Sa 1362/1--1 and an ESI senior research fellowship}
\\ Horng-Tzer Yau${}^3$\thanks{Partially supported by NSF grant
DMS-0307295 and MacArthur Fellowship.} \\
\\
\small
${}^1\;$Institute of Mathematics, University of Munich, 
\\
\small
Theresienstr. 39, D-80333 Munich, Germany
\\
\small
${}^2\;$Max--Planck Institute for Mathematics, Inselstr.\  22, 04103 Leipzig,
and
\\ 
\small
Theoretical Physics, University of Leipzig, Postfach 100920, 04009 Leipzig, 
Germany
\\
\small
${}^3\;$Department of Mathematics, Harvard University, Cambridge
MA-02138, USA \\ }

\maketitle

\abstract{ 
We consider  random Schr\"odinger equations
on $\bZ^d$ for $d\ge 3$
with identically distributed random potential.
Denote by $\lambda$ the coupling constant  and $\psi_t$ the solution
with initial data $\psi_0$. The space and time variables
scale as $x\sim \lambda^{-2 -\kappa/2}, t \sim \lambda^{-2 -\kappa}$
with $0< \kappa < \kappa_0(d)$.
We prove that, in the limit $\lambda \to 0$,
the expectation of the Wigner distribution of $\psi_t$  converges weakly to a
solution of a heat equation
in the space variable $x$ for arbitrary $L^2$  initial data.
The diffusion coefficient is uniquely determined
by the kinetic energy associated to the momentum $v$.

This work is an extension to the lattice case of our previous result
in the continuum \cite{ESYI}, \cite{ESYII}. 
Due to the non-convexity of the level surfaces
of the dispersion relation, the estimates 
of several Feynman graphs are more involved.}



\section{Introduction}

We consider the time evolution of the Anderson model \cite{A}
given by the random Schr\"odinger equation
\be
    i\partial_t\psi_t(x) = H\psi_t(x), \qquad
  \psi_t \in \ell^2(\bZ^d), \;\; t\in \bR
\label{sch}
\ee
on the $d$-dimensional square lattice $\bZ^d$.
The Hamiltonian is given by
\be\label{H}
   H = -\frac{1}{2} \Delta + \lambda V_\om  \; ,
\ee
where $\lambda>0$ is the coupling constant.
The kinetic energy operator on $\ell^2(\mathbb Z^d)$
 is given by
\be\label{Delta}
    (\Delta f)(x): = 2d \; f(x)-\sum_{|e|=1}f(x+e)
\ee
and the random potential is given by
\be\label{ranpot}
        V_\om(x) = \sum_{\a\in \bZ^d} V_\a(x) \qquad
        V_\a(x):= v_\a \; \delta_{x,\a} 
\ee
where $\{ v_\a\; : \; \a\in \bZ^d\}$ are i.i.d. random variables and 
$\delta_{x,\alpha} = {\bf 1}\{x=\a\}$ is the usual 
Kronecker delta function and ${\bf 1}\{ \cdot \}$ 
is the characteristic function.  
We will work in $d=3$ dimension, but our results and proofs
extend to any $d\ge3$ 
in a straight-forward manner.

We  study the long time evolution of the
equation \eqref{sch}. For 
a large coupling constant, $\lambda \ge \lambda_0$,
the spectrum of $H$ is almost surely pure point and
the dynamics is localized \cite{A, FS, AM}.
It is conjectured, but not yet proven, that
the spectrum is absolutely continuous and the dynamics
is diffusive 
if $\lambda <\lambda_0$  is sufficiently small.
We  will investigate the dynamics in this
regime  in the scaling limit,
when time  diverges as $\lambda\to 0$.

Up to time scales $ t\sim \lambda^{-2}$
the dynamics is kinetic, typically with  a finite  number of 
collisions. The evolution on macroscopic space scales,
$x\sim\lambda^{-2}$, is given by a linear Boltzmann equation
\cite{Sp1, EY, Ch, LS}. As the long time limit of the Boltzmann
equation is the heat equation, the quantum evolution 
on scales $t\gg \lambda^{-2}$
is expected to exhibit diffusive behavior.
In this paper we prove this statement
up to time scale $t\sim \lambda^{-2-\kappa}$
with a positive $\kappa$.

We have proved the same statement (with a somewhat bigger $\kappa$)
for a random Schr\"odinger operator
in the continuum, $\bR^d$ \cite{ESYI, ESYII}. 
The history of this problem and related works are summarized
in \cite{ESYI} and will not be repeated here.

Our reasons to extend this work to the lattice case are:
 (i) to show that the methods initiated in \cite{EY, E},
and later developed in \cite{ESYI, ESYII} for longer time scales,
 can be applied to a lattice setting as well;
(ii) to make a connection with   the extended
state conjecture of the
Anderson model.

Anderson (de)localization is a large distance phenomenon,
thus no physical difference is expected between the lattice
and continuum models.  The  localization
proofs, however,  are  typically  simpler in the lattice models
because of certain technical difficulties due to
the ultraviolet regime in the continuum model.

We now explain briefly the differences between
the continuum and lattice models
in our analysis of the delocalization regime.
The finite momentum space is an advantage
in this regime as well; the artificial
large momentum cutoffs introduced 
for the continuum model in \cite{ESYI}
are not necessary here. Moreover, the
computation of the main term is more direct
since the Boltzmann collision kernel is homogeneous
on the energy shells (compare \eqref{def:sigma} below with 
(2.19) of \cite{ESYI}). In particular, the diffusion coefficient
can be computed explicitly. 

However, an important technical estimate
is considerably more involved
for the lattice case. The complication
stems from  the non-convexity of the isoenergy
surfaces of the lattice Laplacian. The isoenergy surfaces
are the level sets, $\Sigma_\a: = \{ p\in \Tor^d
 \; : \; e(p)=\a\}$,
of the dispersion relation
\be
e(p):= \sum_{j=1}^d \big(1-\cos (2\pi p^{(j)})\big), \qquad p=(p^{(1)},
\ldots , p^{(d)})\in \Tor^d:= \Big[ -\frac{1}{2}, \frac{1}{2}\Big]^d \; .
\label{ep}
\ee

\begin{figure}
\begin{center}
\epsfig{file=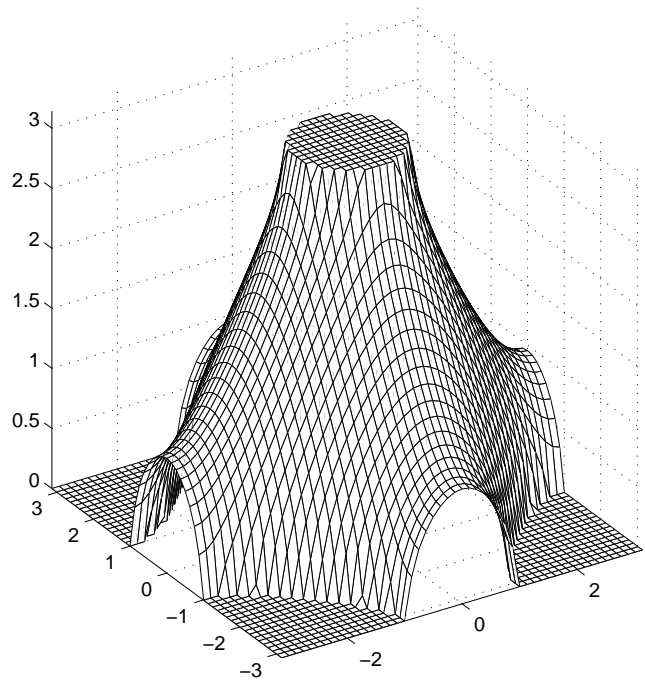,scale=.75}
\end{center}
\caption{Level set of $e(p)=\alpha$ for $2<\a<4$}\label{fig:levelset}
\end{figure}

Our approach heavily uses estimates that integrals of 
resolvent functions, $|\a - e(p)+i\eta|^{-1}$, $\eta\ll 1$ are
concentrated near $\Sigma_\a$. 
In particular, our most involved estimate
(Four Denominator Lemma \ref{lemma:4dee}) 
translates naturally to a specific decay property
of the Fourier transform of 
measures supported on the hypersurfaces $\Sigma_\a$.
Such decay bounds are readily available for 
surfaces with non-vanishing Gauss curvature, but
we were unable to find in the literature the necessary estimate
for surfaces with Gauss curvature
vanishing along a submanifold. 
Since in the energy range $\a \in (2, 2d-2)$,
the Gaussian curvature of the level sets $\Sigma_\a$ 
vanish along a codimension one submanifold 
(see Figure \ref{fig:levelset}),
we had to prove this estimate separately 
\cite{ESYFour}.  Another related bound (Two Denominator 
Lemma \ref{lemma:newrow}) will be proven in this paper.

The analogous bounds in the continuum model,
with dispersion relation $e(p)=\frac{1}{2} p^2$, are 
much easier. The actual proofs (e.g. Proposition 2.3
in \cite{EE} or Lemma A.1 in \cite{ESYI}) use the explicit
form of $e(p)$, but 
the key fact is that the level sets of  the dispersion 
relation are convex.

Most importantly, the
Two-Denominator Lemma is much stronger in the
continuum case; the estimate in Lemma \ref{lemma:newrow}
carries a diverging factor 
$\eta^{-3/4-\kappa}$. The corresponding bound
(formula (10.28) in \cite{ESYI}) is only logarithmically
divergent in $d=3$. This makes the estimates of the error
terms in \cite{ESYII} much easier.

We mention that the difficulty related to the
non-convexity of the level sets is present
in the proofs of the Boltzmann equation from
a lattice model \cite{Ch, LS} as well.
On the kinetic time scale, however,  weaker
bounds (called crossing estimates)
were sufficient (Lemma 7.8 of \cite{Ch}
or Assumption (DR4) of \cite{LS}).

The main inputs
we use from our other works are:
(i) the stopping rules 
(Section 3 of \cite{ESYII}); 
(ii) the basic integration procedure for non-repetition
Feynman graphs
(Sections 8-10 of \cite{ESYI});
(iii) the organization of the estimates 
on the error terms (Section 4 of \cite{ESYII}); and
(iv) the Four Denominator Lemma 
proved in \cite{ESYFour}.
The new ingredients are: (i) the estimates of the error terms
(with the Four Denominator Lemma replacing the strong
form of the Two Denominator Lemma not available in
the lattice case); (ii) the computation of the main
term and the explicit diffusion coefficient.

\section{Statement of the main result}
\setcounter{equation}{0}

\subsection{Notations}

In this paper we consider the random Schr\"odinger
operator \eqref{H} acting on $\ell^2(\bZ^d)$.
The kinetic energy operator is given by the
discrete Laplacian \eqref{Delta}
and $V_\omega(x)$ is the random potential \eqref{ranpot}.
Denote the
moments of the single-site
random potential by $m_k=\bE \; v_\a^{k}$. 
We  assume that $m_1=m_3=m_5=0$, $m_6 < \infty$,
and $m_2=1$ by normalization.
Universal constants will be denoted by $C$ 
and their value may vary from line to
line.

On the lattice $(\delta\bZ)^d$, $\delta>0$, 
we introduce the notation
\be
    \int_{(\delta\bZ)^d} (\cdots) \rd x
: = \delta^d \sum_{x\in (\delta\bZ)^d} (\cdots) \; .
\label{def:x}
\ee
On the dual space $(\Tor/\delta)^d$,
 with $\Tor:= [-\frac{1}{2}, \frac{1}{2}]$, the integration refers
to the usual Lebesgue integral
\be
  \int_{(\Tor/\delta)^d}  (\cdots)  \rd p \; .
\label{def:leb}
\ee
We will use these formulas mostly for $\delta=1$
in which case  we will drop the subscripts
$(\delta\bZ)^d$ and  $(\Tor/\delta)^d$ 
indicating the integration domains in \eqref{def:x}, \eqref{def:leb}.
The letters   $x,y,z$ will always denote lattice variables, while
$p, q, r, u, v, w$ will be reserved for $d$-dimensional momentum
variables on the torus. This notation will distinguish 
between the two integrations (\ref{def:x}) and (\ref{def:leb}).
Note that as $\delta\to0$, both integrals converge
to the standard Lebesgue integral on $\bR^d$.

For any $f\in \ell^2((\delta\bZ)^d)$
the  Fourier transform is given by
\be
    \wh f(p) \equiv (\Fou f) (p) := \int_{(\delta\bZ)^d}
  e^{-2\pi ip\cdot x} f(x) \rd x
 =\delta^d \sum_{x\in (\delta \bZ)^d}
    e^{-2\pi ip\cdot x} f(x) \;,
\label{FF}
\ee
where $p=(p^{(1)},\dots,p^{(d)})\in(\Tor/\delta)^d$.
The
 inverse Fourier
transform is given by
$$
    (\Fou^{-1}\wh g)(x)
    =  \int_{(\Tor/\delta)^d}   \wh g(p) e^{ 2\pi i p\cdot x}
    \rd p \; .
$$
The notations $\Fou(\cdot)$ and $\wh{(\cdot)}$ will be used
according to convenience. For functions defined on the
phase space, $f(x,v)$, with  $x\in (\delta\bZ)^d$, $v\in (\Tor/\delta)^d$,
the Fourier transform will always be taken only in the space
variable, i.e.
$$
  \wh f(\xi, v): = \int_{(\delta\bZ)^d} e^{-2\pi i \xi\cdot x} f(x, v)
  \rd x, \quad \xi\in  (\Tor/\delta)^d\; .
$$
We also remark that addition and subtraction of momenta
will always be defined on the torus, i.e. with periodic boundary conditions.

The Fourier transform of the kinetic energy operator \eqref{Delta}
on $\ell^2(\bZ^d)$
  is given by
$$
    (\Fou \Delta f)(p) =   - 2e(p) \wh f(p) \; ,
$$
where $e(p)$ is the dispersion law defined in \eqref{ep}.
For $h:\Tor^d\to\bC$ and an energy value $e\in [0,2d]$
 we  introduce the notation
\be\label{coar}
   [ h ](e):=
   \int h(v) \delta(e-e(v))\rd v : = \int_{\Sigma_e } h(q)
 \; \frac{\rd \nu(q)}{|\nabla      e(q)|}
\ee
where $\rd \nu(q)=\rd\nu_e(q)$ 
is the restriction of the $d$-dimensional Lebesgue measure
to the level surface 
$$
\Sigma_e:=\{ q\; : \; e(q)=e\}\subset \Tor^d.
$$
 By the co-area formula it holds that
\be
    \int_0^{2d} [ h ](e) \rd e   = \int h(v) \rd v \; .
\label{coa}
\ee
We define the projection to the energy space of the free Laplacian by 
\be\label{F}
\langle \, h(v) \, \rangle_e : = \frac{[h](e)}{ \Phi(e)}\; ,
\quad\mbox{ where }\quad
\Phi(e): = [1](e)=\int \delta(e-e(u))\rd u \;.
\ee
Define the {\it Wigner transform} of a function $\psi\in
\ell^2(\mathbb Z^d)$ 
\be
        W_\psi(x,v): =2^d\sum_{y,z\in \bZ^d\atop y+z=2x} 
        e^{2\pi iv\cdot (y-z)} \overline{\psi(y)}\psi(z), \qquad x\in \Big(
   \frac{\bZ}{2}\Big)^d, \; v\in \Tor^d \; .
\label{WP}
\ee
Notice that the position variable $x$ takes values on  the refined
lattice $(\bZ/2)^d$. It is easy to compute the Fourier transform
of $ W_\psi(x,v)$ in  $x$ using (\ref{FF}) with $\delta=1/2$,
and one obtains
\be
     \wh W_\psi(\xi, v)
    =\overline{\widehat
     \psi\Big(v-\frac{\xi}{2}\Big)}
        \widehat\psi\Big(v+\frac{\xi}{2}\Big) \; , \qquad \xi\in (2\Tor)^d, 
  \; v\in \Tor^d \; .
\label{FW}
\ee
Note that $\wh W_\psi(\xi, v)$ is periodic with respect to the
double torus $(2\Tor)^d$.
It is easy to check that $W_\psi(x,v)$ reproduces the
correct marginals in the following sense
$$
   \int W_\psi(x,v) \rd v =\left\{ 
\begin{array}{cl} 2^d |\psi(x)|^2 & \mbox{if} \;\;  x\in \bZ^d \\
0 & \mbox{if} \;\; x\in (\bZ/2)^d\setminus \bZ^d
\end{array} \right.
$$
\be
   \int_{(\bZ/2)^d}  W_\psi(x,v)\rd x=
   2^{-d}\sum_{x\in (\bZ/2)^d} W_\psi(x,v) = |\wh\psi(v)|^2\; ,
\label{marg}
\ee
and in particular 
$$
      \int_{(\bZ/2)^d}  \int  W_\psi(x,v) \rd v\rd x =\|\psi\|^2\; .
$$
Recall that momentum integrations on unspecified domains
are always considered on $\Tor^d$.
Define the rescaled Wigner distribution as
\be
        W^\e_\psi (X, V) : = \e^{-d}W_\psi\Big( \frac{X}{\e}, V\Big)\; ,
        \quad
    X\in (\e\bZ/2)^d, \; V\in \Tor^d \; .
\label{wig}
\ee
Its  Fourier transform in $X$ is given by
$$
     \widehat W^\e_{\psi}(\xi, V) = \cF \big( W^\e_\psi (\;
     \cdot \; , V)\big)(\xi)=
     \overline{\widehat \psi \Big(V-\frac{\e \xi}{2}\Big)}
        \widehat \psi \Big( V+ \frac{\e \xi}{2}\Big) \; , 
\quad \xi\in (2\Tor/\e)^d, \;
   V\in \Tor^d \; .
$$
To test the rescaled Wigner transform against a Schwarz
function $\cO(x, v)$ on $\bR^d\times \Tor^d$,
 we introduce the 
scalar products:
\begin{align}
  \langle \, \cO,    W_{\psi}^\e \rangle: &
=  \int_{(\e \bZ/2)^d}  \rd X  \int
 \rd v  \; \cO(X, v)   W_{\psi}^\e(X, v)
\nonumber \\
  \langle \, \wh \cO,   \wh W_{\psi}^\e \rangle: &=
   \int_{(2\Tor/\e)^d} \rd \xi \int \rd v \; \wh \cO(\xi, v) 
\ov{\wh W_{\psi}^\e}(\e \xi , v)  \;  .
\label{bracket}
\end{align}
By unitarity of the Fourier transform,   
$\langle \, \cO,    W_{\psi}^\e \rangle =  
\langle \, \wh \cO,   \wh W_{\psi}^\e \rangle$.

One may, of course, also define the Wigner transform of a function 
$\psi\in \ell^2(\bZ^d)$ by first defining
$\wh W_\psi(\xi, v)$ for $\xi\in (2\Tor)^d$ by (\ref{FW}),
 then taking the inverse Fourier transform in $\xi$ 
to reproduce the lattice function defined in (\ref{WP}).
An interpretation of the Wigner function 
as a distribution starting from (\ref{FW}) for any $\xi\in \bR^d$
was given in \cite{LS}.

\subsection{Main theorem}

The weak coupling limit is defined by the following scaling:
\be \label{wcl}
\cT:=\e t,\quad
\cX:= \e x, \quad \e= \lambda^2\; .
\ee
It was proved \cite{Ch} that in the limit $\e\to 0$
the Wigner distribution $W^\e_{\psi_{ \cT/\e}} (\cX, \cV)$
converges weakly to the solution of the Boltzmann equation
\eqn\label{B}
        &&\partial_\cT F_\cT(\cX,V) + 
 \sin  (2\pi V)\cdot \nabla_\cX F_\cT(\cX,V)
        \nonumber\\
        &&\hspace{2cm}=
       2\pi \int \rd U \delta(e(U)-e(V)) \Big[ F_\cT(\cX,U)
         -  F_\cT(\cX,V) \Big]\; ,
\eeqn
with velocity vector $ \sin  (2\pi V) := (  \sin  (2\pi V_1),
 \sin (2\pi V_2), \sin  (2\pi V_3)) =\frac{1}{2\pi}\nabla e(V)$.
Here $\cF_\cT(\cX, \cV)$ is the time dependent
limiting phase space density with $\cX\in \bR^d$, $\cV\in \Tor^d$.
Note that the Boltzmann equation can be viewed as
 the generator of a Markovian semigroup on phase space.
In particular, all correlation effects become negligible
in this scaling limit.

In this paper we consider the long time scaling
\be\label{scale}
x= \lambda^{-\kappa/2-2}X = \e^{-1} X, \quad t =
\lambda^{-\kappa-2} T = \e^{-1} \lambda^{-\kappa/2} T,
\quad \e = \lambda^{\kappa/2+2} \; 
\ee
with $\kappa > 0$. 
Our main result is the following theorem.

\begin{theorem}\label{main}
Let $d=3$ and $\psi_0 \in \ell^2(\bZ^d)$ be an initial wave function.
Let $\psi(t)=\psi_{t,\om}^\lambda$ solve the Schr\"odinger equation
(\ref{sch}). Let $\cO(x, v)$ be a Schwarz
function on $\bR^d\times \Tor^d$.
For almost all energies
 $e\in [0,2d]$, $\big[  |\wh \psi_0(v)|^2 \big](e)$ is finite
and for these energies 
let $f$ be the solution to the heat equation
\be
\partial_T f(T, X, e) = \; \nabla_X\cdot D(e)
\nabla_X f(T, X, e) \label{eq:heat}
\ee
with the initial condition
$$
     f(0, X, e): = \delta(X) \Big[ |\wh \psi_0(v)|^2 \Big](e)
$$
and the diffusion matrix
$D$
    \be D_{ij}(e):=  \frac{
    \big\langle \,  \sin  (2\pi v^{(i)}) \cdot
 \sin (2\pi v^{(j)})    \, \big\rangle_e}{2\pi\; \Phi(e)}
\; \qquad i,j=1,2,3.\label{diffconst}
\ee
Then for $\kappa<\kappa_0$ and $\e$ and $\lambda$ related by \eqref{scale},
the Wigner distribution  satisfies
\be
\lim_{\e \to 0} \int_{(\e\bZ/2)^d} \!\! 
\rd X \! \int \! \rd v \; \cO(X, v)  \bE
 W^\e_{\psi(\lambda^{-\kappa-2} T)} (X, v)
= \int_{\bR^d} \rd X \int \!\! \rd v \; \cO(X, v) f(T, X, e(v)) \;, 
\label{fint}
\ee
and the limit is uniform on $T\in [0, T_0]$ with any fixed $T_0$.
In $d=3$ dimensions, one can choose $\kappa_0=1/9800$.
\end{theorem}

{\it Remarks.}
(i) We stated the Theorem and will carry out the proof in $d=3$
for simplicity but our method
 works for any $d \ge 3$.

\medskip

(ii) The analogous theorem for the continuum model was proved
in \cite{ESYI} and \cite{ESYII} with a somewhat larger
 threshold for $\kappa$. The threshold 
is obtained from technical estimates, it has
no physical relevance and it can be improved
with a more careful analysis.

\medskip

(iii) By the symmetry of the measure $\langle \cdot\rangle_e$
under each sign flip $v_j\to -v_j$ and by the permutational
symmetry of the coordinate axes, we see that $D(e)$ is a constant
times  the identity matrix:
$$
    D_{ij} (e) = D_e \; \delta_{ij}, \qquad D_e:=
\frac{ \big\langle  \sin^2 (2\pi v^{(1)})
    \, \big\rangle_e}{2\pi\; \Phi(e)}  \; ,
$$
in particular we see that the diffusion is nondegenerate.

\medskip

(iv) The diffusion matrix can also be obtained from the long time
limit of the Boltzmann equation (\ref{B}) (see also
the explanation to Figure 1 in \cite{ESYI}). For any fixed energy $e$,
let
\be
   L_e f(v): = \int \rd u \; \sigma(u, v) [ f(u)-f(v)], \qquad e(v)=e\; ,
\label{Lgen}
\ee
be the generator of the momentum jump process on $\Sigma_e$
with the uniform stationary measure $\langle \cdot \rangle_e$.
The collision kernel is 
\be
\sigma(U,V):=2\pi\delta( e(U)-e(V)).
\label{def:sigma}
\ee
The diffusion matrix in general is  given by the
velocity autocorrelation function
\be
  D_{ij}(e)=\int_0^\infty \rd t  \; 
   \big\langle \sin (2\pi v^{(i)}(t))\cdot \sin (2\pi v^{(j)}(0))\big\rangle_e \; ,
\label{Dij}
\ee
where $v(t)$ is the process generated by $L_e$.
Since for any fixed $V$ the  collision kernel
$\sigma(U,V)$, 
is  symmetric on the energy surface in the $U$ variable,
the correlation between $v(t)$ and $v(0)$  vanishes after
the first jump and we obtain  (\ref{diffconst}), by using
$$
    \int\rd u\; \sigma(u, v)  = 2\pi \Phi(e)\; , \qquad e(v)=e\;.
$$

\section{Preparations}
\setcounter{equation}{0}

\subsection{Renormalization}

The purpose of this procedure is to include immediate recollisions
with the same obstacle into  the propagator itself.
Without renormalization, these graphs
are exponentially large (``divergent''), but
their sum is finite. Renormalization removes this
instability and the analysis
of the resulting Feynman graphs will become simpler.

The self-energy operator is given by the multiplication operator
in momentum space
\be
 \theta(p) := \Theta(e(p)), \qquad    \Theta (\alpha): = \lim_{\e \to 0+ }
 \Theta_\e (\alpha) \; ,
\label{eq:thetalim}
\ee
where
\be\label{theta}
    \Theta_\e (\alpha): =  \int \frac  { \rd q} { \alpha- e(q) + i \e} \; .
\ee
The existence of the limit and  related properties of $\Theta$
are proved in Lemma \ref{le:A1}

We rewrite the Hamiltonian as
$$
    H= H_0 + \wt V,
$$
where
\be\label{renH}
  H_0:=\omega(p):= e(p) +\lambda^2 \theta(p), \qquad \wt V := \lambda
    V -\lambda^2 \theta(p)\; .
\ee
We note that our renormalization is only an approximation to
the standard self-consistent renormalization  given
by the solution to the equation
\be
     \om(p) = e(p) + \lambda^2 \lim_{\e\to 0+0}
     \int \frac  { \rd q } { \om(p)- \om(q) + i \e} \; .
\label{eq:selfc}
\ee
Due to our truncation procedure, the definition (\ref{eq:thetalim})
is sufficient and is more convenient for us.

We also need a few properties of the function $\om(p)$.
First, $\om(p)$
is symmetric under the permutation of the coordinate axes.
Second,
$e(p)$ is symmetric for reflections onto any coordinate axis:
$e(p^{(1)},  \ldots, p^{(d)})=e(\pm p^{(1)}, \ldots,  \pm p^{(d)})$
for any choices of the
signs and these properties are inherited by $\theta(p)$ and $\om(p)$.
Moreover, $e({\bf \frac{1}{2}} - p)= 2d - e(p)$, where
${\bf \frac{1}{2}}:=(\frac{1}{2}, \frac{1}{2}, \ldots, \frac{1}{2})$.
Therefore $\Theta(\alpha) = -\ov{\Theta}(2d-\alpha)$ and
\be
    \theta({\bf \frac{1}{2}}-p)=-\ov{\theta(p)}\; .
\label{eq:thetasim}
\ee
These relations allow us to restrict our attention to the subdomain
\be
    \cD: = \Big\{ p=(p^{(1)}, p^{(2)}, \ldots , p^{(d)})\in \Tor^d\; : \; 0\leq
    p^{(1)} \leq p^{(2)} \leq \ldots \leq p^{(d)}
    \; ,
      \sum_j p^{(j)}\leq \frac{d}{4} \Big\}
\label{def:cD}
\ee
of the momentum space.
On this domain
we have the following estimate on $\theta(p)$:
\begin{lemma}\label{lemma:theta}
For any $d\ge 3$ there exist universal positive constants $c_1, c_2$ such that
\be
    |\theta(p)|\leq c_2 \; ,
\label{eq:thetaest}
\ee
\be
    - c_2|p|^{d-2}\leq \mbox{Im}\;\theta(p) \leq -c_1|p|^{d-2}
\label{eq:imthetaest}
\ee
for any $p\in \cD$.
\end{lemma}
Using this bound and the symmetry properties of $\om(p)$ above, 
we easily arrive at
\begin{corollary}\label{cor:theta} For $d\ge 3$ we have
\be
    \mbox{Im} \; \om (p) \le - c_3\lambda^2 [D(p)]^{d-2} ,\qquad p\in \Tor^d
\label{eq:lowim}
\ee
for some $c_3>0$, where
$$
    D(p): = \min \Big\{
     |p-v|\; : \; v =0 \;\; \mbox{or} \;\; v=\Big(\pm \frac{1}{2}, \;
 \pm \frac{1}{2}, \ldots, \pm \frac{1}{2}\Big) \; \Big\}
$$
is the distance between $p$ and the set consisting
of  the origin and the vertices of $\Tor^d$. $\;\;\Box$
\end{corollary}
We remark that this estimate would produce logarithmic corrections
in $d=2$ dimensions, this is one of the reasons why the proof is
somewhat simpler in $d\ge 3$.

\medskip

{\it Proof of Lemma \ref{lemma:theta}.}
By the co-area formula (\ref{coa}) and recalling the definition (\ref{F})
we can write $\Theta_\e(\alpha)$ as
\begin{equation}\label{co-area}
   \Theta_\e(\alpha) =  \int_{0}^{2d} \frac{\rd s}{\a - s +i\e} \; \Phi(s) \; ,
   \qquad \mbox{with}\quad
   \Phi(s)=\int_{ \Sigma_s}
   \frac{\rd\nu(q)}{|\nabla e(q)|}\; .
\end{equation}
Because of the symmetries of $e(q)$, it is sufficient to study
$$
    \wt\Phi(s):=\int_{ \Sigma_s\cap \cD}
   \frac{\rd\nu(q)}{|\nabla e(q)|}\; ,
$$
i.e. where the integration is restricted to $\cD$.

The critical values of $e(q)$  are the even integers, $2m$,
between 0 and $d$, 
and within $\cD$ they correspond to the critical points
$$
p_m= \Big( 0,0, \ldots, 0, \underbrace{\frac{1}{2}, \ldots ,
\frac{1}{2}}_{m \;\;\mbox{times}} \; \Big)
$$
with $m\leq d/2$.
If $s$ is away from a neighborhood of the critical values 
of $e(q)$, then $\Phi(s)$ is a bounded
function with a strictly positive lower bound.

If $s$ is near a critical value, $s=2m + \beta$ with $\beta$ sufficiently 
small,
 then we can write
$$
    \wt \Phi(s) = \int_{ \Sigma_s\cap \cD}
   \frac{\chi_m(q)}{|\nabla e(q)|}\; \rd\nu(q) + \int_{ \Sigma_s\cap \cD}
   \frac{1-\chi_m(q)}{|\nabla e(q)|} \; \rd\nu(q)
$$
where $\chi_m$ is a smooth cutoff function around the critical point $p_m$.
The second integral is a regular function in $s$ that is separated away from 0.
The first integral can be brought into the following normal form
by a smooth local coordinate transformation:
$$
    \Phi^\#(\beta):=\int_{C(\beta)} \frac{\chi(q)}{|q|} \; \rd q \; ,
$$
where $\chi$ is a smooth cutoff function around 0 and
$$
   C(\beta):=\Big\{ q\in \bR^d   \;  : \; \sum_{j=1}^{d-m} [q^{(j)}]^2
   -  \sum_{j=d-m+1}^{d} [q^{(j)}]^2 = \beta  \; \Big\} \; .
$$
It is a straightforward calculation to see the following behavior of
the function $\Phi^\#>0$ for  small $\beta$
\begin{align}\label{eq:Phi}
   \Phi^\#(\beta) =O\Big(\beta^{\frac{d}{2}-1}\Big)
   ,\qquad \frac{\rd}{\rd\beta}\Phi^\# (\beta)= O\Big(
 \beta^{\frac{d}{2}-2}\Big) \;\;\;
\mbox{if}\;\;\;  m=0 \\
\Phi^\#(\beta) =O(1)
   ,\qquad \frac{\rd}{\rd\beta}\Phi^\# (\beta)= O(1) \;\;\;
 \mbox{if} \quad 1\leq m \leq d/2. \nonumber
\end{align}
{F}rom these estimates,  $\Phi$ is differentiable  away from 0,
bounded everywhere 
and $\Phi(s)$ behaves as $s^{1/2}$ near 0 in $d=3$ dimensions. For higher
dimensions $\Phi'$ is bounded.
 From the formula
\eqref{co-area}, we can rewrite $\Theta$ as
\begin{equation}
   \Theta_\e(\alpha) =  \int^{\alpha}_{\alpha-2d} \frac{\rd s }{s +i\e} \;
   \Phi(\alpha -s) \; .
\end{equation}
Using the above properties of $\Phi$
we can take the limit $\e \to 0$ and define
 $\Theta (\alpha):=\lim_{\e\to0+0} \Theta_\e(\alpha)$.
If we write $\Theta(e)=  {\cal R} (e)- i{\cal I}(e)$, 
where ${\cal R} (e)$ and
${\cal I}(e)$ are real functions,
and recall  $ Im (x+  i 0)^{-1} = - \pi \delta(x)$,  we have
\be
    \mbox{Im} \,\Theta (\alpha) = -\pi\int \delta(e(q)-\alpha) \rd q
    =-\pi \Phi(\alpha) \; .
\label{eq:opt} \ee Using the properties of $\Phi$, we can also
check that for any $0\leq e \leq d$
\be\label{2.1}
      \Phi(e)= f(e) e^{\frac{d}{2}-1}, \qquad 
 {\cal R} (e)=a(e) (1+e^{\frac{d}{2}-1}), 
 \qquad  {\cal I} (e)=b(e) e^{\frac{d}{2}-1}
\ee
where $a, b, f$ are bounded functions, 
uniformly separated away from zero and $f(e)>0$.
 We also have $\cI(\alpha)=\pi \Phi(\alpha)$.
The estimates in
Lemma \ref{lemma:theta} then follow from the fact that $e(p)\ge c
p^2$ on $\cD$.
Later we will also need the bounds
\be\label{2.2}
      |\Phi'(e)| \le C (1+  e^{\frac{d}{2}-2})
      \qquad |\Theta'(e)| \leq | {\cal R} '(e)| + 
| {\cal I} '(e)| \le C (1+e^{\frac{d}{2}-2}) \; ,
\ee
that can be proven by a similar analysis. $\;\;\;\Box$

The following lemma collects some estimates on the renormalized  propagators
we shall use to prove Theorem~\ref{main}.
Its proof will be given in the Appendix. These are technical bounds and their
meanings will become clear when they are used.

\begin{lemma}\label{le:opt}
Suppose that $\lambda^2 \ge \eta \ge \lambda^{2+ 4 \kappa}$ with
$\kappa \le 1/12$. Then we have,
\be
    \sup_\alpha\int \frac{\rd p}{|\alpha - \om(p)+ i\eta|}
    \leq C |\log\lambda| \; 
\label{eq:logest}
\ee
and for $0\le a<1$
\be
    \sup_\alpha \int \frac{\rd p}{|\a -\om(p) + i\eta|^{2-a}}
 \leq C_a\lambda^{-2(1-a)} .
\label{eq:2aint}
\ee
For $a=0$, the following more precise estimate holds.
There exists a universal constant $C_0$ such that
\be
\begin{split}
  \sup_\alpha \int \frac{ \lambda^{2} \; \rd p  }{|\alpha-\ov\om(p)-i\eta|^{2}} & \leq
    1+ C_0\lambda^{1-12\kappa}  \\
\sup_{\a,\beta, r} \int \frac {\lambda^2\; \rd p}{ |\a - \ov\om(p+r)
      -i\eta|
     \; |\beta - \om(p-r)  +i\eta|} \;
     & \leq 1+ C_0\lambda^{1-12\kappa}\; .
\label{eq:ladderint}
\end{split}
\ee
\end{lemma}

\subsection{Truncation}\label{sec:trun}

For any real number $\a$ we define
\be
\tri \alpha \tri :=  \min\{ |\alpha|, |\alpha -2|, |\a-3|, 
|\a -4|, |\a-6|\} \; 
\label{def:tria}
\ee
in the $d=3$ dimensional model.
The values $0,2,4,6$ are the critical values of $e(p)$.
The value $\a=3$ is special, for which the
level surface $\{ e(p) = 3\}$ has  a flat point.
For $\tri\a\tri$ separated away from zero, the following
key estimate holds (the proof is given in \cite{ESYFour}):

\begin{lemma}\label{lemma:4dee}  [Four Denominator Lemma]
For any $\Lambda>\eta$ there exists $C_\Lambda$ such that
for any $\alpha \in [0,6]$ with $\tri \a\tri \ge \Lambda$,
$$
  \int\frac{\rd p\rd q\rd r}{|\a -e(p)+i\eta||\a -e(q)+i\eta| 
|\a -e(r)+i\eta||\a -e(p-q+r+u)+i\eta|} \leq C_\Lambda
|\log\eta|^{14}
$$
uniformly in $u$.  $\;\;\Box$
\end{lemma}
 In general,
in $d\ge 3$ dimensions, $\tri \a\tri$ is the minimum of $|\a -d|$ and
of all $|\a - 2m|$, $0\leq m\leq d$.

We will prove the main Theorem \ref{main} under the assumption
that the initial wave function in Fourier space is continuously differentiable,
$\|\wh\psi_0\|_{C^1}<\infty$, and
it satisfies the following
condition: There exists $\Lambda>0$ such that
\be
    \tri e(p)\tri  \ge 3\Lambda  \qquad \mbox{on the support of} \;\; 
   \wh\psi_0 \; .
\label{suppsi}
\ee

Once the theorem is proven for such initial data, we can
easily extend it for the general case. 
Since $e(p)$ is a Morse function,
for any positive $\Lambda>0$ we can define a smooth cutoff function
$0\leq \chi^\Lambda\leq 1$ on $\Tor^d$ with the property that
$$
       \tri e(p)\tri \ge 3\Lambda  \qquad \mbox{on the support of} \;\; 
   \chi^\Lambda
$$
and
$$
   \lim_{\Lambda\to0}  \int |1-\chi^\Lambda(p)|^2\rd p =0 \; .
$$
Moreover, we assume that $\chi^\Lambda \ge \chi^{\Lambda'}$
pointwise  if $\Lambda\leq 
\Lambda'$.
The  wave function will be decomposed as
$$ 
    \psi(t) = \psi_1(t) +\psi_2(t) \; ,
$$
where $\psi_1, \psi_2$ are defined in Fourier space as
$$
 \wh\psi_1(t):= e^{-itH}(\chi^\Lambda \wh\psi_0)
   \qquad \wh\psi_2(t):= e^{-itH} [ (1-\chi^\Lambda) \wh\psi_0] \; .
$$

The Wigner transform enjoys the following continuity property:
if the random wave function is decomposed as $\psi=\psi_1+\psi_2$, then
\be
   \Big| \bE\langle \wh\cO, \wh W_\psi^\e\rangle -
    \bE\langle \wh\cO, \wh W_{\psi_1}^\e\rangle\Big|
    \leq \Big(\int_{(2\Tor/\e)^d} \sup_v |\wh\cO(\xi, v)| \rd\xi \Big)
    \sqrt{\bE \big[\| \psi_1 \|^2 +\|\psi_2\|^2\big] \cdot \bE \| \psi_2 \|^2}
\label{wigcont}
\ee
(see Section 2.1 of \cite{EY}, but due to a misprint,
the $\|\psi_2\|^2$ term was erroneously omitted).
Since
$$
   \|\psi_2(t)\|=
   \| e^{-itH} [ (1-\chi^\Lambda) \wh\psi_0]\|_2 =
    \|(1-\chi^\Lambda)\wh\psi_0\|_2\; ,
$$
by monotone convergence
we see that
$$ 
   \lim_{\Lambda\to0} \Big| \bE\langle \wh\cO, \wh W_{\psi(t)}^\e\rangle -
    \bE\langle \wh\cO, \wh W_{\psi_1(t)}^\e\rangle\Big| =0
$$
uniformly in $t$ (and thus in $\e$).
This means that the truncation procedure is continuous
on the  left hand side of
(\ref{fint}).

Similarly, on the side of the limiting heat equation, we
can define $f^\Lambda(T, X, e)$ to be the solution to 
(\ref{eq:heat}) with initial data $f^\Lambda(0, X, e): = 
\delta(X)\big[  |\chi^\Lambda\wh \psi_0|^2\big](e)$.
Clearly $\big[  |\chi^\Lambda\wh \psi_0|^2\big](e)$
monotonically converges to $\big[  |\wh \psi_0|^2\big](e)$
as $\Lambda\to0$ for any $e$ such that $\tri e\tri \neq 0$.
Therefore $f^\Lambda(0, X, e)$ converges to $f(0, X, e)$
in $L^1(dX)$, and thus the same statement holds for the solution
$$
   f^\Lambda(T, X, e) \to f(T, X, e) \qquad \mbox{in} \;\; L^1(dX)
$$
for almost all $e$ and uniformly in $T$. But then the right hand side of
(\ref{fint}) is also continuous as $\Lambda\to0$.
 
The condition, $\|\wh\psi_0\|_{C^1}<\infty$ can also
be removed by an analogous truncation argument, see
Section 3.2 of \cite{ESYI} for details.

\section{The proof of the Main Theorem~\ref{main}}
\setcounter{equation}{0}

The structure of the proof
 is the same in the continuous and in the lattice case,
so this section is almost identical 
to Sections 4 and 5 of \cite{ESYI}.
There are three minor differences
in the structure.
First, 
in \cite{ESYI} the single site
potentials were indexed by the finite
set $\{1,2, \ldots, M\}$, while here they are indexed by
$\bZ^d$ \eqref{ranpot}. Second, in the continuous
case the problem was first restricted to a finite box
(Section 3.3 \cite{ESYI}) and this restriction
was removed at the end of the analysis. This complication is
absent here. Finally, the  Boltzmann collision kernel
 contains an additional factor
in the continuous case (compare \eqref{def:sigma} with
the corresponding formula (2.19) of \cite{ESYI}).

We expand the unitary kernel of $H= H_0 + \wt V$ (see \eqref{renH})
 by the Duhamel formula.  
For any fixed integer $N\ge 1$
\begin{equation}\label{duh}
        \psi_t : = e^{-itH}\psi_0 = \sum_{n=0}^{N-1} \psi_n (t)
     + \Psi_{N}(t) \; ,
\end{equation}
with
\be
        \psi_n(t) : = (-i)^n\int_0^t [\rd s_j]_1^{n+1} \; \;
    e^{-is_{n+1}H_0}\wt V e^{-is_nH_0}\wt V\ldots
       \wt V e^{-is_1 H_0}\psi_0
\label{eq:psin}
\ee
being the fully expanded terms
and
\be
        \Psi_{N} (t): = (-i) \int_0^t \rd s \, e^{-i(t-s)H}
   \wt V \psi_{N-1}(s)
\label{eq:PsiN}
\ee
is the  non-fully expanded or error term. We used the shorthand notation
$$
    \int_0^t [\rd s_j]_1^n : = \int_0^t\ldots \int_0^t
 \Big(\prod_{j=1}^n \rd s_j\Big)
        \delta\Big( t- \sum_{j=1}^n s_j\Big) \; .
$$

Since each potential $\wt V$ in (\ref{eq:psin}), (\ref{eq:PsiN})
is a summation itself, $\wt V=
-\lambda^2\theta(p)+\sum_\a \lambda V_\a$,
 both of these terms in (\ref{eq:psin}) and (\ref{eq:PsiN})
are actually big summations over so-called
 elementary wavefunctions,
which are characterized by their collision history, i.e. by a sequence
of obstacles and, occasionally, by $\theta(p)$.
Denote by
$\tGamma_n$, $n\le\infty$,  the set of sequences
\be\label{Gamman}
\tgamma = (\tgamma_1, \tgamma_2, \ldots , \tgamma_n), \qquad
\tgamma_j\in \bZ^d\cup \{ \vartheta\}
\ee
and by $ W_\tgamma$ the associated  potential
$$
        W_\tgamma :=  \left\{ \begin{array}{cll} \lambda V_\tgamma & \qquad
\mbox{if} \quad & \tgamma\in\bZ^d \\
     - \lambda^2 \theta(p)  & \qquad \mbox{if} \quad & \tgamma =\vartheta
 \; . \end{array} \right.
$$
The tilde refers to the fact that the additional $\{ \vartheta\}$
symbol is also allowed.
 An element $\tgamma \in \bZ^d\cup \{ \vartheta\}$
is identified with the potential 
$V_\tgamma$ and it is called {\it potential label}
if $\tgamma\in\bZ^d$, otherwise it is called a $\vartheta$-label.
Potential labels carry a factor $\lambda$, 
$\vartheta$-labels carry a factor $\lambda^2$.

For any $\tgamma\in\tGamma_n$ we define the
following fully expanded wavefunction with truncation
\be
    \psi_{*t, \tgamma}: = (-i)^{n-1}\int_0^t [\rd s_j]_1^{n} \; \; W_{\tgamma_n}
    e^{-is_nH_0} W_{\tgamma_{n-1}} \ldots
    e^{-is_2H_0} W_{\tgamma_1} e^{-is_1H_0} \psi_0
\label{eq:trunc}
\ee
and without truncation
\be
    \psi_{t, \tgamma}: =  (-i)^{n}\int_0^t [\rd s_j]_1^{n+1} \; \;
    e^{-is_{n+1}H_0} W_{\tgamma_n}
    e^{-is_nH_0} W_{\tgamma_{n-1}} \ldots
    e^{-is_2H_0} W_{\tgamma_1} e^{-is_1H_0} \psi_0\; .
\label{eq:exp}
\ee
In the notation the star $(*)$ will always refer to truncated functions.
Note that
$$
    \psi_{t,\tgamma} = (-i)\int_0^t \rd s \; e^{-i(t-s)H_0}
    \psi_{*s,\tgamma}\; .
$$

Every term (\ref{eq:exp})
 along the expansion procedure is characterized by its order
$n$ and by a sequence $\tgamma\in\tGamma_n$.
The main term is given by the sequences
that contain different potential labels only. Their set is defined as
\be
    \Gamma_k^{nr}:=\Big\{
    \gamma = (\gamma_1, \ldots , \gamma_k) \; : \;
   \gamma_j\in\bZ^d,\;
 \gamma_i\neq \gamma_j \; \mbox{if} \; i\neq j\Big\} \; .
\label{nr}
\ee
Let
$$
   \psi_{t,k}^{nr}:= \sum_{\gamma\in \Gamma_k^{nr}} \psi_{t,\gamma} 
$$
denote the corresponding elementary wave functions. 

The typical
number of collisions up to time $t$ is of order $\lambda^2t$. 
To allow us for some room, we set 
\be
     K := [\lambda^{-\delta}(\lambda^2 t)]\; , 
\label{def:K}
\ee
($[ \; \cdot \; ]$ denotes integer part),
where $\delta=\delta(\kappa)>0$ is 
a small positive number to be fixed later on.
$K$ will serve as an upper threshold for the number of
collisions in the expansion.

\bigskip

The proof of the Main Theorem~\ref{main} is divided into three theorems.
 The first one states that all terms other than $\psi_{t,k}^{nr}$,
$0\leq k< K$,
are negligible:

\begin{theorem} [$L^2$-estimate of the error terms]\label{7.1}
Let $t=O(\lambda^{-2-\kappa})$ and $K$ given by \eqref{def:K}.
If $\kappa < \kappa_0(d)$ and $\delta$ is sufficiently small (depending only
on $\kappa$),  then
$$
   \lim_{\lambda\to0}\bE 
\Big\| \psi_t -\sum_{k=0}^{K-1} \psi_{t,k}^{nr} \Big\|^2 =0 \; .
$$
In $d=3$ dimensions, one can choose $\kappa_0(3)=\frac{1}{9800}$.
\end{theorem}

 The second  theorem 
gives an explicit formula
for the main terms,  $\psi_{t,k}^{nr}$, expressed in terms
of the {\it ladder diagrams}.
 We introduce the {\it  renormalized propagator}
$$
        R_\eta(\a, v):= \frac{1}{\alpha - \om(v) + i\eta} \; .
$$

\begin{theorem} [Only the ladder diagram contributes] \label{thm:L2}
Let $\kappa<\frac{1}{144}$,  $\e=\lambda^{2+\kappa/2}$,
$\eta=\lambda^{2+\kappa}$, $t=O(\lambda^{-2-\kappa})$,
 and $K$ given by \eqref{def:K}.
 For a sufficiently small 
positive $\delta$  and for  any $1\leq k < K$ we have
\be
   \bE \|\psi_{t,k}^{ nr}\|^2=  V_\lambda(t, k)
    + O\Big( \lambda^{\frac{1}{16}-9\kappa-O(\delta)}\Big)
\label{eq:L2bound}
\ee
\be  
\langle \wh \cO, \bE\wh W^\e_{\psi_{t,k}^{ nr}}
\rangle=  W_\lambda(t, k, \cO)
    + O\Big( \lambda^{\frac{1}{16}-9\kappa-O(\delta)}\Big)
\label{eq:Wbound}
\ee
as $\lambda\ll 1$. Here
\begin{align}
   V_\lambda(t, k): = & \frac{\lambda^{2k}e^{2t\eta}}{(2\pi)^2}
  \iint_{-\infty}^\infty \rd\a\rd\beta  \; e^{i(\a-\beta)t} 
 \int\Big(  \prod_{j=1}^{k+1}  \rd p_j \Big) \; 
|\wh\psi_0(p_1)|^2
\nonumber\\
&\times \prod_{j=1}^{k+1} \ov{ R_\eta(\a, p_j)} R_\eta(\beta, p_j)\;,
\label{ladder} \\
W_\lambda(t, k, \cO): = & \frac{\lambda^{2k}e^{2t\eta}}{(2\pi)^2}
  \iint_{-\infty}^\infty \rd\a\rd\beta  
\; e^{i(\a-\beta)t} \int_{(2\Tor/\e)^d}
 \rd \xi
\int  \Big( \prod_{j=1}^{k+1} \rd v_j \;  \Big)
\wh\cO(\xi, v_{k+1})\overline{\wh W_{\psi_0}^\e}(\xi, v_1) 
  \nonumber \\
 & \times 
\prod_{j=1}^{k+1} 
  \ov{R_\eta\Big(\a, v_j +\frac{\e\xi}{2}\Big)}
   R_\eta\Big(\beta, v_j -\frac{\e\xi}{2}\Big).
\label{9.51}
\end{align}
\end{theorem}
We   adopt the
notation $O(\delta)$ in the exponent of $\lambda$. This always
means $\mbox{(const.)}\delta$ 
with  universal, explicitly computable positive constants
that depend on $\kappa$ and that can be easily computed
 from the proof.
We note that the definition \eqref{9.51} does not apply literally
 to the free evolution term $k=0$; this term is defined
separately: 
\be
   W_\lambda(t, k=0,\cO)  : = \int_{(2\Tor/\e)^d} \rd\xi \int\rd v\;
    e^{it\e  \xi\cdot \nabla e(v)}\; e^{2t\lambda^2 {\scriptsize \mbox{Im}}\,
 \theta (v)}\;
 \wh\cO(\xi, v)\ov{\wh W_{\psi_0}}(\e\xi, v) \; .
\label{xi0}
\ee

\medskip

The third theorem identifies the limit of $\sum_k W_\lambda(t, k,\cO)$
as $\lambda\to0$
with the solution to the heat equation. 

\begin{theorem} [The ladder diagram converges to the heat equation]
\label{thm:laddheat}
Under the conditions of Theorem~\ref{thm:L2} and 
$t=\lambda^{-2-\kappa}T$, we have 
\be
  \lim_{\lambda\to0} \sum_{k=0}^{K-1} W_\lambda(t, k,\cO) = 
  \int \rd X  \int \rd v \; \cO(X, v)f(T, X, e(v))\; ,
\label{eq:laddheat}
\ee 
where $f$ is the solution to the heat equation \eqref{eq:heat}.
\end{theorem}

{\it Proof of the Main Theorem \ref{main} using
Theorems~\ref{7.1}, \ref{thm:L2}    and \ref{thm:laddheat}.}
We compute the expectation of the rescaled Wigner transform, 
$\bE W^\e_t = \bE W^\e_{\psi_t}$,
 tested against
a Schwarz  function, $\langle \cO,\bE W^\e_t\rangle$ 
(see \eqref{bracket}). By combining Theorem~\ref{7.1}
with the $L^2$-continuity of the Wigner transform \eqref{wigcont}, 
it is sufficient to compute the Wigner transform of
$\psi(t, K):= \sum_{k=0}^{K-1} \psi_{t,k}^{nr}$.
The Wigner transform $W_{\psi(t, K)}$ 
contains a
 double summation
$$
     W_{\psi(t, K)} =\sum_{k,k'=0}^{K-1}\ov {\psi^{ nr}_{t,k}}
 (\cdots)
 \psi^{ nr}_{t,k'}(\cdots) \; .
$$
The potential labels are not repeated
within $\ov\psi$ and $\psi$. Moreover, 
 the expectation of a single potential in \eqref{eq:exp} is zero.
Thus the potential labels in the $\psi$ and $\ov\psi$ must pair,
in particular 
 taking expectation reduces 
this double sum to a single sum over $k$
$$
   \bE \,  W_{\psi(t, K)} =\sum_{k=0}^{K-1}
\bE\,  W_{\psi^{  nr}_{t,k}}\; .
$$ 
 By using \eqref{eq:Wbound}
and  \eqref{eq:laddheat} together with $K= O(\lambda^{-\kappa-\delta})$,
we obtain Theorem~\ref{main}. $\;\;\Box$

\section{Stopping rules}\label{sec:stop}
\setcounter{equation}{0}

The Duhamel expansion \eqref{duh} allows for the flexibility 
 at every expansion step $N$ to decide if the full
evolution $e^{-i(t-s)H}$ in \eqref{eq:PsiN} is expanded further
or not. The decision is based upon the collision history
of the expanded terms. The stopping rules  organize the expansion.
The basic idea is to expand up to the identification of the main terms,
but not to expand error terms unnecessarily further. 
The stopping rules are identical in the continuous and lattice
cases and they were given in Section 3 \cite{ESYII} in full
details. Here we only summarize the concepts informally
and refer to \cite{ESYII} for the precise definitions.

\medskip

\begin{figure}
\begin{center}
\epsfig{file=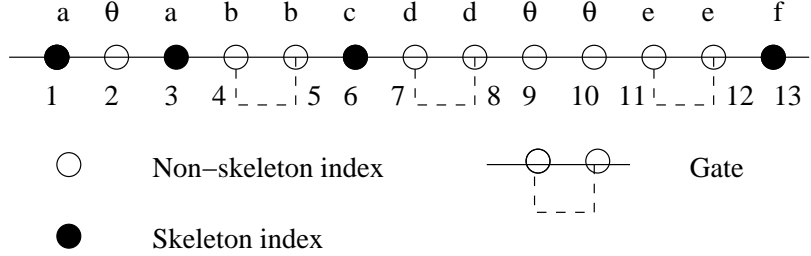,scale=1}
\end{center}
\caption{Skeleton indices and gates for 
 $\tgamma=(a, \vartheta, a, b,b, c, d,d,
\vartheta, \vartheta, e,e, f)$.}\label{fig:skel}
\end{figure}

In a sequence $\tgamma\in \tGamma_n$
we identify the {\it immediate recollisions}
  inductively starting from $\tgamma_1$
(due to their graphical picture, they are also called {\it gates}).
The gates must involve potential labels and not $\vartheta$.
Any potential
label which does not belong to a gate will be called {\it skeleton
label}. The index $j$ of a skeleton label $\gamma_j$
in $\tgamma$ is called {\it skeleton index}. The set of skeleton indices is
$S(\tgamma)$. 
The $\vartheta$ terms are never
part of the skeleton. 
This definition is recursive so we can
identify skeleton indices successively along the expansion procedure.
Notice that the last skeleton index may become a gate index at the next step.
Fig. \ref{fig:skel} shows an example
for these concepts, the formal definition is given in 
Definition 3.1 \cite{ESYII}.
The number of skeleton indices will be denoted by
$ k(\tgamma):= |S(\tgamma)|$. Let $t(\tgamma)$ denote
the number of $\vartheta$-labels in $\tgamma$. 
Recalling that potential labels carry a factor $\lambda$
and $\vartheta$-labels carry a factor of $\lambda^2$,
we let 
\be
\label{r}
    r(\tgamma):= \frac{1}{2}[n-k(\tgamma)] + t(\tgamma)
\ee
denote the total $\lambda^2$-power collected from non-skeleton indices.

\medskip

Sequences where the only repetitions in potential labels
occur within the gates are called {\it non-repetitive}
sequences. A special case of them are the sequences in
$\Gamma_k^{nr}$ \eqref{nr} that contain no gates or $\vartheta$-labels.
The repetitive sequences are divided into the following categories.
If  two non-neighboring skeleton labels coincide,
then the collision history includes a
 {\it genuine (non-immediate) recollision}.
If a skeleton label coincides with a gate label, then we have a {\it triple
collision} of the same obstacle. If two neighboring skeleton labels coincide
and are not immediate recollisions because there are gates or $\theta$'s
in between, then we have a {\it nest}. The precise definitions
are given in Definition 3.3 \cite{ESYII}.

\bigskip\bigskip
\centerline{\epsffile{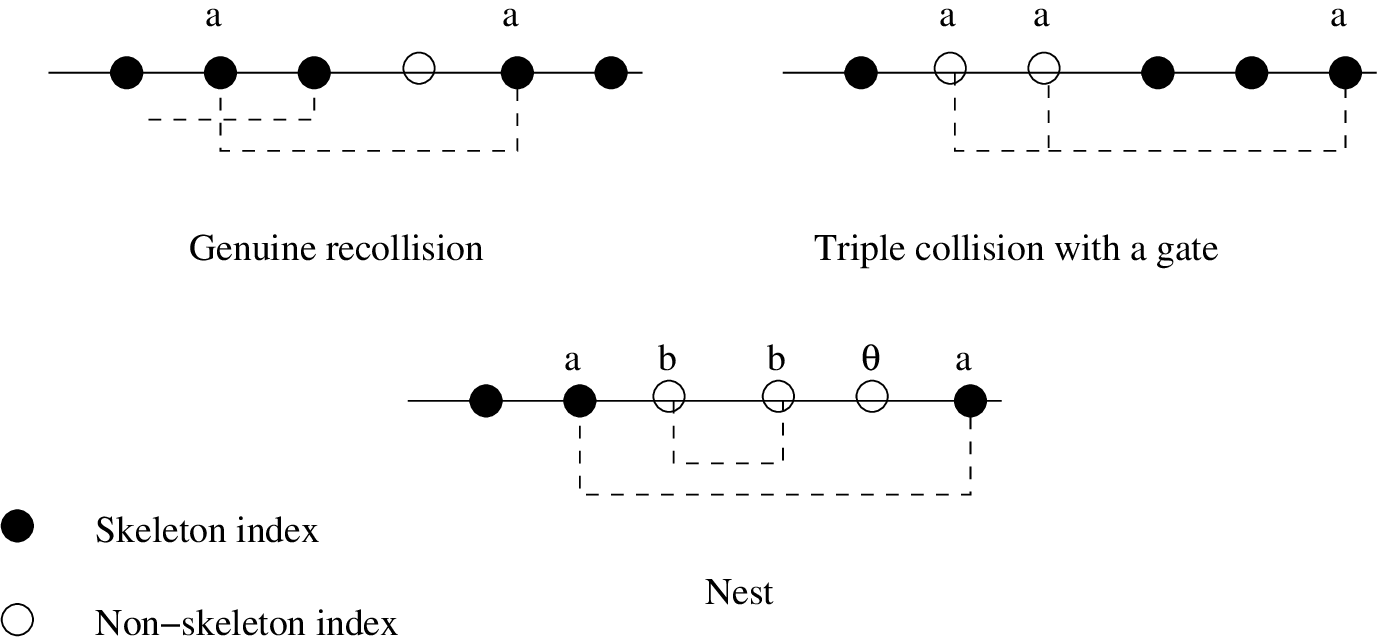}}

\bigskip

We stop the expansion at an elementary truncated wavefunction (\ref{eq:trunc})
characterized by $\tgamma$,
if any of the following happens:

\medskip

$\bullet$  The number of skeleton indices  in $\tgamma$ reaches $K$.
We denote  the sum of the truncated elementary non-repetitive
wave functions up to time $s$ with at most one
$\lambda^2$ power from the non-skeleton indices or $\theta$'s
and with $K$ skeleton indices by
$\psi_{*s, K}^{(\leq 1),nr}$. The superscript $(\leq 1)$ refers
the number of collected $\lambda^2$ powers.

\medskip

$\bullet$  We have collected $\lambda^4$ from non-skeleton labels.
We denoted by
$\psi_{*s,k}^{(2),last}$ the sum of the truncated elementary wave functions
up to time $s$ with two
$\lambda^2$ power from the non-skeleton indices
(the word {\it last} indicates that the last $\lambda$ power was collected
at the last collision).

\medskip

$\bullet$  We observe a repeated skeleton label, i.e., a recollision
or a nest. The corresponding wave functions are denoted by
$\psi_{*s,k}^{(\leq 1),rec}, \psi_{*s,k}^{(\leq 1),nest}$.

\medskip

$\bullet$  We observe three identical
potential labels, i.e., a triple collision.
The corresponding wave functions are denoted by
$\psi_{*s,k}^{(\leq 1),tri}$.

\medskip

Finally, $\psi_{t,k}^{(\leq 1),nr}$ denotes the sum of non-repetitive
elementary wavefunctions without truncation (i.e. up to time $t$)
 with at most one
$\lambda^2$ power from the non-skeleton indices or $\theta$'s
and with $k$ skeleton indices. In particular, the non-repetition
wave functions without gates or $\vartheta$ (denoted by $\psi_{t,k}^{nr}$
above) contribute to this sum. For convenience, we will rename
$\psi_{t,k}^{(0),nr}:=\psi_{t,k}^{nr}$ indicating explicitly the
number of $\lambda^2$-powers from non-skeleton labels, $r=0$.

\medskip

This stopping rule gives rise to the following representation
that is intuitively clear.
The detailed  proof is given in Proposition 3.2 of \cite{ESYII}.

\begin{proposition}\label{prop:duh}[Duhamel formula] For any $K\ge 1$
we have
\be
\psi_t=e^{-itH}\psi_0 = \sum_{k=0}^{K-1} \psi_{t,k}^{(\leq 1),nr}\qquad
\qquad \qquad\qquad \qquad \qquad \qquad \qquad \qquad
\label{eq:duha}
\ee
$$
\qquad \qquad -i \int_0^t\rd s \; e^{-i(t-s)H}\Bigg\{
\psi_{*s, K}^{(\leq 1),nr}+\sum_{k=0}^{K}
\Big( \psi_{*s,k}^{(2),last} + \psi_{*s,k}^{(\leq 1),rec}+
\psi_{*s,k}^{(\leq 1),nest}+\psi_{*s,k}^{(\leq 1),tri}\Big)
\Bigg\} \; .
$$
\end{proposition}

\section{Graphical representation}
\setcounter{equation}{0}

For Theorems~\ref{7.1} and \ref{thm:L2}, we need to compute
expectations of quadratic functionals of elementary wavefunctions
$\psi_{t,\tgamma}$. This computation is organized by 
 Feynman graphs. In the continuum
model, the computation was shown for the non-repetition
terms in details in Section 6 of \cite{ESYI} and
the precise definitions of the Feynman graphs were given in Section 7
of \cite{ESYI}. For the lattice model, 
the Feynman graphs are very similar but somewhat simpler
because the ultraviolet regime is absent and 
the Boltzmann collision kernel  \eqref{def:sigma}
depends only on the energy.
 We will not repeat
the complete arguments here, but we introduce the necessary
modifications for the lattice case.

\subsection{Circle graphs and their values}

We start with an oriented circle graph with
 two distinguished vertices denoted by $0$, $0^*$.
The number of vertices is $N$.
The vertex set is $\cV$, the set of oriented edges  is $\cL(\cV)$.
For $v\in \cV$ we use the notation $v-1$ and  $v+1$ for
the vertex right before and after $v$ in the circular ordering.
We also denote $e_{v-}=(v-1,v)$ and $e_{v+}=(v, v+1)$ the edge
right before and after the vertex $v$, respectively.
For each $e\in\cL(\cC_N)$ we introduce a momentum $w_e$ and
a real number $\alpha_e$ associated to this edge.
The collection of momenta is denoted by $\bw=\{ w_e\; : \;
e\in \cL(\cV)\}$ and $\rd\bw =\otimes_e \; \rd w_e$ is the
Lebesgue measure. The notation $v\sim e$ indicates that
the edge $e$ and adjacent to the vertex $v$.

Let $\bP = \{ P_\mu \; : \; \mu \in I\}$ 
be a partition of the set $\cV \setminus\{0, 0^*\}$
$$
        \cV\setminus\{0, 0^*\} = \bigcup_{\mu\in I}P_\mu \; ,
$$
(all $P_\mu$ are pairwise disjoint and non-empty),
 where $I=I(\bP)$ is the index set to label the nonempty sets in
the partition. Let $m(\bP):= |I(\bP)|$. We will call the sets $P_\mu$
$\bP$-{\it lumps} or simply {\it lumps}.
Two elements of $v,v'\in \cV$ are called $\bP$-equivalent
if $v$, $v'$ belong to the same lump of $\bP$.

We will assign a variable (called {\it auxiliary momenta}),
 $u_\mu\in \Tor^d$, $\mu \in I(\bP)$,
to each lump.  We will always assume that the
auxiliary momenta add up to 0
\be
    \sum_{\mu\in I(\bP)} u_\mu =0 \; .
\label{sumu}
\ee

 The set of all partitions of the vertex
set $\cV \setminus\{0, 0^*\}$  is denoted by $\cP_\cV$.
For any $P\subset\cV$ we
let
$$
L_+(P) : = \{ (v,v+1)\in\cL (\cV)\; : \; v+1\not\in P, \; v\in P\}
$$
denote the set of  edges that go out of $P$, with respect
to the orientation of the circle graph,
 and similarly
 $L_-(P)$ denote the set of 
 edges that go into $P$. We set $L(P):= L_+(P) \cup L_-(P)$.

For any $\xi\in \bR^d$, 
define the following product of delta functions
 \be
   \Delta_\xi(\bP, \bw, \bu): = \delta \Big( \xi+
 \sum_{e\in L_\pm( \{ 0^*\} )} \pm w_e\Big)
\prod_{\mu\in I(\bP)} \delta\Big( \sum_{e\in L_\pm(P_\mu)}
    \pm w_e - u_\mu\Big) \; ,
\label{def:Delta}
\ee
where $\bu: = \{ u_\mu\; : \; \mu\in I(\bP)\} \in \Tor^d$ is
a set of auxiliary momenta.
The sign $\pm$ indicates that momenta $w_e$ is added or subtracted
depending whether the edge $e$ is outgoing or incoming.
We also recall that all momentum variables live on
the torus, in particular  addition is also defined on $\Tor^d$.
 The
dependence on $\xi$ will be mostly omitted from the notation;
$\Delta=\Delta_\xi$. All 
estimates will be uniform in $\xi$.

Summing up all arguments of these delta functions and
using (\ref{sumu}) we see that these delta functions
force the two momenta corresponding to the
two edges adjacent to $0$ to differ by $\xi$: $w_e-w_{e'}=\xi$
for $e\in L_+(\{ 0\})$, $e'\in L_-(\{ 0\})$. This will
correspond to the momentum shift in the Wigner transform
\eqref{FW}.

With these notations, we define for any $\bP\in \cP_\cV$ and $\eta>0$ the 
{\bf $E$-value of the partition}
\be
    E(\bP, \bu,\balpha): =\lambda^{N-2}
    \int \prod_{e\in \cL(\cV)}   \frac{\rd w_e}{|\alpha_e- \om(w_e)+i\eta|}
   \; \Delta(\bP, \bw, \bu) |\wh\psi_0(w_{e_{0+}})| 
|\wh\psi_0(w_{e_{0-}})|  \; .
\label{def:E}
\ee
The prefactor $\lambda^{N-2}$ is due to the fact that in
the applications all but two distinguished  
vertices, $0, 0^*$, will carry a factor $\lambda$.
The $E$-value depends on the parameters $\lambda$ and $\eta$
but this is omitted from the notation. In the applications, 
the regularization  will be mostly chosen
as $\eta= \lambda^{2+\kappa}$.

We will also need a slight modification of this definition, indicated
by a lower star in the notation:
\be
    E_{*}(\bP, \bu,\balpha): =\lambda^{N-2} \!\!
    \int \!\! \prod_{e\in \cL(\cV)} \rd w_e \!\! \prod_{e\in \cL(\cV)\atop
      e\not\in L( \{ 0^*\} )}   \frac{1}{|\alpha_e- \om(w_e)+i\eta|}
    \; \Delta(\bP, \bw, \bu)  |\wh\psi_0(w_{e_{0+}})| 
|\wh\psi_0(w_{e_{0-}})| \;.
\label{def:E*}
\ee
The only difference is that the denominators carrying the momenta associated
to edges that are adjacent to $0^*$ are not present in $E_{*\eta}$.
We will call $E_{*}$ the {\bf truncation} of $E$.
We will see that Feynman diagrams arising
from the perturbation expansion can naturally be
estimated by quantities of the form (\ref{def:E}) or (\ref{def:E*}).

The formulas \eqref{def:E} and \eqref{def:E*} are the
lattice analogues of (7.5) and (7.6) of \cite{ESYI} in
the continuous model, but the momentum cutoffs, the 
polynomially decaying factors for the special set $\cG$
 and the non-trivial collision kernel $\wh B$  are absent.

\bigskip

Following the continuum model,
we also define four operations on a partition given on the vertex set of a
circle graph and we estimate how the the $E$-value changes. 
The estimates are somewhat simpler for the lattice case, 
so here we just summarize the results and prove 
an additional estimate (see \eqref{eq:tru} below). For the details see 
 Lemma 9.5 of \cite{ESYI} and Appendix C of \cite{ESYII}.

\begin{lemma} [Operation I. Breaking up lumps] \label{lemma:breakup}
Given $\bP=\{P_\mu\; :\; \mu \in I(\bP)\}
\in \cP_\cV$, we break up one of the lumps into two smaller nonempty lumps;
 $P_\nu = P_{\nu'} \cup P_{\nu''}$ with
$P_{\nu'}\cap P_{\nu''}=\emptyset$. Let
$\bP^*=\{ P_{\nu'},  P_{\nu''},
 P_\mu  \;  : \; \mu\in I(\bP)\setminus \{ \nu\}\}$ denote
the new partition. Then
\be
     E_{(*)}( \bP, \bu, \balpha) \leq \int \rd r \; E_{(*)}
    (\bP^*, \bu^*(r,\nu),
     \balpha)\; ,
\label{eq:OpI}
\ee
where the new set of momenta $ \bu^*=\bu^*(r,\nu)$ is given
by $u^*_\mu:=u_\mu$, $\mu\in I(\bP)\setminus \{\nu\}$
and $u^*_{\nu'}= u_\nu-r$, $u^*_{\nu''} = r$. In particular,
\be
\sup_\bu  E_{(*)}( \bP, \bu, \balpha)\leq
 \sup_\bu E_{(*)}( \bP^*, \bu, \balpha) \;. \qquad \Box
\label{opIest}
\ee
\end{lemma}
The notation $E_{(*)}$ simultaneously refers to $E$ and  $E_*$, i.e.
to formulas with and
without truncation.

\begin{lemma} [Operation II. Removal of a single vertex]
\label{lemma:remove}
Let $v\in \cV \setminus\{0, 0^*\}$ be a vertex and let $\bP\in \cP_\cV$
such that $P_\sigma = \{ v\}$ for some $\sigma\in  I(\bP)$,
i.e. the single element set
$\{ v \}$ is a lump.
 Define $\cV^*:=\cV\setminus\{ v\}$, $\cL(\cV^*):=\cL(\cV)
\cup \{ (v-1, v+1)\} \setminus \{
(v-1,v), (v, v+1)\}$, i.e. we simply remove the vertex $v$ from the 
graph and connect the vertices $v-1, v+1$. Let 
$\bP^* :=\bP\setminus \{ \; \{ v \} \; \} \in \cP_{\cV^*}$, then
\be
   \sup_\bu E_{(*)}(\bP, \bu,\balpha) \leq \lambda\eta^{-1}\sup_{\bu^*}
 E_{(*)}
    (\bP^*, \bu^*, \balpha) \; .
\label{eq:nontr}
\ee

 Furthermore, if $v$ and $0^*$ are neighbors
in the graph, then we have the following stronger estimate for the truncated
value:
\be
   \sup_\bu  E_{*}(\bP, \bu,\balpha) \leq  \lambda \; (3d+ |\alpha_e|)
\sup_{\bu^*} E
    (\bP^*, \bu^*, \balpha) \; ,
\label{eq:tru}
\ee
where $e$ is the edge connecting $0^*$ with its  neighbor other than $v$.

Finally, if both neighbors of $0^*$, $v\neq v'$, form single
lumps in $\bP$, then both of these lumps can be simultaneously
removed to obtain a partition $\bP^*: = \bP\setminus\{ \{ v\}, \{ v'\} \}$
with the estimate
\be
   \sup_\bu  E_{*}(\bP, \bu,\balpha) \leq  \lambda^2 \; \sup_{\bu^*} E
    (\bP^*, \bu^*, \balpha) \; .
\label{eq:2tru}
\ee
\end{lemma}

{\it Proof.} Estimates \eqref{eq:nontr} and \eqref{eq:2tru} were
proven in Appendix C of \cite{ESYII}.
For the proof of \eqref{eq:tru}, notice that if
 $v+1$ or $v-1$ is $0^*$,
then the denominator
$|\alpha_{e} - \om(w_{e})+i\eta|^{-1}$ is not present in
the truncated value
for $e=(v-1,v)$ or $e=(v, v+1)$. Therefore the proof of  \eqref{eq:nontr}
can be repeated without paying the $\eta^{-1}$ price.
The extra factor estimates the missing denominator $|\a_e -\om(w_e)+i\eta|\leq
|\a_e| + 3d$.  $ \;\;\;\Box$
\bigskip

\begin{lemma}
[Operation III. Removal of   half of a gate]
\label{lemma:removehalfgate}
Let $v, v+1\in \cV \setminus\{0, 0^*\}$ form a gate
in  a partition $\bP\in \cP_\cV$, i.e.
$P_\sigma = \{ v, v+1\}$ for some $\sigma\in  I(\bP)$.
 Define $\cV^*:=\cV\setminus\{ v+1\}$,
$\cL(\cV^*):=\cL(\cV)\cup \{ (v, v+2)\} \setminus \{
 (v, v+1), (v+1, v+2)\}$, i.e. we simply remove the vertex $ v+1$
 from the circle graph with the adjacent edges
and add a new edge between the vertices $v, v+2$.
Let $\bP^*\in \cP_{\cV^*}$ be $\bP$
after  simply replacing  the lump $\{ v,  v+1\} $ with $\{v\}$. Then
$$
    E_{(*)}(\bP, \bu,\balpha) \leq \lambda |\log\eta| \; E_{(*)}
    (\bP^*, \bu, \balpha) \; . \qquad \Box
$$
\end{lemma}

\begin{lemma}
[Operation IV.: Removal of a gate] \label{lemma:removegate}
Let $v, v+1\in \cV \setminus\{0, 0^*\}$ form a gate in
 $\bP\in \cP_\cV$, i.e.
$P_\sigma = \{ v, v+1\}$ for some $\sigma\in  I(\bP)$.
 Define $\cV^*:=\cV\setminus\{ v,v+1\}$,
$\cL(\cV^*):=\cL(\cV)\cup \{ (v-1, v+2)\} \setminus \{
(v-1,v), (v, v+1), (v+1, v+2)\}$, i.e. we simply remove the gate.
Let $\bP^*\in \cP_{\cV^*}$ be $\bP$
after  removing the lump $\{ v, v+1\} $. Then
$$
   \sup_\bu E_{(*)}(\bP, \bu,\balpha) \leq \lambda^2\eta^{-1}|\log\eta| \;
  \sup_{\bu^*} E_{(*)}
    (\bP^*, \bu^*, \balpha) \; . \qquad  \Box
$$
\end{lemma}

\bigskip
\subsection{Feynman graphs}\label{sec:FG}

Feynman graphs are special circle graphs that naturally 
arise in the perturbation expansion. 
Consider the cyclically ordered set $\cV_{n,n'}: =
\{0, 1, 2, \ldots, n, 0^*, \tilde n', \wt{n'-1},
 \ldots ,\tilde 1 \}$ and view this as the
 vertex set of an oriented circle graph on $N=n+n'+2$
 vertices. We set $I_n:=\{ 1, 2, \ldots n\}$
and $\wt I_{n'} :=\{ \wt 1, \wt 2, \ldots, \wt n'\}$
and the vertex set can be identified with $\cV_{n,n'}= I_n\cup \wt I_{n'}
\cup\{ 0, 0^*\}$.
The set of edges $\cL(\cV_{n,n'})$ is partitioned into $\cL(\cV_{n,n'}) 
= \cL\cup
\wt\cL$ such that $\cL$ contains the edges between $I_n\cup \{0,0^*\}$
and $\wt\cL$ contains the edges between $\wt I_{n'}\cup \{0,0^*\}$.

Let $\cP_{n,n'}$ be the set of partitions $\bP$
of the set $I_n \cup \wt I_{n'}$. Let $G=G(\bP)$ be the set of edges that
enter a  single lump
 and let $g(\bP):= |G(\bP)|$ be its cardinality.
In case of $n=n'$, we will use the shorter notation 
$\cV_n=\cV_{n,n}$, $\cP_n=\cP_{n,n}$
etc. In our applications we always have $|n-n'|\leq 4$ and $n,n'\leq K$.

Let $Q:\bR^d\to \bC$ be an arbitrary function that will
represent the momentum dependence of the observable. For
convenience, we can assume that $\| Q\|_\infty \leq 1$.
For $\a, \beta\in\bR$, $\bP\in \cP_{n,n'}$,  and we define
\be
    V_\xi(\bP,\a,\beta): =\lambda^{n+n'+g(\bP)}
\int \rd\bw 
    \prod_{e\in\cL} \frac{1}{\alpha- \ov\om(w_e) - i\eta} \prod_{e\in\wt\cL}
        \frac{1}{\beta- {\om}(w_e) + i\eta}
\label{def:Vlong}
\ee
$$
    \times  \Delta_\xi(\bP, \bw, \bu\equiv 0)
  \prod_{e \in \cL\cap G}[- \ov{\theta(w_e)}]
 \prod_{e \in \wt\cL\cap G}[- \theta(w_e)] \;
\ov{\wh\psi(w_{e_{0+}})}\wh\psi(w_{e_{0-}}) Q\Big[ \frac{1}{2}
  (w_{e_{0^*-}} +w_{e_{0^*+}}) \, \Big]\;,
$$
with $\bw: =\{ w_e\; : \; e\in \cL\cup\wt\cL\}$. The subscript $\xi$
will mostly be omitted.

The truncated version, $V_{*\xi}(\bP,\a,\beta)$, is defined analogously
but those $\alpha$ and $\beta$ denominators are
removed that correspond to $e \in L( \{ 0^*\} )$.
We set
\be
        V_{(*)}(\bP): = \frac{e^{2t\eta}}{(2\pi)^2} \iint_{-4d}^{4d}
        \rd\alpha\rd\beta \; e^{it(\alpha-\beta)} V_{(*)}(\bP,\a,\beta)
\label{def:Vshort}
\ee
and
$$
        E_{(*)}(\bP,\bu):=\frac{e^{2t\eta}}{(2\pi)^2}
        \iint_{-4d}^{4d} \rd\alpha\rd\beta \;  E_{(*)}(\bP,\bu,\balpha) \; ,
$$
where $\balpha$ in $ E_{(*)}(\bP,\bu,\balpha)$
is defined as $\alpha_e = \alpha$ for $e\in\cL$ and
 $\alpha_e:=\beta$ for $e\in\wt\cL$.
 We will call these numbers the $V$-value and
$E$-value of the partition $\bP$, or sometimes, 
of the corresponding Feynman graph. Strictly speaking, they
depend on the vector $\xi$ and on the function $Q$ as well;
when necessary, we will use the notations $V_{(*)\xi} (\bP,  Q)$,
$E_{(*)\xi} (\bP, \bu, Q)$, etc.

Clearly
\be
        |V_{(*)}(\bP)|\leq (C\lambda )^{g(\bP)}\;
        E_{(*)}(\bP,\bu\equiv 0) \; .
\label{eq:VleqE}
\ee
If $\bu\equiv 0$, 
we will use the notation $E_{(*)}(\bP):=E_{(*)}(\bP, \bu \equiv 0)$.

As we will see, in the graphical representation of the Duhamel expansion
what
we really need  is
\be
        V_{(*)}^\circ(\bP): = \frac{e^{2t\eta}}{(2\pi)^2} \iint_\bR
        \rd\alpha\rd\beta \; e^{it(\alpha-\beta)} V_{(*)}(\bP,\a,\beta)\; ,
\label{def:Vshortcirc}
\ee
i.e. a version of $ V_{(*)}(\bP)$ with unrestricted $\rd\a \rd \beta$
integration (the circle superscript will refer to the unrestricted
integration). However, the difference is negligible even after
summing them up for all partitions. 

\begin{lemma}\label{lemma:VV} Assuming that $\eta\ge \lambda^{2+4\kappa}$
and $1\leq n+n' \leq 2K$,
we have
\be
   \sum_{\bP\in \cP_{n,n'}}  \Big| V_{(*)}(\bP)-  V_{(*)}^\circ(\bP)
\Big| 
  = O(\lambda^{\frac{1}{2}(n+n')}) \; .
\label{def:EMout}
\ee
The same result holds if $V_{(*)}(\bP)$ were defined
by restricting the $\alpha,\beta$-integral to
any domain that contains
 $[-4d, 4d]\times [-4d, 4d]$.
 \end{lemma}

{\it Proof.} Outside of the regime $|\a|, |\beta|\leq 4d$, 
at least either the
 denominators  with $\alpha$ or with
 $\beta$ in (\ref{def:Vlong}) are uniformly bounded
since $|\om(p)|\leq 2d+1$ for small $\lambda$. The other denominators
can be integrated out  at the expense of
 $(C|\log \lambda|)^{\max(n,n')+1}$
by using (\ref{eq:logest}).
The contribution to
 $V_{(*)}(\bP)$ from the complement of $|\alpha|, |\beta|\leq 4d$ is
therefore bounded by
$$
  \Bigg| V_{(*)}(\bP)-  V_{(*)}^\circ(\bP)\Bigg|\leq
   \lambda^{n+n'+g(\bP)}(C|\log\lambda|)^{\max(n,n')+1}\; ,
$$
if $\lambda$ is small. Since the total number of partitions,
$|\cP_{n,n'}|$, is bounded by  $(n+n')^{n+n'}$ and $n+n'\leq 2K =
O(\lambda^{-\kappa-\delta})$, we obtain  \eqref{def:EMout}. $\;\;\Box$

\medskip

Sometimes we will use a numerical labelling of the edges, see
Fig,~\ref{fig:relab}.
In this case, we label the edge between $(j-1, j)$
 by $e_j$, the edge between $(\wt j, \wt{j-1})$
by $e_{\tilde j}$. At the special vertices $0, 0^*$ we
denote the edges as follows: $e_{n+1}:=(n, 0^*)$,
$e_{\wt{n'+1}}:=(0^*, \tilde n')$, $e_1=(0,1)$
 and $e_{\wt 1}:=(\wt 1, 0)$.
Therefore the
edge set $\cL=\cL(\cV_{n,n'})$ is identified with the index
set $I_{n+1} \cup \wt I_{n'+1}$ and
we set $p_j: = w_{e_j}$, $\tp_j: = w_{e_{\wt j}}$.
These two notations will sometimes  be used in parallel.
Note that
 we always
have
\be
    p_1- \tp_1=\xi\; .
\label{peqp}
\ee

\bef\bec
\epsfig{file=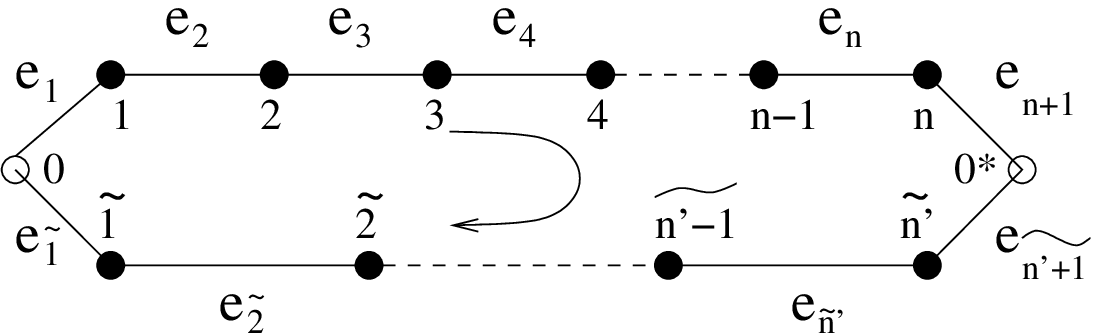,scale=.8}
\eec
\caption{Vertex and edge labels}\label{fig:relab}
\eef

\subsection{Non-repetition Feynman graphs}\label{sec:nrF}

Let $\cA_n$ be the set of partitions of $I_n:=
\{ 1,2, \ldots, n\}$,
i.e. $\bA =\{ A_\mu \; : \; \mu\in I(\bA) \} \in\cA_n$
if $\cup_{\mu\in I(\bA)} A_\mu =I_n$
 and the elements of $\bA$ are
disjoint.  The sets in the partition are labelled
by the index set $I(\bA)$ and let $m(\bA)=|I(\bA)|$
denote the number of elements in
$\bA$. The elements of the partition $\bA$ will also
be called {\it lumps}. A lump is {\it trivial} if it
has only one element. The trivial partition,
where every lump is trivial, is denoted by $\bA_0$.

A partition $\bP\in \cP_n$ of $I_n\cup\wt I_n$ is called {\bf even} if
for any $P_\mu\in \bP$ we have $|P_\mu\cap I_n|=|P_\mu\cap
\wt I_n|$. In particular, in an even
partition there are no single lumps,
 $G(\bP)=\emptyset$.

Let $\fS_n$ be the set of permutations on $I_n$ and
let $id$ be the identity permutation.
Note that $\bA\in \cA_n$ and $\sigma\in \fS_n$,
uniquely determine an even partition in $\bP(\bA, \sigma) \in \cP_n$,
by $I(\bP):=I(\bA)$  and $P_\mu : = A_\mu \cup \sigma(A_\mu)$.
Conversely, given an even partition $\bP\in \cP_n$,
we can define its projection onto $I_n$, $\bA:=\pi(\bP)\in \cA_n$,
 by $I(\bA):=I(\bP)$ and $A_\mu: = P_\mu\cap I_n$.
We let
$$
        \fS_n(\bP): = \{ \sigma\in \fS_n\; : \; \bP( \pi(\bP),\sigma)= \bP\}
$$
be the set of permutations that are {\bf compatible} with
a given even partition $\bP$. In other words, $\sigma\in \fS_n(\bP)$
if for each $i\in I_n$  the pair $(i, \sigma(i))$ belongs
to the same $\bP$-lump.
Clearly
\be
        |\fS_n(\bP)|= \prod_{\mu\in I(\bP)} \Big( \frac{|P_\mu|}{2}\Big)!
        = \prod_{\mu\in I(\pi(\bA))} |A_\mu|\, ! \; .
\label{eq:Pk}
\ee
We will use the notation
\be
V_{(*)}(\bA, \sigma, Q): = V_{(*)}(\bP(\bA, \sigma), Q)
\label{def:VsigmaA}
\ee
and similarly for $E_{(*)g}$ and $V^\circ_{(*)}$. 
In the proofs, $Q$ will be omitted. We also introduce
\be
       c(\bA): =\prod_{\nu\in I(\bA)} c(|A_\nu|) \; ,
\ee
where $c(n)$ are the coefficients of the connected graph formula
defined in \eqref{def:cn}.
With these notations we can state the representation
of the non-repetition terms as a summation over
Feynman diagrams.

\begin{proposition}\label{prop:lump} With $Q\equiv 1$ and  $\xi=0$ we have
\be
   \bE \|\psi_{t,k}^{nr}\|^2
    = \sum_{\sigma\in\fS_k}
    \sum_{\bA\in\cA_k}c(\bA) V_{\xi=0}^\circ(\bA, \sigma, Q\equiv 1)
\label{eq:psicM}
\ee
and with $Q_\xi(v): = \wh\cO(\xi, v)$  we have
\be
\bE \langle \wh \cO, 
\wh W^\e_{\psi_{t,k}^{  nr}}\rangle
    = \sum_{\sigma\in\fS_k}
    \sum_{\bA\in\cA_k} c(\bA) \int_{(2\Tor/\e)^d} \rd\xi \; 
V_{\e\xi}^\circ(\bA, \sigma, Q_\xi)   \; .
\label{eq:WcM}
\ee
\end{proposition}

The proof is essentially given in 
Section 6 and Proposition 7.2 of \cite{ESYI}
with a few minor modifications. The restriction to the
finite box $\Lambda_L$ is absent.
The expectation
value of the potentials in the expansion of  $\bE \|\psi_{t,k}^{nr}\|^2$
is given by
\be
    \bE \prod_{j=1}^k  \overline{\wh V_{\gamma_j}(p_{j+1}-p_{j})}
     \wh V_{\gamma_j}( \tp_{\sigma(j)+1}-\tp_{\sigma(j)})=
    \sum_{\gamma_1, \ldots, \gamma_k\in\bZ^d\atop \gamma_i\neq\gamma_j}
   \prod_{j=1}^k e^{2\pi i\gamma_j(p_{j+1}-p_{j}-
   (\tp_{\sigma(j)+1}-\tp_{\sigma(j)}))}\; .
\label{eq:treegraph}\ee
instead of (6.4) of \cite{ESYI}. The following identity was proven
in Lemma 6.1 \cite{ESYI}:
\begin{lemma}
For any fixed $k$,
\be
    \sum_{\gamma_1, \ldots ,\gamma_k\in \bZ^d\atop \gamma_i\neq\gamma_j}
    \prod_{j=1}^k e^{2\pi i q_j\gamma_j} = \sum_{\bA \in\cA_k}
    \prod_{\nu\in I(\bA)} c(|A_\nu|)\delta\Big(\sum_{\ell \in A_\nu}
     q_\ell\Big)
\label{eq:conngr}
\ee
with
\be
     c(n): = \sum_{\Gamma\subset K_n\atop\Gamma \; connected}
     (-1)^{|\Gamma|}\; ,
\label{def:cn}
\ee
where $K_n$ denotes the complete graph on $n$ vertices
and $|\Gamma|$ denotes the number of edges in
the subgraph $\Gamma$. 
The following estimate holds
\be
     |c(n)|\leq n^{n-2} \; . \qquad \Box
\label{eq:ajest}\ee
\end{lemma}

\section{Proof of Theorem~\ref{thm:L2}}
\label{sec:nonrep}
\setcounter{equation}{0}

We recall that the $E$- and $V$-values of the partitions depend on
the parameters $\lambda, t, \xi$ and  $k$; a fact
that is not explicitly
included in the notation. In Sections \ref{sec:nonrep} and
\ref{sec:error}  we will always assume the following
relations
\be
    \eta=\lambda^{2+\kappa}, \quad
 t= \lambda^{-2-\kappa} T,\quad T\in [0,T_0],\quad
   K=[\lambda^{-\delta}(\lambda^2t)],
 \quad k< K,
\label{param}
\ee
with a sufficiently small positive $\delta>0$  that is
independent of $\lambda$ but depends on $\kappa$.
All estimates will be uniform in $\xi$ and in $T\in [0,T_0]$.

\subsection{Estimates on graphs with high degree}

We recall the key definition of the {\it degree} of a permutation
$\sigma\in \fS_k$ from Definition 8.3 of \cite{ESYI}.
Let $\sigma$ act on $I_k=\{1, 2, \ldots ,k\}$ and let $\tsi$
be its extension to $\{0, 1, \ldots , k+1\}$, by $\tsi(0)=0$,
$\tsi(k+1)=k+1$, otherwise $\tsi(i)=\sigma(i)$. An 
index $j\in I:=\{1, 2, \ldots k\}$ is
called ladder index of $\sigma$ if 
$\sigma(j)-1\in \{ \tsi(j-1), \tsi(j+1)\}$.
 Let $I_\ell$ be the
set of ladder indices and $\ell=\ell(\sigma):=|I_\ell|$.
Finally, the degree of $\sigma$ is defined as
$$
   \mbox{deg} (\sigma): = k-\ell(\sigma) \; .
$$

Starting with \eqref{eq:psicM}, we
notice that the graph with  the trivial partition $\bA_0$ 
and with the identity permutation on $I_k$ 
gives the main term in Theorem~\ref{thm:L2}
since
$$
   V_\lambda(t, k) = V^\circ_{\xi=0}(\bA_0, id) \; , \qquad c(\bA_0)=1\; .
$$
This graph is  called the {\bf ladder graph}
(Fig.~\ref{fig:ladd}).

\bef\bec
\epsfig{file=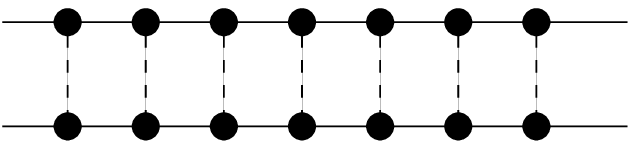, scale=.8}
\eec
\caption{Ladder graph}\label{fig:ladd}
\eef

To prove that all other graphs are negligible,
we first replace $V^\circ(\cdots)$ with $V(\cdots)$; the error
is negligible by Lemma \ref{lemma:VV}. 
We  then first estimate $ V(\bA, \sigma)$ 
for the trivial partition  $\bA=\bA_0$, where every lump has one element.
Since $|V(\bA, \sigma)|\le  E(\bA, \sigma, \bu\equiv 0)$, the 
following bound is the key estimate:

\begin{theorem}\label{thm:Vsi} Assume \eqref{param} with
$\kappa < \frac{1}{64}$ and let $\sigma\in\fS_k$.
 Then the $E$-value of the graph of the trivial partition
with permutation $\sigma$ is estimated by
\be
    \sup_\bu E_{(*)}(\bA_0, \sigma,\bu) \leq
    C \Big(\lambda^{\frac{1}{16}-4\kappa} 
\Big)^{\de(\sigma)}|\log\lambda|^2
\label{eq:Vsi}
\ee
if $\lambda\ll 1$.
\end{theorem}
The proof is essentially given in Section 10 of \cite{ESYI} with
a few modifications that we will explain in Section \ref{sec:mainproof} below.
This theorem is complemented by the following combinatorial lemma
that was proved in \cite{ESYI} (Lemma 8.5):

\begin{lemma}\label{lemma:comb}
Let $k \leq O(\lambda^{-\kappa-\delta})$, $D\ge0$ integer, and let
 $\gamma>\kappa+\delta$ be fixed.
 Then
\be
    \sum_{\sigma\in\fS_k\atop \dee(\sigma)\ge D} \lambda^{\gamma 
  \,\de(\sigma)} \leq
    O\Big( \lambda^{D(\gamma-\kappa-\delta)}\Big)
\label{eq:lambdasum}
\ee
for $\lambda\ll 1$. $\;\;\Box$
\end{lemma}

Since ${\rm deg}(\sigma)\ge 2$ if $\sigma\neq id$,
from Theorem \ref{thm:Vsi} and Lemma \ref{lemma:comb} we
immediately obtain:

\begin{proposition}\label{prop:Vsi} Assuming \eqref{param} with
$\kappa < \frac{1}{80}$ we have
\be
    \sum_{\sigma\in \fS_k\atop\sigma\neq id}  |V(\bA_0, \sigma)| \leq
    O\Big(\lambda^{\frac{1}{8}-10\kappa-O(\delta)}\Big)   
 \;\qquad
\label{eq:Vsisum}
\ee
 for $\lambda\ll 1$.
$\Box$
\end{proposition}

For the general case  $\bA\neq\bA_0$, we need to recall
the notion of {\it joint degree} of a permutation $\sigma\in \fS_k$
and a partition $\bA\in\cA_k$
from Definition 9.1 of \cite{ESYI}.
Let $S(\bA)$ be the union of non-trivial lumps in $\bA$, and let
$s(\bA):= |S(\bA)|$ be its cardinality. The joint degree of a pair
$(\sigma,\bA)$ is given by
$$
   q(\sigma, \bA):= \max \Big\{ \mbox{deg}(\sigma), \;
 \frac{1}{2} s(\bA)\Big\}\; .
$$
The following statement was proved in Lemma 9.3 \cite{ESYI} 
(recall the definition of the compatibility and that of
the projection $\pi(\bP)$ 
from Section \ref{sec:nrF}).

\begin{lemma}\label{lemma:ds}
For any even partition $\bP\in \cP_k$
there exists a compatible permutation $\wh\sigma=\wh\sigma(\bP)\in \fS_k(\bP)$
such that
\be
     {\rm deg}(\wh\sigma)\ge \frac{1}{2} s(\pi(\bP)) \; .  \qquad \Box
\label{eq:ds}
\ee
\end{lemma}

The following corollary shows that the estimate
of a general partition can be reduced to that of
a trivial partition with the help of Lemma \ref{lemma:ds}.
The proof is somewhat simpler
than Corollary 9.4 of \cite{ESYI} 
in the continuous case.

\begin{corollary}\label{cor:jointdeg}
Given $\sigma\in \fS_k$ and $\bA\in \cA_k$,  we have, for $\kappa<
\frac{1}{64}$
\be
     \sup_{ \bu} E_{(*)}
   (\bA, \sigma,\bu) \leq C|\log \lambda|^2
   \Big(\lambda^{\frac{1}{16}-4\kappa} 
     \Big)^{q(\bA, \sigma)}  \; .
\label{eq:jointdeg}
\ee
\end{corollary}

{\it Proof of Corollary \ref{cor:jointdeg}.}   We define
a permutation $\s^*:=\s^*(\bA, \s)$ as $\s^*:=\s$ if
 ${\rm deg}(\sigma)\ge \frac{1}{2} s(\bA)$,
and $\sigma^* := \wh\s(\bP(\bA, \s))$ otherwise.
By Lemma \ref{lemma:ds} we have  ${\rm deg}(\sigma^*)= q(\bA, \s)$.
Clearly $\bP(\bA, \sigma) = \bP(\bA, \sigma^*)$, in particular,
$E_{(*)}(\bA, \sigma,\bu) = E_{(*)}(\bA, \sigma^*,\bu)$.
By Operation I. we can artificially break
up all non-trivial lumps in $\bA$
 and use the auxiliary momenta to account for
the additional Kirchoff rules. Using the estimate 
\eqref{eq:OpI} and that $\int \rd r =1$, we immediately see
that
$$
 \sup_{ \bu} E_{(*)}   (\bA, \sigma^*,\bu) \leq
\sup_{ \bu} E_{(*)}   (\bA_0, \sigma^*,\bu) \; ,
$$
and Theorem \ref{thm:Vsi} completes the proof. $\;\;\Box$.

Finally, we have the following bound on the summation
of general non-repetition graphs. 
The proof is the same as the proof of Proposition 9.2 of \cite{ESYI}
but  the estimate \eqref{eq:jointdeg} replaces  Corollary 9.4 
of \cite{ESYI} in the argument:

\begin{proposition}\label{prop:sumup} We assume \eqref{param}.
Let $D\ge0$ and $s\ge 2$ be
 given integers, let $q:= \max\{ D,
\frac{1}{2}s\}$. For any $\kappa< \frac{1}{144}$
 and $\delta\leq \delta(\kappa)$
sufficiently small,
we have
\be
\sum_{\sigma\in \fS_k\atop \dee (\sigma)\ge D}
   \sum_{\bA\in \cA_k\atop s(\bA)\ge s}
   \sup_{\bu} E_{(*)}
   (\bA, \sigma,\bu)|c(\bA)|\leq C \Big( \lambda^{ \frac{1}{16}-
9\kappa
 -O(\delta)}\Big)^q |\log \lambda|^2\; .  \;\; \Box   
\label{eq:sumup}
\ee
\end{proposition}

Finally, the proof of Theorem \ref{thm:L2} follows from 
Proposition \ref{prop:sumup} exactly as the proof of Theorem 5.2
in the continuous case explained in Section 7.3 of \cite{ESYI}
(Proposition \ref{prop:lump} replaces Proposition 7.2
of \cite{ESYI} and Lemma \ref{lemma:VV} replaces
Lemma 7.1 of \cite{ESYI}). $\;\;\Box$

\subsection{Proof of Theorem \ref{thm:Vsi}}\label{sec:mainproof}

The proof of Theorem \ref{thm:Vsi} follows the integration
scheme presented in Section 10 of \cite{ESYI} with a few modifications.
Apart from the finite momentum space, the finite cutoff in
the $\a,\beta$ variables and the simpler collision
kernel, the main difference is that the 
Two Denominator Lemma  is  weaker
in the lattice case (compare Lemma \ref{lemma:newrow}
below with Lemma 10.5 in \cite{ESYI}). This results in
weaker $\lambda$-exponents 
in the estimates and eventually a smaller threshold $\kappa_0$
in the main result.

We define
\be
   \tri q \tri : = \eta+\min \{ |q - \gamma^{(j)}|, \; j=1, 2, \ldots , 2^d\}
\label{tri}
\ee
with $\gamma^{(j)} = \frac{1}{2} a(j)$ and $a(j) = 
(a_1(j), a_2(j), \ldots , a_d(j))$
is the dyadic expansion of $j-1$. In other words, $\tri q\tri-\eta$ is
the minimal distance of $q$ from the critical points of $e(p)$ (measured on
the torus $\Tor^d$). This is not a norm, but it satisfies the
triangle inequality, $\tri p +q\tri \leq \tri p\tri+\tri q\tri$.

For any index set $I'\subset I:=\{ 1, 2,\ldots,
k+1\}$, any $|I'|\times(k+1)$
matrix $M$ and any vector $\fb=(b_1, b_2, \ldots , b_{k+1})\in \bR^{k+1}$,
we define
\bey
    E(I', M, \fb): &=& \lambda^{2k}\sup_{\tbu,v}
    \iint_{-4d}^{4d} \rd\alpha \rd\beta \sup_{p_j\; : \; j\not\in I'}
    \int \prod_{j\in I'}\rd p_j \; \frac{1}{\tri \fb \cdot \bp + v\tri} 
 \nonumber\\
&& 
\times\Bigg(\prod_{i\in I'}
    \frac{1}{|\alpha-\ov\om(p_i)-i\eta| \, \big|\beta -\om\big(
    \sum_{j=1}^{k+1} M_{ij}p_j +\tu_i\big)+i\eta\big|}\Bigg),
\label{eq:EE}
\eey
where $v\in \bR^3$ is 
an additional  dummy momentum and 
$$
   \fb \cdot \bp := b_1 p_1+ b_2p_2 +\ldots + b_{k+1}p_{k+1} \in \bR^3 \; .
$$
We will use the notation $E(I', M, \emptyset)$ defined exactly 
as \eqref{eq:EE} but without the factor $\tri \fb\cdot \bp + v\tri^{-1}$
in the integrand and without the supremum over $v$. We will
refer to this case as choosing the ``empty vector'' $\fb=\emptyset$.
Notice that
the definition \eqref{eq:EE}
 is somewhat simpler than the corresponding (10.14)
of \cite{ESYI}.

For any permutation $\sigma\in \fS_k$ we associate a
$(k+1)\times (k+1)$ matrix $M(\sigma)$ according to (8.7)
of \cite{ESYI}. This matrix encodes the momentum dependencies
in the delta function $\Delta$ in the $E$-value of 
the partition $\bP(\bA_0, \sigma)$. We have
\be
\sup_\bu E(\bA_0, \sigma,\bu) \leq  \| \wh \psi_0\|_\infty^2 \,
E( I, M(\sigma), \emptyset) \; .
\label{suppe}
\ee
The truncated version,  $E_*$,  can be estimated by
the untruncated one
$ E_{*}(\ldots )\leq C E(\ldots)$
 because in the regime
$|\alpha|, |\beta|\leq 4d$ every propagator is bounded from
below. 

\bigskip

An easy estimate is available for $ E_{(*)}( I, M(\sigma), \emptyset)$
by first separating all but one $\a$ and $\beta$ denominator
by a Schwarz inequality ($k\ge 1$):
\begin{align}
 E( I, M(\sigma), \emptyset) \leq &
\lambda^{2k}\sup_{\tbu}  \int \rd \bp 
    \iint_{-4d}^{4d} \frac{\rd\alpha \rd\beta}{ |\alpha-\ov\om(p_1)-i\eta|
   |\beta - \om (p_1+\tu_1) + i\eta|}  \label{schw}\\
& \times\Bigg[   \prod_{j=2}^{k+1} 
  \frac{1}{|\a- \ov\om(p_j) -i\eta|^2} + 
 \prod_{j=2}^{k+1} 
  \frac{1}{|\a- \om(q_j) +i\eta|^2} \Bigg] \;, \nonumber
\end{align}
where $q_i= \sum_j M_{ij}(\sigma) p_j + \tu_i$. The estimate for $E_*$ has one
less denominator.
Since $M(\sigma)$ is invertible with determinant $\pm 1$
(Proposition 8.2 of \cite{ESYI}), 
 the contributions
of the two terms in the square bracket are identical. To estimate
the first term, we can integrate out all $p_j$ variables, $j=2, 3,\ldots,
k+1$, by using \eqref{eq:ladderint}, yielding a
factor $(1+ C_0\lambda^{1-12\kappa})^k
= O(1)$ since $k\leq K \ll \lambda^{-1+12\kappa}$. 
After
integrating $\a,\beta$ and finally $p_1$, we obtain
\begin{align}
     \sup_{\sigma\in \fS_k}  E( I, M(\sigma), \emptyset) 
   \leq  & C|\log\lambda|^2
\label{etriv}\\
 \sup_{\sigma\in \fS_k}  E_*( I, M(\sigma), \emptyset) 
   \leq  & C\lambda^2|\log\lambda|^2 \; .
\nonumber
\end{align}
Note that the squared denominators can be integrated out in
arbitrary order, unlike in the continuum case (Section 10.1.2
of \cite{ESYI}).

\bigskip

To obtain a bound of the form $\lambda^{(const.) \de(\sigma)}$
for $E( I, M(\sigma), \emptyset) $, one has to gain a $\lambda$-power
from the non-ladder variables. This requires a successive integration
procedure described in Section 10.3 of \cite{ESYI}. We will
not repeat here the formal procedure, but just mention the basic idea.
The difficulty is that for a general $\sigma$ each variable $p_j$
appears in many denominators in \eqref{eq:EE}. To break this 
complicated dependence structure,
a set of carefully selected $\beta$-denominators in \eqref{eq:EE} are
estimated trivially by $\eta^{-1}$. Performing then the $p_j$-integrations
in a specific order,  the remaining
$\beta$ denominators can be integrated out (together with
the $\a$-denominators) without losing further $\eta^{-1}$ factors.
Each integration involves only two propagators
(Two Denominator Lemma \ref{lemma:newrow}).

In principle, if
the propagators corresponding to the ladder indices
are estimated  by a Schwarz inequality argument
\eqref{schw},
  one should gain $\lambda^2$ from each non-ladder
index. This would give a bound
of order $\lambda^{2 \, \de (\sigma)}$ in \eqref{eq:Vsi}, modulo
logarithmic corrections.
Unfortunately,  point singularities
may arise from the repeated application of Lemma \ref{lemma:newrow};
this necessitates the factor $\tri \fb\cdot \bp +v\tri^{-1}$
in \eqref{eq:EE}.
To avoid  the accumulation of point singularities 
 during the integration procedure,
we estimate trivially not only the $\beta$-denominators with 
ladder indices, but several other ones as well. 
This accounts for the smaller power in \eqref{eq:Vsi}.

The integration procedure removes the $\beta$-denominators
one by one.
 Each $\beta$-denominator in \eqref{eq:EE}
is labelled by an index $i\in I$ and we will treat them
in increasing order of the index $i$.
 The  index set $I=\{ 1, 2, \ldots , k+1\}$ is partitioned
into six disjoint subsets, 
\be
I= I_p\cup I_v \cup I_\ell\cup I_{cs}\cup I_{uc}
\cup I_{last}
\label{parti}
\ee
 described in Definition 8.3 and  
Definition 10.3 of \cite{ESYI}.

 To bookkeep the integrations, in Section 10.3 of \cite{ESYI} we defined
a sequence of matrices, $M^{(h)}$, a sequence of index sets, $I^{(h)}$,
and a sequence of vectors, $\fb^{(h)}$,
for $h=1,2, \ldots, k+1$. We set $E(h):= E(I^{(h)}, M^{(h)}, \fb^{(h)})$.
Initially $I^{(0)}= I$, $M^{(0)}=M(\sigma)$ and $\fb ^{(0)}=\emptyset$,
so from \eqref{suppe}
\be
   \sup_\bu E(\bA_0, \sigma,\bu) \leq  \| \wh \psi_0\|_\infty^2 E(0)\; .
\label{first}
\ee
As $h=1,2, \ldots$ 
increases, in each step we estimate $E(h-1)$ in terms of $E(h)$. 
The estimate depends on the set where $h\in I$ falls into
according to the partition \eqref{parti}.
The actual estimates are 
somewhat different in the lattice case than the corresponding
bounds (10.19), (10.20), (10.25), (10.29), (10.32) and (10.35)
 of \cite{ESYI} in
the continuous case. We will list the results only, the proofs
are analogous to the arguments in \cite{ESYI}.

In {\it Case 1,} $h\in I_p$, we have
\be
    E( h-1) \leq \eta^{-1}  E(h) \; .
\label{case1}
\ee

In {\it Case 2,} $h \in I_\ell$, we consider  $h, h+1, \ldots, 
h+\tau-1\in I_\ell$  a maximal sequence of consecutive ladder 
indices (i.e. $h-1,\;  h+\tau\not\in I_\ell$ with some $\tau\ge1$), 
then
\be
   E( h-1)  
  \leq C\lambda^{-2\tau}  E(h+\tau-1) \; .
\label{case2}
\ee
The proof is easier here in the lattice case: since the form factor
$\wh B$ is absent in \eqref{eq:EE}, each ladder index can be integrated
out independently by using \eqref{eq:ladderint}. Note that
the constant in \eqref{case2} is independent of $\tau$.

In {\it Case 3,} $h\in I_{us}$, we use \eqref{eq:logest} 
and \eqref{logpo} from the Appendix, i.e., the lattice
versions of (10.23) and (10.24) from  \cite{ESYI}:
\be
  E(h-1)\leq C\eta^{-1-\kappa/2}|\log\eta|^3 E(h) \leq
 C\eta^{-1-\kappa} E(h) \; .
\label{case3}
\ee

In {\it Case 4,} $h\in I_{cs}$, we need the following
lattice version of Lemma 10.5 in \cite{ESYI} that will be
proved in the Appendix:

\begin{lemma}\label{lemma:newrow} [Two Denominator Lemma] For 
 $\eta=\lambda^{2+\kappa}$ we have
\be
    \sup_{|\alpha|, |\beta|\leq 4d}\sup_r \int
    \frac{1}{|\alpha - \ov\om(p)-i\eta|\; |\beta - \om(p+q)+i\eta|}
    \frac{1}{\tri p-r\tri } \;   \rd p \;
    \leq \frac{C\eta^{-7/8-\kappa}}{\tri q\tri}
\label{eq:dp}
\ee
Without the point singularity we have
\be
    \sup_{|\alpha|, |\beta|\leq 4d} \int
    \frac{\rd p}{|\alpha - \ov\om(p)-i\eta|\; |\beta - \om(p+q)+i\eta|}
       \leq \frac{C\eta^{-3/4-\kappa}}{\tri q\tri }\; .
\label{eq:nopo}
\ee
Finally we have  
\be
    \int_{-4d}^{4d} \rd \alpha \int
    \frac{\rd p}{|\alpha - \ov\om(p)-i\eta|\; |\a - \om(p+q)+i\eta|}
       \leq \frac{C\eta^{-1/2-\kappa}}{\tri q\tri }\; .
\label{eq:bsquare}
\ee
\end{lemma}
Notice that these bounds are weaker than the ones given in
 Lemma 10.5 \cite{ESYI}
 which themselves are not optimal.
For example,  the factor $\eta^{-7/8-\kappa}$ in \eqref{eq:dp}
can be improved to $\eta^{-1/2}$ in the analogous estimate
for the continuum model.

Using this bound and following the argument of Case 4 in Section 10.3
of \cite{ESYI}, we have
\be 
   E(h-1) 
\leq C\eta^{-7/8-\kappa }     E(h)\; .
\label{case4}
\ee

In {\it Case 5,} $h\in I_v$, we use
$$
   \sup_{|\a|, |\beta|\leq 4d} \sup_r \int\int 
 \frac{\rd p\rd p'}{|\a - \ov\om(p)-i\eta|\; | \a - \ov\om(p')-i\eta|
\; |\beta - \om(p-p'+q)+i\eta|} \frac{1}{\tri p-r\tri} 
$$
$$\leq 
C\eta^{-7/8-2\kappa}
$$
instead of (10.31) of \cite{ESYI}.
This inequality follows from \eqref{eq:dp} and \eqref{logpo}. The same
bound holds if the point singularity is of
the form $\tri p\pm p' -r\tri^{-1}$ (compare with
(10.32) of \cite{ESYI}) or if there is no point singularity
at all. Following the argument of Case 5 in \cite{ESYI}, we 
obtain
\be 
   E(h-1) 
\leq C\eta^{-7/8-2\kappa }     E(h)\; .
\label{case5}
\ee
if $h\in I_v$.

In the last step, $h=k+1\in I_{last}$, we can estimate 
$E(k+1)$ by directly integrating out $\a$ and $\beta$ 
similarly to (10.35) of \cite{ESYI} to obtain
\be
   E(k+1) \leq C\lambda^{2k}|\log\eta|^2 \; .
\label{case6}\ee

Combining \eqref{first}, \eqref{case1}--\eqref{case3} and
\eqref{case4}--\eqref{case6}, and using that Case 2 has been applied
not more than $k-\ell=\mbox{deg}(\sigma)$-times (see \cite{ESYI}),
  we obtain
$$
\sup_\bu E(\bA_0, \sigma,\bu) \leq  C^{k-\ell} 
\lambda^{2(k-\ell)} \eta^{-|I_p|-(1+\kappa)|I_{us}|
-(\frac{7}{8}+\kappa)|I_{cs}|
-(\frac{7}{8}+2\kappa)|I_{v}|}
|\log\eta|^2 \; .
$$
Using that $|I_p|=|I_v|\ge 1$, $|I_{us}|\leq v+1$ from \cite{ESYI}
 and that $|I_p|+|I_v|+ |I_{us}|+ |I_{cs}|=
k-\ell$, we obtain
$$
\sup_\bu E(\bA_0, \sigma,\bu) \leq 
C\big( \lambda^{\frac{1}{16}-4\kappa}\big)^{\de(\sigma)}|\log\eta|^2
$$
if $\kappa<1/64$ and $\lambda$ is sufficiently small.
This proves Theorem  \ref{thm:Vsi}.
 $\;\;\Box$

\bigskip

\section{Error terms: Proof of Theorem \ref{7.1}}\label{sec:error}
\setcounter{equation}{0}

The main contribution to the wave function $\psi_t$ in (\ref{eq:duha})
comes from the fully expanded non-recollision terms with $r=0$, i.e
$\psi_{t,k}^{(0), nr}$. Here we show that the contribution of all
other terms is negligible. 
Our result can be summarized in the following Theorem which will be
proven in Sections \ref{sec:error} and \ref{sec:cases}.

\begin{theorem}\label{thm:error}
 Assume \eqref{param} with
 $\kappa < 1/9800$ and 
a sufficiently small $\delta$.  Then, as $\lambda\to0$,
\be
    \bE \| \psi_{*t, k}^{(r), \#} \|^2 = o(\lambda^{4 + 2\kappa +2\delta })
\label{eq:noladerr}
\ee
for the following choices of
the parameters: $\{\#=rec, r=0,1\}$; 
 $\{ \#= nest, tri, r=1\}$ or  $\{ \#=last, r=2\}$.
Furthermore, for $k=K$ and  $r=0,1$,
\be
    \bE \| \psi_{*t, K}^{(r), nr} \|^2 =
 o(\lambda^{4 + 2\kappa +2\delta}) \;
\label{eq:laderr}
\ee
and for $k<K$
\be
    \bE \| \psi_{t, k}^{(1), nr} \|^2 = o(\lambda^{2\kappa +2\delta}) \; .
\label{eq:laderr1}
\ee
\end{theorem}
From this Theorem and from \eqref{eq:duha}, Theorem~\ref{7.1}
easily follows by using the unitarity estimate on the truncation
of the Duhamel formula
\be
     \Big\| \int_0^t \rd s \; e^{-i(t-s)H} \psi_s \Big\|^2
   \leq t^2 \sup_{0\leq s\leq t} \|\psi_s\|^2 \; .
\label{unit}
\ee
Note that this estimate effectively loses a factor of $t$
by neglecting the oscillation on the left hand side.
(See Section 4 of \cite{ESYII}  for more details.)

\bigskip

{\it Proof of Theorem \ref{thm:error}.}  We start  with a resummation and
symmetrization for the {\it core indices} that identify the non-repetitive 
potential labels in a sequence of collisions $\gamma$. We repeat
the Definition 4.2 \cite{ESYII} here:

\begin{definition}[Core of a sequence]\label{def:core}
Let $\tgamma \in \tGamma_n$,
 then
 the set of {\bf core indices} of $\tgamma$ is defined as
$$
  I_n^{core}(\tgamma):= \Big\{ j\in S(\tgamma) \; : \; \tgamma_j \neq \tgamma_i, \mbox{for}
  \; i\neq j\Big\}
$$
and we set $c(\tgamma)= |I_n^{core}(\tgamma)|$.
The corresponding $\tgamma_j$ labels are called {\bf core labels}.
The subsequence of core labels form an element in $\Gamma_c^{nr}$, i.e. a sequence
of different $\bZ^d$-labels.
The elements of
$$
I_n^{nc}(\tgamma):= I_n\setminus [I_n^{core}(\tgamma) \cup I_n^\theta(\tgamma)]
$$
are called {\bf non-core potential indices}.
\end{definition}

{\it In other words, the core indices are those skeleton indices  that
do not participate in any recollision, gate, triple collision or
nest.} Given our stopping rule, the number of non-core potential
indices and $\theta$-indices together is at most 4.
The number of core indices $c$ is related to the number of
 skeleton indices $k$ as follows
\be
c: = \left\{  \begin{array} {c@{\quad \mbox{if}\quad}l}
   k & \#=nr, last \\
   k-1 & \#=triple\\
   k-2 & \#=nest,rec
\end{array}\right.
\label{def:c}
\ee

\subsection{Resummation and symmetrization}

Let $\tau =\tau(\tgamma) =(\tau_1, \ldots , \tau_c)\in \tGamma_c^{nr}$
denote the core labels of the sequence $\tgamma$. 
We rewrite each error term by first summing over the core labels.
When computing $\bE \| \psi^{(r), \#}_{(*)t,k}\|^2
=\bE \int \ov{\psi} (\ldots) \psi(\ldots)$ by using
the expansions of both $\psi$ and $\ov\psi$,
the core labels of $\psi$ and $\ov\psi$ are exactly paired,
the pairing is given by a permutation $\sigma\in \fS_c$.
The location of non-core indices within a sequence is encoded 
by a {\it location code}, $w$. The set of location codes, $W$,
depends on  $\#$, $c$  and $r$. Having specified the location
of the $r$ gates/$\theta$-indices, we introduce another
code, $h\in \{ g, \theta\}^r$, called {\it gate-code},
to specify whether there is a  gate or a $\theta$ at the given location.
By using a Schwarz inequality, we symmetrize for the location 
codes in the estimate of $\bE \| \psi^{(r), \#}_{(*)t,k}\|^2$, 
so both $\psi$ and $\ov\psi$ have the same location code, $w$.
The two gate codes, $h,h'$, corresponding to $\psi$ and $\ov\psi$
are not symmetrized, since we still have to exploit
the cancellation between the gates and $\theta$'s and 
this effect would disappear after a Schwarz inequality.
See Sections 4.2 and 4.3  \cite{ESYII} for the details.

For a given $\#$, $r$, a given number of core indices $c$,
a given permutation of core indices, $\sigma\in \fS_c$, a given
location code $w$ and for given gate codes, $h, h'$
we define a partition $\bD_0$ of the joint index set $I_n\cup \wt I_{n'}$
of $\psi_n\ov\psi_{n'}$. The partition $\bD_0$ lumps 
{\it exactly} those indices that are {\it required} 
to carry the same potential label by the prescribed structure.

Some of the non-core indices may have further coincidences
(e.g. two gates in the expansion may incidentally share the same
potential label, creating a lump of four elements).
This defines a new partition, $\bD$, that is the coarsening
of $\bD_0$, in notation: $\bD\succ \bD_0$. Since the total
number of non-core indices is bounded, the number of possible
$\bD$ is also bounded. The $\vartheta$-indices remain always single.
Most lumps of $\bD$ have two elements, and some gates may form
quartets or sextets, their number is denoted by $\varrho_4(\bD)$
and $\varrho_6(\bD)$. Higher lumps do not appear.

Let $\bD^*\subset \bD$  denote the collection of non-single elements
 of $\bD$. A distinct potential label is selected for
each element of $\bD^*$, in particular the connected graph
formula is applied for the index set $\bD^*$.
Let $\bA \in\cA(\bD^*)$ be a partition of the set $\bD^*$. We
define $\bP(\bA, \bD)\in\cP_{n,n'}$
 to be the partition of $I_n\cup \wt I_{n'}$
whose lumps are given by the equivalence relation that
two elements of $I_{n}\cup \wt I_{n'}$ are
$\bP(\bA, \bD)$-equivalent
if their $\bD$-lump(s) are $\bA$-equivalent. The single lumps of $\bD$ remain
single in $\bP$ (these are the $\theta$ indices).

With all these notations, the following bound was proven
in Proposition 4.3 of \cite{ESYII}:

\begin{proposition} Under the choices
of parameters $\#$, $r$ in Theorem~\ref{thm:error}, 
let $c$ be given by (\ref{def:c}), and let
$W=W^{(r), \#}_c$ be the set of location codes.
Then
\be
\bE \|\psi_{(*)t, k}^{(r), \#} \|^2 \leq
    |W| \sum_{w\in W} \sum_{\s\in \fS_c}
   \sum_{h,h'\in \{g,\theta\}^r} \sum_{\bD\succ\bD_0}
\underline{m}^{\varrho(\bD)}
\sum_{\bA\in \cA(\bD^*)}c(\bA) V_{(*)}^\circ( \bP(\bA, \bD)) \; ,
\label{eq:decS}
\ee
where $m_k:= \bE \, v_\a^k$ are the moments of the
single site random potential in \eqref{ranpot} and
$\underline{m}^{\varrho(\bD)}:=
 m_4^{\varrho_4(\bD)}m_6^{\varrho_6(\bD)}$.
$\;\;\Box$
\end{proposition}
The hidden parameters in the definition of the $V$-value 
of $\bP(\bA, \bD)$ (see  \eqref{def:Vlong}--\eqref{def:Vshort})
will be chosen $\xi=0$ and $Q\equiv 1$ thoroughout
the entire section.

\subsection{Splitting into high and low complexity regimes}

For a coarsening $\bD\succ\bD_0$, the partition
$\bD^*$ contains all core elements of $\bD_0$.
Any partition $\bA\in\cA(\bD^*)$ can be naturally restricted
onto the core elements and can thus be identified with 
a partition
of $I_c$ (see Section 4.4 \cite{ESYII} for  a precise definition).
 We denote this restricted partition by $\wh\bA$.

The sum (\ref{eq:decS}) will be split into two parts and estimated 
differently.  We set a threshold $q\ge 3$ for the joint 
degree $q(\wh \bA, \s)$ of $\sigma$ and  $\wh\bA$ and
obtain
\be
   \bE \|\psi_{(*)t, k}^{ (r), \#} \|^2 
\leq (I) + (II) + O(\lambda^{1/2})
 \label{I+II}
\ee
with
\be
   (I) := |W|
   (c+4)^4 \sum_{w\in W}\sum_{h,h' } \sum_{\s\in \fS_c} \sum_{\bD\succ\bD_0}
   \underline{m}^{\varrho(\bD)}
   \!\!\!\!\!\sum_{\bA'\in \cA_c\atop
     q( \bA', \s)\ge q }
   \sup_\bA \Big\{ | V_{(*)}( \bP(\bA, \bD)) c(\bA) | \; : \;
\wh\bA =\bA' \Big\}
\label{eq:I}
\ee
where the supremum is over all possible $\bA\in \cA(\bD^*)$
whose restriction $\wh\bA$
is the given partition $\bA'$;
and
\be
 (II) :=   |W| \sum_{w\in W} \sum_{\s\in \fS_c} \Bigg|
   \sum_{h,h'\in \{g,\theta\}^r} \sum_{\bD\succ\bD_0}
\underline{m}^{\varrho(\bD)}
\sum_{\bA\in \cA(\bD^*)\atop q( \wh\bA, \s)< q} V_{(*)}
( \bP(\bA, \bD)) c(\bA) \Bigg| \; .
\label{eq:II}
\ee
The error term $O(\lambda^{1/2})$ comes from replacing 
$V_{(*)}^\circ(\cdots)$ with $V_{(*)}(\cdots)$ by using 
 Lemma~\ref{lemma:VV}.

The high complexity regime, term (I), can be estimated
by using Proposition~\ref{prop:sumup}. By applying Operation I
(Lemma~\ref{lemma:breakup}),  we can break up all non-core lumps
into single lumps,  we then can remove them by Operation II
(Lemma~\ref{lemma:remove})..
The remaining partition is identified with
$\bP(\wh \bA, \sigma)$ on the core indices, $I^{core}_n\cup 
I^{core}_{n'}$.

We lose at most a factor $(\lambda\eta^{-1})^8$ 
since Operation II is applied at most 8 times.
Unlike in the continuous case, \cite{ESYII}, Operation I
does not cost extra factors. Following the argument 
of Section 4.5 of \cite{ESYII} and using
the bound from Proposition~\ref{prop:sumup}
instead of its continuous version (Proposition 
9.2. of \cite{ESYI}) used in \cite{ESYII},
 we see that 
(I) in \eqref{eq:I} is negligible in the sense of
Theorem \ref{thm:error}
if
\be
   \kappa < \frac{q-192}{144q +288}
\label{eq:qlow}
\ee
and $\delta\leq \delta(\kappa)$ is chosen sufficiently small.

\bigskip

In the low complexity regime we will use the special
structure given by the recollisions, nests, triple collisions
and gates. The combinatorics in  (\ref{eq:II})
was estimated in Lemma 4.5 \cite{ESYII} which we recall here:

\begin{lemma}\label{lemma:smallcomb}
 For any  $q\in \bN$,  $c\leq K$
 and structure type $\#$,
 we have
$$
     \sup_{w, h, h'}\sum_{\sigma \in \fS_c}
     \sup_{\bD\succ\bD_0}\sum_{\bA \in \cA(\bD^*)\atop q( \wh\bA,\s)<q}
     |c(\bA)| \leq (CqK)^{3q+3}
$$
where we recall that $\bD_0$ depends on $(\#, c, \s, w, h, h')$. $\;\;\Box$
\end{lemma}

\bigskip

The size of the individual terms in (\ref{eq:II})
are estimated in the following Proposition
whose proof will be given in Section \ref{sec:cases}.
This is the lattice version of Proposition 4.6 \cite{ESYII}
and most of these estimates substantially differ from
their continuous counterparts. The proof is given
in Section~\ref{sec:cases}.

\begin{proposition}\label{prop:i-iv} 
We assume \eqref{param} and we assume
that the initial condition $\wh\psi_0$
satisfies (\ref{suppsi}) for some $\Lambda>0$.
 Let
 $\sigma\in \fS_c$, $w\in W_c^{(r), \#}$,
 $h, h'\in \{ g, \theta\}^r$, 
where $\#$ and $r$ vary in the different cases and let
$\bD\succ\bD_0(\#, c, \sigma, w, h, h')$,

1) Let $\bA\in \cA(\bD^*)$ such that
$q(\wh\bA, \sigma )< q$. Then the following estimates hold:

(1a) [Many collisions] Let $\#=nr$, $r=0,1$ and $c=K$,  then
\be
     |V_{*} (\bP(\bA, \bD))|\leq  C^q\lambda^{\frac{\delta}{2}K} \; .
\label{eq:manyev}
\ee

(1b) [Recollision]. Let $\#=rec$, $r=0,1$, then
\be
    |V_{*} (\bP(\bA, \bD))|
\leq C^q\lambda^{\frac{17}{4}-10\kappa}|\log\lambda|^{O(1)}
\label{eq:recv}
\ee

(1c) [Triple collision] Let $\#=triple$, $r=1$, then
\be
    |V_{*} (\bB(\bA, \bD))|\leq  C^q\lambda^{6} |\log\lambda|^{O(1)}
\label{eq:triplev}
\ee

2) Now let $\bA'\in \cA_c$ be given. Then the following hold:

(2a) [Non-repetition with a gate] Let $\#=nr$, $r=1$, then
\be
     \sup_{\sigma, w} \Big|\sum_{h,h'\in \{ g, \theta\}^r}
     \sum_{\bD\succ\bD_0} \sum_{\bA\in \cA(\bD^*) \atop \wh\bA=\bA'}
 V (\bP(\bA, \bD)) c(\bA)\Big|\leq C\lambda^{\frac{1}{16}-10\kappa
-O(\delta)} \; .
\label{eq:nrev}
\ee

(2b) [Last] Let $\#=last$, $r=2$, then
\be
     \sup_{\sigma, w} \Big|\sum_{h,h'\in \{ g, \theta\}^r}
     \sum_{\bD\succ\bD_0} \sum_{\bA\in \cA(\bD^*) \atop \wh\bA=\bA'}
 V_{*} (\bB(\bA, \bD)) c(\bA)\Big|\leq C
\lambda^{5-\frac{9}{2}\kappa}|\log\lambda|^{O(1)}
\label{eq:lastev}
\ee

(2c) [Nest] Let $\#=nest$, $r=1$, then
\be
     \sup_{\sigma, w} \Big|\sum_{h,h'\in \{ g, \theta\}}
     \sum_{\bD\succ\bD_0} \sum_{\bA\in \cA(\bD^*) \atop \wh\bA=\bA'}
 V_{*} (\bB(\bA, \bD)) c(\bA)\Big|\leq
  C\lambda^{\frac{17}{4}-10\kappa}|\log\lambda|^{O(1)} \; .
\label{eq:nestev}
\ee
All constants $C$ depend on $\Lambda$.
\end{proposition}

Combining  Lemma
\ref{lemma:smallcomb} with these estimates, and  using $|W|\leq K^2$,
we obtain that the  contributions of the error terms from (II) to
$ \bE \|\psi_{(*)t, k}^{(r), \#} \|^2 $ (see (\ref{I+II}))
satisfy the bounds in Theorem~\ref{thm:error} if
\be
\kappa < \frac{1}{48q+368}
\label{eq:qup}
\ee
and $\delta$ is sufficiently small.
Combining this with (\ref{eq:qlow}) and optimizing,
we see that the biggest $\kappa$ that still guarantees a solution for $q$
is around $\kappa\sim 1/9800$ with $q=195$.
$\;\;\;\Box$.

\section{Proof of Proposition \ref{prop:i-iv}}\label{sec:cases}
\setcounter{equation}{0}

\subsection{Many collisions}

The proof of Case (1a) is very similar to the proof of Case (1a)
of Proposition 4.6 of \cite{ESYII};  actually the estimates
are sharper because Operation I does not carry an
additional factor $\Lambda=O(\lambda^{-2d\kappa-O(\delta)})$
(compare Lemma 9.5 of \cite{ESYI} with  \eqref{opIest}). 
The proof of the key estimate,
$$
    \int \rd \bp |K(t,\bp,k)|^2 \leq (C_a \lambda^{-2+\delta a})^{k-1}
$$
($0\leq a < 1$, $k\ge T\lambda^{-\kappa-\delta}$, $t=T\lambda^{-2-\kappa}$,
see Lemma 5.1 of \cite{ESYII}),
is also simpler in the discrete case since no additional large-momentum
cutoff factors need to be introduced.
The only estimate that is somewhat weaker in the discrete case 
than its continuum counterpart 
is  \eqref{def:EMout} 
(compare with Lemma 7.1 of \cite{ESYI}),
but for $n+n'\ge 2K$ it still gives a negligible error.
We do not repeat the details here.

\subsection{Recollision and triple collision}

The estimates for (1b) and (1c) substantially differ from
their continuum counterparts (Proposition 4.6 of \cite{ESYII}).
The main reason is that the two-denominator bound \eqref{eq:nopo}
is weaker by a factor $\eta^{-3/4-O(\kappa)}$ than 
its continuum version, (10.28) of \cite{ESYI}.
Taking into account the $\lambda^2$-factor associated with
these two denominators, each application of the two-denominator
bound will improve the estimate by $\lambda^2\eta^{-3/4-O(\kappa)}
=\eta^{1/4-O(\kappa)}$ compared with the ladder term.
The corresponding improvement in the continuum case is
$\eta^{1-O(\kappa)}$.

As we demonstrated in \cite{ESYII}, a recollision
Feynman graph can be evaluated by applying the two-denominator bound twice
(recall that due to symmetrization, the recollision graphs
have  two recollision; one on the $\psi$- and one on the
$\ov\psi$-side of the graph). The corresponding gain,
$(\eta^{1-O(\kappa)})^2$, has easily beaten the additional
factor $t$ from the truncation \eqref{unit}.
In the discrete case, the recollision gain, $(\eta^{1/4-O(\kappa)})^2$,
would not be sufficient. With the estimate \eqref{eq:nopo} at hand,
one would need to expand at least up to five recollisions,
which would mean many more terms to estimate individually.
Since this route is quite lengthy, although certainly possible,
we follow a different argument based upon the following
\begin{lemma}\label{lemma:4de}
(i) For arbitrary $\a$, it holds that
$$
\sup_{u}
\int_{\Tor^3}\frac{\rd p\rd q\rd r}{|\a -\ov\om(p)-i\eta||\a -\ov\om(q)-i\eta| 
|\a -\ov\om(r)-i\eta||\a -\ov\om(p-q+r-u)-i\eta|} 
$$
\be
\leq C \lambda^{-3/4-3\kappa}|\log\eta|^{4} \; .
\label{weak4d}
\ee

(ii)  For any $\Lambda>\eta$, there exists $C_\Lambda$ such that
for any $\alpha \in [0,6]$ with $\tri \a\tri \ge \Lambda$
(recall the definition from (\ref{def:tria})), 
 we have
$$
\sup_u\int\frac{\rd p\rd q\rd r}{|\a -\ov\om(p)-i\eta|
|\a -\ov\om(q)-i\eta| 
|\a -\ov\om(r)-i\eta||\a -\ov\om(p-q+r-u)-i\eta|} 
$$
\be
\leq C_\Lambda
\lambda^{-4\kappa}|\log\eta|^{14} \; .
\label{strong4d}
\ee
\end{lemma}

{\it Proof.} 
The first statement (\ref{weak4d})
 in Lemma \ref{lemma:4de} is a direct consequence
of  (\ref{resex}),  (\ref{nopont}), 
(\ref{logpo})
and  (\ref{eq:logest}).
For (\ref{strong4d}), we first use (\ref{resex})
then we use the Four Denominator Lemma \ref{lemma:4dee}. $\;\;\Box$

\bigskip

Now we start the proof of (1b)--(1c) of Proposition~\ref{prop:i-iv}.
Following Section 5.2 of \cite{ESYII}, Operations I, II and IV
can be used to break up the partition into a trivial
one and remove all gates and $\theta$'s. The total cost is at most
$\lambda^{-2\kappa}|\log\lambda|^2$. Then 
(\ref{eq:recv}) and \eqref{eq:triplev} follow from
the Propositions \ref{prop:rec} and \ref{prop:rec1}
exactly as 
the proof of Cases (1b), (1c) in Proposition 4.6 in \cite{ESYII}
followed from Propositions 5.2 and 5.3 of \cite{ESYII}.

\begin{proposition}\label{prop:rec}
Consider the Feynman graph on the vertex set $\cV_{k}$, $k\ge 3$,
choose numbers $a, b, a', b' \in I_k$ such that
$b-a\ge 2$, $b'-a'\ge 2$. Let $\sigma$ be a bijection between
$I_k\setminus \{ a, b\}$ and $\wt I_k\setminus \{ a', b'\}$.
Let $\bP^*$ be the partition on the set $I_k\cup \wt I_{k}$
consisting of the lumps $\{ j, \sigma(j)\}$, $j\in I_k\setminus \{ a, b\}$
and $\{ a, b\}$, $\{a', b'\}$ (Fig.~\ref{fig:recdis}).
 We assume (\ref{suppsi}).
Then
for the truncated version
\be
      \sup_\bu E_{*}(\bP^*, \bu)
      \leq C\lambda^{\frac{17}{4}-8\kappa}|\log\lambda|^{O(1)} \; .
\label{eq:recesttr}
\ee
\end{proposition}

\bef\bec
\epsfig{file=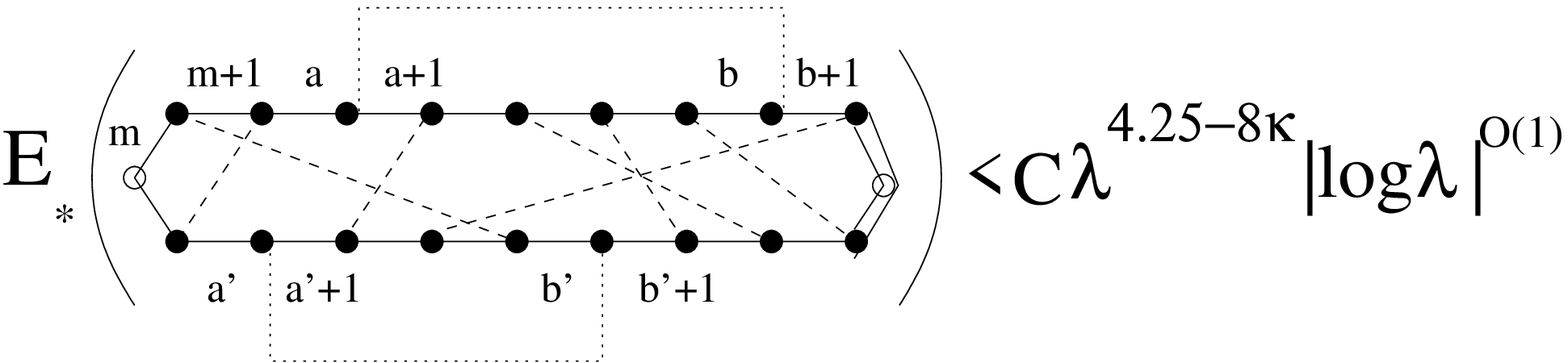,scale=0.7}
\eec
\caption{Estimate of a two-sided recollision graph}
\label{fig:recdis}
\bec
\epsfig{file=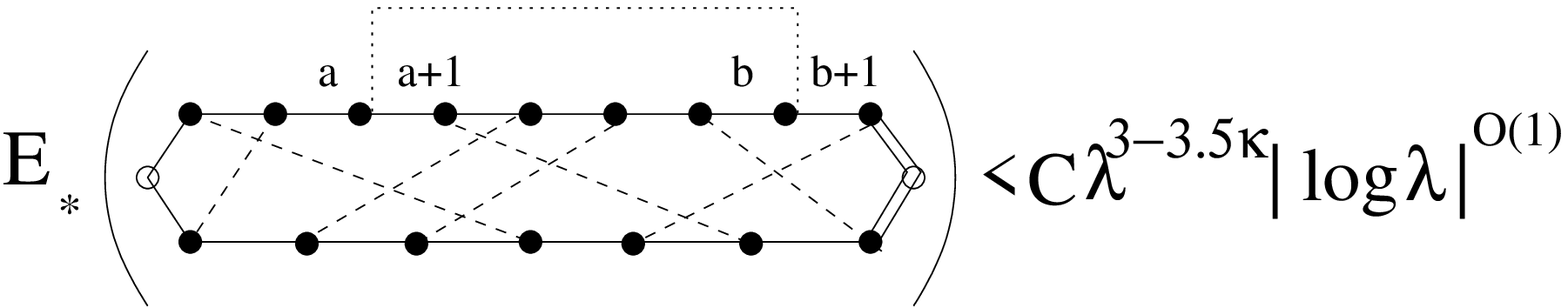,scale=0.7}
\eec
\caption{Estimate of a one-sided recollision graph}
\label{fig:rec1dis}
\eef
We also need a ``one-sided'' version of this estimate (Fig.~\ref{fig:rec1dis}).

\begin{proposition}\label{prop:rec1}
Consider the Feynman graph on the vertex set $\cV_{k, k-2}$, $k\ge 3$,
choose numbers $a, b \in I_k$ such that
$b-a\ge 2$. Let $\sigma$ be a bijection between
$I_k\setminus \{ a, b\}$ and $\wt I_{k-2}$.
Let $\bP^*$ be the partition on the set $I_k\cup \wt I_{k-2}$
consisting of the lumps $\{ j, \sigma(j)\}$, $j\in I_k\setminus \{ a, b\}$
and $\{ a, b\}$. We assume (\ref{suppsi}).
Then
\be
     \sup_\bu E(\bP^*, \bu) \leq C\lambda^{1-\frac{7}{2}\kappa}
   |\log\lambda|^{O(1)} \; ,
\label{eq:halfrecest}
\ee
and
 for the truncated version
\be
      \sup_\bu E_{*}(\bP^*, \bu) \leq C\lambda^{3-\frac{7}{2}
\kappa}
   |\log\lambda|^{O(1)} \; .
\label{eq:halfrecesttr}
\ee
\end{proposition}

{\it Proof of Propostion \ref{prop:rec}.}
 We use $\bp=(p_1, \ldots , p_{k+1})$ and their tilde-counterparts to
denote the momenta to express
$$
   E_{*}(\bP^*, \bu) = 
  \int_{-4d}^{4d} \rd \alpha\rd\beta\; \Xi(\alpha, \beta)
$$
with
$$
   \Xi(\alpha, \beta):=\lambda^{2k}
 \int \rd\bp\rd\tbp \prod_{j=1}^{k} \frac{1}{|\a-\ov\om(p_j)-i\eta|}
   \frac{1}{|\beta-\om(\tp_j)+i\eta|}
$$
$$
  \times \delta(p_{k+1}-\tp_{k+1}) \delta\Big(p_{a+1}-p_a
   + (p_{b+1}-p_{b}) -u_a\Big)  \delta\Big(-\tp_{a'+1}+\tp_{a'}
   - (\tp_{b'+1}-\tp_{b'}) -\tu_{a'} \Big)
$$
$$
   \times \prod_{j=1\atop j\neq a, b}^{k}
    \delta\Big(p_{j+1}-p_j
   - (\tp_{\s(j)+1}-\tp_{\s(j)}) - u_j\Big)\; |\wh\psi_0(p_1)|^2
$$
where the $\bu$-momenta are labelled as
 $\bu = ( u_1, \ldots , u_{b-1}, u_{b+1}, u_{k},
\tu_{a'})$.

We first prove the case $k=3$, when
$$
   E_{*}(\bP^*, \bu) = \lambda^{6}
  \int_{-4d}^{4d} \rd \alpha\rd\beta
   \int\rd \tp_2\int   \frac{\rd p_1}{|\a-\ov\om(p_1)-i\eta|}
\frac{\rd p_2}{|\a-\ov\om(p_2)-i\eta|}\frac{\rd p_3}{|\a-\ov\om(p_3)-i\eta|}
$$
$$
  \times \frac{1}{|\beta-\om(p_1)+i\eta|}
 \frac{1}{|\beta-\om(\tp_2)+i\eta|}
 \frac{1}{|\beta-\om(\tp_2+p_3-p_2+u_1')+i\eta|}\; .
$$
The integral over $\tp_2$ is performed by 
using 
(\ref{eq:nopo})
\be
     \int\frac{\rd p}{|\beta-\ov\om(p-v)+i\eta|
     \; |\beta-\ov\om(p)+i\eta|}  
\leq \frac{C\eta^{-3/4-\kappa} |\log \eta|^2}{\tri v\tri }
\label{eq:square}
\ee
collecting $\tri p_3-p_2+u_1'\tri^{-1}$. Then $\rd\beta$,
$\rd p_2$, $\rd p_3, \rd p_1$ 
can be performed (in this order) using (\ref{logpo}) to 
integrate the point singularity. The result is
$$
    E_{*}(\bP^*, \bu) \leq C \lambda^{\frac{9}{2}-4\kappa}
  |\log\lambda|^{O(1)} \; .
$$
From now on we can assume that $k\ge 4$.

The integration will be split into four domains,
depending on whether $\tri \a \tri$ and $\tri \beta\tri$
are smaller or bigger than $\Lambda$, where we recall
that on the support of $\wh\psi_0(p)$ we have $\tri \om(p)\tri \ge 2\Lambda$.
By symmetry we can assume that $\tri \a\tri \ge \tri\beta\tri$
and we have effectively two cases:
$$
    E_{*}(\bP^*, \bu)\leq  (I) + 2(II),
$$
where
$$
    (I):= \int_{-4d}^{4d} \rd\a \rd\beta
    \;{\bf 1}(\tri a\tri \ge \Lambda, \tri \beta\tri\ge\Lambda) \Xi(\alpha, \beta)
$$
$$
    (II):= \int_{-4d}^{4d} \rd\a \rd\beta
    \;{\bf 1}(\tri \beta\tri\le\Lambda) \Xi(\alpha, \beta)\; .
$$

\bigskip

\underline{\it Term (I):  $\tri a \tri\ge \Lambda$, $\tri\beta\tri\ge\Lambda$.}

\bigskip

Similarly to the proof of Proposition  5.2 in \cite{ESYII},
we partition the set of all $\bp, \tbp$ momenta into two subsets of
size $k+1$ each:
$$
    A: = \{ p_1, p_2, \ldots , p_{b-1}, p_{b+1}, \ldots p_{k+1}, \tp_{b'}\}
$$
$$
    B:= \{ \tp_1, \tp_2, \ldots , \tp_{b'-1}, \tp_{b'+1}, \ldots 
\tp_{k+1}, p_{b}\}\; .
$$
Note that sets $A$ and $B$ are obtained by exchanging
$p_b$ and $\tb_{b'}$ in the sets $\{ \bp\}$ and $\{\tbp\}$.

It is straightforward to check that
 all $A$-momenta can be uniquely
expressed in terms of linear combinations of the $B$-momenta
(plus the $\bu$-momenta) and conversely.  In particular
$$
    p_{b-1} = p_{b} - (\tp_{\s(b-1)+1}-\tp_{\s(b-1)}) -u_{b}
$$
$$
   \tp_{b'-1} = \tp_{b'} - (p_{m+1} -p_m) +
   u_m\quad \mbox{with} \qquad m:= \s^{-1}(b'-1) \; .
$$
The letters on Fig.~\ref{fig:recdis}
 indicate the indices of the correspoding $p$ or $\tp$
momenta.

By the Schwarz inequality, we have
\be
     \prod_{j=1}^{k} \frac{1}{|\a-\ov\om(p_j)-i\eta|}
   \frac{1}{|\beta-\om(\tp_j)+i\eta|}  \leq  \frac{1}{2}\big[(a) + (b)\big]
\label{abest}
\ee
$$
   (a):=\prod_{j=1, b-1, b}
   \frac{1}{|\a-\ov\om(p_j)-i\eta|} \frac{1}{|\beta-\om(\tp_{j'})+i\eta|}
   \prod_{j=2\atop j\neq b-1, b}^{k} \frac{1}{|\a-\ov\om(p_j)-i\eta|^2}
$$
$$
    (b):=\prod_{j=1, b-1, b}
   \frac{1}{|\a-\ov\om(p_j)-i\eta|} \frac{1}{|\beta-\om(\tp_{j'})+i\eta|}
\prod_{j=2\atop j\neq b'-1, b'}^{k}
\frac{1}{|\beta-\om(\tp_j)+i\eta|^2}  \; .
$$
(with a little abuse of notations we used $j'$ for $1$,  $b'-1$ and $b'$
when $j=1, b-1$ and $b$, respectively).

Integrating out all $B$-momenta in (a) and all $A$-momenta in (b), we
have (with $m:= \s^{-1}(b'-1)$)
$$
   (I)\leq 2 \lambda^{2k}
  \int_{\tri\a\tri\ge\Lambda\atop |\a|\leq 4d}\rd \alpha
 \int_{\tri\beta\tri\ge\Lambda\atop |\beta|\leq 4d} \rd\beta
   \int \rd \tp_{b'}\Big(\prod_{j=1\atop j\neq b}^{k+1}
   \rd p_j\Big)
   \frac{1}{|\a-\ov\om(p_1)-i\eta|}
   \frac{1}{|\beta-\om(p_1)+i\eta|} \; |\wh\psi_0(p_1)|^2
$$
\be
    \times  \frac{1}{|\a-\ov\om(p_{b-1})-i\eta|}
   \frac{1}{|\beta-\om(\tp_{b'} - (p_{m+1} -p_m) +
   u_m)+i\eta|}
\label{eq:scww}
\ee
$$
   \times  \frac{1}{|\a-\ov\om(p_{b+1}-p_a+p_{a+1}- u_a)-i\eta|}
   \frac{1}{|\beta-\om(\tp_{b'})+i\eta|}\prod_{j=2\atop j\neq b-1, b}^{k}
    \frac{1}{|\a-\ov\om(p_j)-i\eta|^2} \; .
$$
The integration of $\tp_{b'}$  is performed by using (\ref{eq:square})
with $v: = p_{m+1}-p_{m}-u_m$.
We also integrate out $\rd\beta$ and collect $C|\log\eta|$.

Now we integrate out all $p_j$'s with $j\neq 1,a,a+1, b-1, b, b+1, m, m+1$,
each collects a factor $\lambda^{-2}(1+C_0\lambda^{1-12\kappa})$
by  using Lemma \ref{le:opt}.
Moreover, we remove the square from the remaining five squared
denominators with $j=a,a+1, b+1, m, m+1$ at the expense of $\eta^{-1}$ each. 
Let $S:=\{ 1, a,a+1, b-1, b+1, m, m+1\}$, then we obtain 
\be
(I)\leq C\lambda^{16}
   \eta^{-5-3/4-\kappa}(1+C_0\lambda^{1-12\kappa})^{k-8} |\log\eta|
   \int_{\tri\a\tri\ge\Lambda\atop |\a|\leq 4d} \rd \alpha
   \int \Big(\prod_{j\in S} \rd p_j\Big) 
\label{Ialpha}
\ee
$$
 \times \frac{1}{\tri p_{m+1}-p_{m}-u_m\tri}
\frac{1}{|\a-\ov\om(p_{b+1}-p_a+p_{a+1}- u_a)-i\eta|}
 \prod_{j\in S\setminus \{k+1\}} \frac{1}{|\a-\ov\om(p_j)-i\eta|}\; .
$$
Strictly speaking, this argument is valid only if
the indices $1, a,a+1, b-1, b, b+1, m, m+1$ are all distinct
and $b,m \neq k$.
If there is some coincidence, then we may use 
Lemma \ref{le:opt} for more $p_j$'s, but save 
an  $\eta^{-1}$ factor each time, so the estimate still holds.
In particular, this is the case when $3\leq k\leq 6$.

Suppose for the moment that $\{ m, m+1\}\not\subset \{ a,a+1, b+1\}$.
In this case, we can integrate out all $p_j$, $j\in 
S\setminus\{ a,a+1, b+1\}$, using (\ref{eq:logest})
and (\ref{logpo}) from the Appendix and the point singularity
disappears at one of the integration.
 We obtain, by using $k\leq K\ll \lambda^{-(1-12\kappa)}$,
$$
(I)\leq C\lambda^{\frac{9}{2}-9\kappa}
   |\log\eta|^4
    \int_{\tri\a\tri\ge\Lambda\atop |\a|\leq 4d} \rd \alpha\;
{\bf 1}(\tri\alpha\tri\ge \Lambda)
   \int\rd p_a\rd p_{a+1}\rd p_{b+1}
$$
$$
 \times \frac{1}{|\a-\ov\om(p_{b+1}-p_a+p_{a+1}- u_a)-i\eta|}
 \prod_{j=a,a+1,b+1\atop j\neq k+1} \frac{1}{|\a-\ov\om(p_j)-i\eta|}\; .
$$
The proof of (\ref{eq:recesttr})
is then completed by using 
 Lemma \ref{lemma:4de}.

Finally, we have to discuss the case when 
$\{ m, m+1\}\subset \{ a,a+1, b+1\}$. This can
happen only if $b-a=2$ and $m=a+1$ (i.e. $\sigma(a+2)=\sigma(b)=b'$).
In this case the point singularity $\tri p_{m+1}-p_m-u_m\tri^{-1}$
would appear as an additional factor
in the integrand with four denominators in Lemma \ref{lemma:4de}.
To avoid this situation, we choose  exchange momenta
different from $p_b, \tp_{b'}$ when defining the sets $A$ and $B$.

If $b'-a'>2$, then we can choose $p_b$ and $\tp_{a'+1}$.
The role of $m$ will be played $m^*:=\sigma^{-1}(a'+1)$.
Since $b'-a'>2$, we have $m\neq m^*$, in particular
$\{ m^*, m^*+1\}\not\subset \{ a,a+1, b+1\}$ and the previous
proof goes through.

If $b'-a'=2$, then we can use $\tp_{a'}$
or $\tp_{b'+1}$ 
as a tilde  exchange momenta instead of $\tp_{b'}$,
 unless $a'=1$ or $b'=k$, 
respectively. The role of  $m$ is played by 
$\sigma^{-1}(a'-1)$ and $\sigma^{-1}(b'+1)$, respectively.

Finally, if $b'-a'=2$, $a'=1$, $b'=k$, then $k=3$ and this case
was already investigated.

\bigskip

\underline{\it Term (II):  $\tri\beta\tri\leq \Lambda$.}

\bigskip

From the support property of $\wh\psi_0$ we know that
\be
       \frac{|\wh\psi_0(p_1)|^2}{|\beta-\om(p_1)+i\eta|} \leq C_\Lambda
\label{bzeta}
\ee
so the $\beta$-denominator with $p_1$ can be eliminated.

We will again use a Schwarz inequality but we keep four $\alpha$-
and four $\beta$-denominators on the first power, the
corresponding index sets are $\cA\subset I_k$ and $\cB\subset I_k$:
\be
     \prod_{j=1}^{k} \frac{1}{|\a-\ov\om(p_j)-i\eta|}
   \frac{1}{|\beta-\om(\tp_j)+i\eta|}  \leq  \lambda^{-\rho}(a) + 
\lambda^{\rho}(b)
\label{qs}
\ee
$$
   (a):=\prod_{j\in\cA}
   \frac{1}{|\a-\ov\om(p_j)-i\eta|} \prod_{j'\in\cB}
   \frac{1}{|\beta-\om(\tp_{j'})+i\eta|}
   \prod_{j\in \cA^c}\frac{1}{|\a-\ov\om(p_j)-i\eta|^2}
$$
$$
    (b):=\prod_{j\in\cA}
   \frac{1}{|\a-\ov\om(p_j)-i\eta|}  \prod_{j'\in\cB}
   \frac{1}{|\beta-\om(\tp_{j'})+i\eta|}
\prod_{j'\in \cB^c}
\frac{1}{|\beta-\om(\tp_j)+i\eta|^2}  \; .
$$
Now we explain how we choose the four-element sets  $\cA$ and $\cB$.

We set $\cA:= \{ 1, 2, b-1, b\}$ if $a\ge 2$.
If $a=1$, then
we set $\cA:=\{ 1, 2, 3, k\}$
For the set $\cB$ we set $\cB:=\{ 1, 2, b'-1, b'\}$ 
if $a'\ge 2$. If $a'=1$ and $b'<k$, then we set 
$\cB:=\{ 1, b'-1, b', k\}$. Finally, if
$a'=1$, $b'=k$, we set again $\cB:=\{ 1, 2, b'-1, b'\}$.
The exchange momentum is always $\tp_{b'}$ from the tilde-variables.
For the non-tilde variables we use $p_b$ if $a\ge 2$ and $p_2$
if $a=1$ as exchange momentum. The sets $A$ and $B$ are defined
as before: $A$ contains all $\bp$-momenta and $B$ contains all
$\tbp$ momenta, except the two exchange momenta.

For simplicity, we will neglect all $u$ momenta 
in the formulas below, it can
be checked that they play no role in the arguments.

First we compute (a), expressing everything in terms of $A$-momenta.
The denominator $|\beta-\om(p_1)+i\eta|^{-1}$
disappears by (\ref{bzeta}). We express
$\tp_2 = p_1 + p_{\sigma^{-1}(1)+1}- p_{\sigma^{-1}(1)}$
(with the understanding that if $p_b$ appears, it has
to be reexpressed as $p_{b+1}-p_a+p_{a+1}$) and we use
$\tp_{b'-1} = \tp_{b'} - (p_{m+1} -p_m)$
(with $m:= \s^{-1}(b'-1)$) as before. 

Suppose first that  $a'\ge 2$.
Then only two of the four
$\beta$-denominators contain $\tp_{b'}$, so we can perform 
the $\tp_{b'}$ integration  by (\ref{eq:square}), collecting
$\eta^{-3/4-\kappa}$. Then the last $\beta$-denominator
is eliminated by the $\rd\beta$-integral. For the $\alpha$-denominators
we proceed similarly as before in (\ref{Ialpha}).
We integrate out all but (at most) three squared denominators 
by Lemma \ref{le:opt} and reduce the square to the first power
in the remaining (at most) three denominators
at the expense of $\eta^{-1}$ each. Finally we 
 use  (\ref{weak4d}).
The result is $(a)\leq \lambda^{5-9\kappa}|\log\lambda|^{O(1)}$.

 If $a'=1$, $b'<k$ then 
we observe that $\tp_k$ is independent of 
$\tp_{b'}$, so
the same argument can be used as for $a'\ge2$.

Now we assume that $a'=1$, $b'=k$, then
both $\tp_2$ and $\tp_{b'-1}$ depend on $\tp_{b'}$.
If $\tri \a\tri \ge \Lambda$, then we estimate the $\tp_2$ denominator
by $\eta^{-1}$, integrate out $\rd\beta \rd\tp_{b'}$ at the expense of
$\eta^{-1/2-2\kappa}$ using (\ref{eq:bsquare}).
After reducing the squares of
the $\alpha$-denominator we can use 
 (\ref{strong4d}).
The result is $\lambda^{5-10\kappa}|\log\lambda|^{O(1)}$.
If $\tri \a\tri \le \Lambda$, then we can also remove the $|\alpha -\ov\om(p_1)
-i\eta|^{-1}$ denominator by (\ref{bzeta}). 
We remove $|\a - \ov\om(p_{b+1}-p_a+p_{a+1})-i\eta|^{-1}$ by
supremum norm and note that this was the only $\alpha$ denominator
that may have contained $p_1$. 
If  $\tp_{b'-1}$ does not depend on $p_1$, then we can
integrate out $\rd p_1$ at the expense of $|\log\lambda|$,
then we integrate $\rd\beta \rd\tp_{b'}$ by (\ref{eq:bsquare})
and finish the argument as before to collect 
$\lambda^{5-10\kappa}|\log\lambda|^{O(1)}$.
If $\tp_{b'-1}$ depends on $p_1$, then one can check that
$$
   \tp_{b'-1} = \tp_{b'} - p_1 + (\ldots)
$$
and
$$
   \tp_2 = p_1-  \tp_{b'} + (\ldots)\;,
$$
where ($\ldots$) refers to further $A$-momenta.
Thus we can change integration variables, 
instead of $\rd p_1 \rd\tp_{b'}$ we consider
$\rd(p_1-\tp_{b'}) \rd\tp_{b'}$. We first integrate
out $\rd\tp_{b'}$ (one $\beta$-denominator is eliminated),
then $\rd\beta \rd(p_1-\tp_{b'})$ using (\ref{eq:bsquare}).
The net result of all cases is
\be
     (a) \leq \lambda^{5-10\kappa}|\log\lambda|^{O(1)}\; .
\label{aest}
\ee

Now we turn to the estimate of (b). We express everything 
in terms of $B$-momenta.
We start with the case $a\ge2$.
Only two of the four $\alpha$-denominators depend on $p_b$,
so we can apply (\ref{eq:square}) to perform the $\rd p_b$
integral to remove them.
Then we estimate the denominator $|\beta - \om(\tp_{b'+1}-\tp_{a'}
+\tp_{a'+1})+i\eta|^{-1}$ by the supremum norm, thus removing
the only $\beta$-denominator that may depend on $\tp_1$.
Finally we use (\ref{eq:bsquare}) to integrate out
$\rd\a \rd \tp_1$ if $p_2$ depend on $\tp_1$, if not, then the estimate
is even better.
 We collect $\lambda^{7/2-6\kappa}|\log\lambda|^{O(1)}$.

Now consider the case $a=1$. We first assume $b<k$, then
$p_k$ is indepedent of $p_2$ (when expressed in terms of $B$-momenta),
so we can use (\ref{eq:square}). Then we again estimate
$|\beta - \om(\tp_{b'+1}-\tp_{a'}
+\tp_{a'+1})+i\eta|^{-1}$ by the supremum norm, thus removing
the only $\beta$-denominator that may depend on $\tp_1$.
If $p_k$ depends on $\tp_1$, we use (\ref{eq:bsquare}) to
integrate  $\rd\a \rd \tp_1$, in the other case the estimate
is better. We again obtain $\lambda^{7/2-6\kappa}|\log\lambda|^{O(1)}$.

Finally, we consider the case $a=1$, $b=k$. 
We estimate  $|\beta - \om(\tp_{b'+1}-\tp_{a'}
+\tp_{a'+1})+i\eta|^{-1}$  by supremum norm
so no $\beta$ denominator can depend on $\tp_{k+1}$. We express
$p_k= \tp_{k+1} - \tp_1+ p_2$. If $p_3$ depends on $\tp_{k+1}$,
then we perform $\rd\tp_{k+1}$ using (\ref{eq:square}),
then we can perform $\rd p_2$ removing one $\alpha$-denominator and
finally we integrate out $\alpha$ to
remove the last $\alpha$-denominators (with $p_1$).
The remaining $\beta$-denominators
are independent and we collect $\lambda^{9/2-4\kappa} |\log\lambda|^{O(1)}$.
If $p_3$ does not depend on $\tp_{k+1}$, then we
integrate out $\rd p_{k+1}$ to remove one $\alpha$-denominator
(with $p_k$)
and collecting $|\log\lambda|$. Then we integrate $\tp_2$
using (\ref{eq:square}) and finally perform the $\alpha$-integration.
The result is again  $\lambda^{9/2-4\kappa} |\log\lambda|^{O(1)}$.

In summary, we obtain
\be
(b)\leq \lambda^{7/2-6\kappa}|\log\lambda|^{O(1)},
\label{best}
\ee
and together with (\ref{aest}) and 
optimizing for $\rho$ in (\ref{qs}) we obtain 
(\ref{eq:recesttr}). This completes the proof of
Proposition \ref{prop:rec}. $\;\;\Box$

\bigskip

{\it Proof of Proposition \ref{prop:rec1}.}
This proof is  similar to the previous
one but simpler. 
We choose
the set of $A$ and $B$ momenta are as follows:
$$
    A: = \{ p_1, p_2, \ldots , p_{b-1}, p_{b+1}, \ldots , p_{k+1}\}\; ,\qquad
    B:= \{ \tp_1, \tp_2, \ldots , \tp_{k-1}, p_{b}\} \;
$$

If $\tri\a\tri\ge\Lambda$, then 
the Schwarz estimate is the following
$$
     \prod_{j=1}^{k+1} \frac{1}{|\a-\ov\om(p_j)-i\eta|}
   \prod_{j=1}^{k-1}\frac{1}{|\beta-\om(\tp_j)+i\eta|}  \leq \lambda^\rho  
 (a) + \lambda^{-\rho}(b)
$$
$$
   (a):=\frac{1}{|\beta-\om(\tp_1)-i\eta|}
  \prod_{j=1, b} \frac{1}{|\a-\ov\om(p_j)-i\eta|}
   \prod_{j=2\atop j\neq  b}^{k+1} \frac{1}{|\a-\ov\om(p_j)-i\eta|^2}
$$
$$
    (b):=\frac{1}{|\beta-\om(\tp_1)-i\eta|}
  \prod_{j=1,  b} \frac{1}{|\a-\ov\om(p_j)-i\eta|}
  \prod_{j=2}^{k-1}\frac{1}{|\beta-\om(\tp_j)+i\eta|^2}\; .
$$
To estimate the integral of $(a)$,
 we express $\tp_1=p_1$
 and $p_b=p_{b+1}- p_a+p_{a+1}
- u_a$, so every term in (a) will depend only on $A$-momenta.
We first integrate $\rd\beta$, then integrate all $p_j$,
$j\neq a, a+1, b+1$ and reduce the square of the
remaining denominators to the first power. Finally
we use  Lemma \ref{lemma:4de}.  The
result is $C\lambda^{\rho-7\kappa}|\log\lambda|^{O(1)}$.
The integral of (b) is even easier, after expressing $p_1=\tp_1$,
we can integrate it out in any order with an estimate
$C\lambda^{2-\rho}|\log\lambda|^{O(1)}$. After optimizing for $\rho$
this gives $\lambda^{1-7\kappa/2}|\log\lambda|^{O(1)}$
as announced in (\ref{eq:halfrecest}).

If $\tri \a \tri \leq\Lambda$, then we use
$$
     \prod_{j=1}^{k+1} \frac{1}{|\a-\ov\om(p_j)-i\eta|}
   \prod_{j=1}^{k-1}\frac{1}{|\beta-\om(\tp_j)+i\eta|}  \leq \lambda^\rho  
 (a) + \lambda^{-\rho}(b)
$$
$$
   (a):=\prod_{j=1,2}\frac{1}{|\beta-\om(\tp_j)-i\eta|}
  \prod_{j=1,b-1, b} \frac{1}{|\a-\ov\om(p_j)-i\eta|}
   \prod_{j=2\atop j\neq b-1, b}^{k+1} \frac{1}{|\a-\ov\om(p_j)-i\eta|^2}
$$
$$
    (b):=\prod_{j=1,2}\frac{1}{|\beta-\om(\tp_j)-i\eta|}
  \prod_{j=1,b-1,  b} \frac{1}{|\a-\ov\om(p_j)-i\eta|}
  \prod_{j=3}^{k-1}\frac{1}{|\beta-\om(\tp_j)+i\eta|^2}\; .
$$
We use (\ref{bzeta}) to eliminate $|\a - \ov\om(p_1)-i\eta|^{-1}$.

In the term (a) we express everything in terms of $A$-momenta.
We estimate $|\a - \ov\om(p_{b+1}-p_a+p_{a+1})-i\eta|^{-1}$
by $\eta^{-1}$. 
Thus no more $\alpha$-denominator depends on $p_1$.
Depending on whether $\tp_2$ depends on $p_1$ or not, we
can use (\ref{eq:bsquare}) or subsequent $\rd p_1$ and $\rd\beta$
integrations to remove all $\beta$-denominators. The remaining
$\alpha$-denominators are independent and we collect
$$
    (a)\leq \lambda^{-1-2\kappa} |\log\lambda|^{O(1)}
$$

In the term (b) we use the $B$-momenta for integration.
Here only two $\alpha$-denominators depend on $p_b$, so we
can use  (\ref{eq:bsquare}) to perform $\rd p_b \rd\a$, then
all $\beta$-denominators are integrated independently. We obtain
$$
   (b)\leq \lambda^{3-2\kappa}  |\log\lambda|^{O(1)} \; .
$$
After optimizing for $\rho$, we obtain 
a bound smaller than (\ref{eq:halfrecest}).

The proof of
(\ref{eq:halfrecesttr}) is the same, but the last
squared denominators are missing, this is where the gain $\lambda^2$
comes from. $\;\;\Box$

\subsection{Cancellation with a gate}

The proof of (2a)--(2c) of Proposition~\ref{prop:i-iv}
depends on a cancellation mechanism between a gate and a 
$\theta$ label. More precisely, if two Feynman graphs
differ only by replacing a gate with a $\theta$-label,
then their sum is by a factor 
$\lambda^2\eta^{-1/2}$ smaller than the $E$-value of
the two partitions individually. Moreover, this
cancellation effect is local in the graph:
if another gate/$\theta$ pair occurs somewhere
else in these graphs, doubling their number,
 then the sum of these four Feynman diagrams
is smaller by a factor $(\lambda^2\eta^{-1/2})^2$.
For the general statement, see Lemma 5.5 in \cite{ESYII}.
Although this lemma is formulated for the continuum model,
taking the collision function $\wh B\equiv 1$
and considering all momemtum integrals in $\Tor^3$,
the proof of Lemma 5.5 goes through for the lattice case as well.

The detailed proofs of (2a)--(2c) follow the arguments
of Section 5.3.2--5.3.4 of \cite{ESYII} line by line
and will not be repeated here. We only point out the three minor differences:

\medskip

(i) The $\lambda$-exponent in the
estimate on $E_{(*)}(A, \sigma, \bu)$ given in
Corollary~\ref{cor:jointdeg} differs from  
its continuum counterpart (9.4) of \cite{ESYI};

\medskip

(ii) In the continuum model, each application of
 Operation I  costs a factor $\lambda^{-2d\kappa-O(\delta)}$
(denoted by $\Lambda$ in \cite{ESYI, ESYII}), this loss is absent here;

\medskip

(iii) The $\lambda$-power
in the estimates \eqref{eq:recesttr}--\eqref{eq:halfrecesttr}
are weaker than their continuum analogues (Propositions 
5.2 and 5.3 of \cite{ESYII}).

\medskip

These changes account for the somewhat different $\lambda$-powers
in (2a)--(2c) of Proposition~\ref{prop:i-iv} compared with 
Proposition 4.6 of \cite{ESYII}. $\;\;\Box$

\bigskip

\section{The main term: Proof of Theorem~\ref{thm:laddheat}}
\label{sec:wigner}
\setcounter{equation}{0}

For simplicity, all results in this section
are written for $d=3$, the calculation for higher dimensions is
similar.
We follow  a different path than in the proof of the
analogous theorem
 in Section 6 of \cite{ESYII}. Due to the uniformity of 
the Boltzmann collision kernel \eqref{def:sigma}, we can circumvent
the reference to the Boltzmann process and we 
identify the heat equation by a
direct computation.

As in Section 6 of \cite{ESYII}, we start with the identity
$$
    W_\lambda(t, k, \cO) = \int_{(2\Tor/\e)^d} V^\circ_{\e\xi}
  \big( \bA_0, \wh\cO(\xi, \cdot)\big)\rd \xi, \qquad k\ge1
$$
with $\bA_0$ being the trivial partition on $I_k$,
where we chose the function $Q(v)$ in the definition
of $V^\circ$ to be $\xi$-dependent, namely $Q(v)= Q_\xi(v):=\wh\cO(\xi, v)$
(see \eqref{def:Vlong} and \eqref{def:Vshortcirc} for definitions).

Analogously to the argument in Section 6 of \cite{ESYII},
 the $\rd\xi$ integration
can be restricted to the regime $\{|\xi|\le\lambda^{-\delta}\}$
with a negligible error (even after summation over $k$):
\be
    \sum_{1\leq k<K}W_\lambda(t, k, \cO) = \sum_{1\leq k<K}
 \Xi^\circ_k +o(1)\; , \qquad
\Xi^\circ_k:=\int^* V^\circ_{\e\xi}
  \big( \bA_0, \wh\cO(\xi, \cdot)\big)\rd \xi \; ,
\label{xires}
\ee
where  we used the notation 
$$
   \int^* \big(\cdots\big)\rd\xi : = \int_{(2\Tor/\e)^d}
 \big(\cdots\big){\bf 1}(|\xi|\leq 
\lambda^{-\delta} )\rd \xi\; .
$$
The bound \eqref{xires} follows from 
Lemma~\ref{lemma:VV}, from
$\big| V_{\e\xi}
  \big( \bA_0, \wh\cO(\xi, \cdot)\big) \big| \leq \|Q_\xi\|_\infty
  \sup_{\xi, \bu} E(\sigma=id, \bu)$,
from the uniform bound  \eqref{suppe}--\eqref{etriv} on
$ E(\sigma=id, \bu)$
 and from the fast decay
of $\| Q_\xi\|_\infty = \sup_v |\wh\cO(\xi, v)|$ in $\xi$.

\medskip
Writing out the definition of $\Xi^\circ_k$ more explicitly, we have
\begin{align}
\Xi_k^\circ= & 
\lambda^{2k}
\iint_\bR \frac{\rd \alpha \rd \beta}{(2\pi)^2}
 \; e^{it(\alpha-\beta)+ 2 t \eta} \int^*  \rd \xi 
\int  \prod_{j=1}^{k+1} \rd v_j \;  
\wh\cO(\xi, v_{k+1})\overline{\wh W_{\psi_0}}(\e\xi, v_1) 
  \nonumber \\
 & \times 
\prod_{j=1}^{k+1} \Big[ 
  \ov{R_\eta\Big(\a, v_j +\frac{\e\xi}{2}\Big)}
   R_\eta\Big(\beta, v_j -\frac{\e\xi}{2}\Big)\Big] \;.
\label{9.511}
\end{align}

The estimates of the error terms were performed with the 
choice $\eta= \lambda^{2+\kappa}$.
However, $\Xi_k^\circ$,  given by \eqref{9.511}, is 
independent of $\eta$.
 Therefore we can change the  value of 
$\eta$ to $\eta:=\lambda^{2+4\kappa}$
for the rest of this calculation and we define
$$
        R(\a, v):=  R_{\eta}(\a, v) \; , \qquad \mbox{with} 
\quad \eta:=\lambda^{2+4\kappa} \; .
$$
We   recall that the restriction of the $\rd\alpha\rd\beta$
integration in \eqref{9.511} to any set that contains 
$\{ \a, \beta \, : \,|\a|, |\beta|\leq 4d \}$
 results in
negligible errors, even after the summation over $k$
(Lemma~\ref{lemma:VV}).
We will consider the set  $D:= \{ (\a, \beta) \, : \,
 |\a + \beta|\leq 8d, |\a-\beta|\leq 8d\,\}$.
We denote by $\Xi_k$ the version of $\Xi_k^\circ$
given by  formula \eqref{9.511} with the $\rd\a\rd\beta$ 
integrals  restricted to $D$,
$$
    \Xi_k : = \lambda^{2k} \iint_{D} \frac{\rd \alpha \rd \beta}{(2\pi)^2}
 \; \Big[ \mbox{Integrand from \eqref{9.511}} \Big] \; ,
$$
then
$$
   \sum_{1\leq k\leq  K} |\Xi_k^\circ- \Xi_k| = o(1) \; .
$$
We also remind the reader  that this
 argument does not apply literally 
to the trivial $k=0$ case, when the $\rd\a\; \rd\beta$ integral in
\eqref{9.511} gives free evolutions and this term is computed
directly:
\be
     \Xi_0: = \int^* \rd\xi\rd v\;
    e^{it\e \xi\cdot \nabla e(v)}\; e^{2t\lambda^2 \mbox{Im}\, \theta (v)}\;
 \wh\cO(\xi, v)\ov{\wh W_{\psi_0}}(\e\xi, v) + o(1)\; .
\label{xi01}
\ee
The error term comes from the error term in $\ov{\om}(v+\e\xi/2)
- \om(v-\e\xi/2) = \e\xi\cdot \nabla e(v) + 2 i \lambda^2 \cI(v) + 
O(\lambda^2\e\xi)+ O(\e^2\xi^2)$. 
By using $t\lambda^2 \to \infty$, the bound \eqref{eq:lowim},
 and the decay of the observable,
 one easily obtains that  $|\Xi_0| = o(1)$.

\bigskip

We start with a crucial technical lemma which is proven 
 in the Appendix. 

\begin{lemma}\label{lemma:opt1}
Let $\kappa < 1/18$ and set $\gamma: = (\a +\beta)/2$.
Let  $\eta$ satisfy
 $\lambda^{2+ 4\kappa}\leq \eta\leq \lambda^{2+\kappa}$.
Then for $|r|\le \lambda^{2+\kappa/4}$ and any $f\in C^1(\Tor^d)$ we have,
\begin{align}\label{eq:optical}
& \int \frac {\lambda^2 f(p) }{\big( \a - \ov\om(p-r)
 - i\eta \big)
     \big(\beta - \om(p+r)  
+i\eta \big)} \; \rd p  \\
&     =   -2\pi i\lambda^2\int  
 \frac{ f(p)\; \delta(e(p)-\gamma)}{ (\a-\beta)
 + 2 (\nabla  e)(p) \cdot r -
       2 i [\lambda^2  {\cal I} (\gamma)+\eta ]} \, 
\rd p + O(\lambda^{1/2-9\kappa})\| f \|_{C^1} \; .
\nonumber
\end{align}
\end{lemma}

\bigskip

We now apply this lemma to compute $\Xi_k$. Introduce new
variables as
 $a:=(\a+\beta)/2$ and $b:=\lambda^{-2}(\a-\beta)$.
Then
\begin{align}\label{9.3}
\lambda^2 \int & \rd v  \Upsilon(\xi, v)\;  
\ov{R\Big(\a, v +\frac{\e\xi}{2}\Big)}
   R\Big(\beta, v -\frac{\e\xi}{2}\Big)  \\
&
= \int
    \frac{-2\pi i \Upsilon(\xi, v)\;
     \delta(e(v)-a)}{  b +  \e\lambda^{-2} (\nabla e)(v)\cdot \xi -
       2 i [ {\cal I} (a)+\lambda^{4\kappa}]} \; \rd v
       + O(\lambda^{1/2-9\kappa})\|\Upsilon\|_{4d, 1}\; .
\nonumber
\end{align}

We now replace the  product of $k+1$ factors in 
the restricted version of \eqref{9.511} one by one. 
We need a $\lambda^2$ factor for each application of  (\ref{9.3}).
The $(k+1)^{\rm st}$ factor $\lambda^2$ comes from the change of variables 
$\rd\alpha \rd \beta = \lambda^2 \rd a \rd b$.
We also define $D^*:=\{ (a,b)\; : \; |a|\leq 4d, \; |b| 
\leq 8d\lambda^{-2} \}$ as the domain $D$ in the new variables.

Introduce the notation
$$
F_1(\xi, v):= \ov{\wh W_{\psi_0}}(\e\xi,v), \; 
\qquad F_{k+1}(\xi, v):= \wh\cO(\xi, v)\;
$$
and $F_j(\xi, v):= 1$ for $j=2,\ldots k$. 
Using \eqref{9.3} with $\Upsilon = F_j$
and using that $\|\cO\|_{C^1}$ and $\| \wh W_{\psi_0} \|_{C^1}$ are bounded, 
 we obtain by a telescopic summation that
\begin{align}\label{9.4}
   \Bigg| \sum_{k<K} \Xi_k- \sum_{k<K} & \int^* \rd \xi
   \int_{D^*}
 \frac{\rd a  \rd b}{(2\pi)^2}  \; e^{it\lambda^2b+ 2 t \eta}
 \nonumber \\
&\times
\Bigg(\prod_{j=1}^{k+1} \int
   \frac{-2\pi i F_j(\xi, v_j) \;
 \delta(e(v_j)-a) }{ b + \e\lambda^{-2} (\nabla e)(v_j)\cdot \xi -
       2 i [\cI (a)+\lambda^{4\kappa}]} \rd v_j \Bigg)  \Bigg| \nonumber \\
&   \leq  \sum_{k<K} \sum_{\ell \leq k}  \cF_{k,\ell} + o(1)
\end{align}
with
\begin{align}
 \cF_{k,\ell}:= O(\lambda^{1/2-9\kappa}) & 
\int^* \rd \xi   \int_{D^*}
 \frac{\rd a  \rd b}{(2\pi)^2} 
   \prod_{j=1}^{\ell-1}  \Bigg(  \int
   \Bigg|  \frac{2\pi i \; \delta(e(v_j)-a)}{b+
 \e\lambda^{-2} (\nabla e)(v_j)\cdot \xi -
       2 i [\cI (a)+\lambda^{4\kappa}]}\Bigg|\rd v_j\Bigg)\nonumber \\
&
  \times \prod_{j=\ell+1}^{k+1} \Bigg( \lambda^2\int
  \Bigg |R\Big(\a, v_j +\frac{\e\xi}{2}\Big)
  R\Big(\beta, v_j -\frac{\e\xi}{2}\Big)\Bigg|  \rd v_j\Bigg) \; .
\label{cf}
\end{align}

The  first factor in  \eqref{cf} is zero if $a\not\in (0,2d)$,
otherwise it can be estimated by
\be
     \int
   \Bigg|  \frac{2\pi i \; \delta(e(v)-a)}{b + \e\lambda^{-2}
 (\nabla e)(v)\cdot \xi -
       2 i [ \cI (a)+\lambda^{4\kappa}]}\Bigg|\rd v\leq
       \frac{1}{\cI(a)}\int \; \pi \; \delta(e(v)-a) \rd v
      = 1\;
\label{trivv}
\ee
by using (\ref{eq:opt}).
The second factor in \eqref{cf}
can by bounded by the Schwarz inequality and by \eqref{eq:ladderint}:
$$
    \lambda^2 \sup_\a \int |R(\a, v)|^2 \rd v 
\leq 1 + O(\lambda^{1-12\kappa}) \; .
$$
Thus the right hand side of \eqref{9.4} vanishes  in the limit $\lambda \to 0$
since the double summation yields only a factor 
$K^2= O(\lambda^{-2\kappa-2\delta})$.

\bigskip

Now we concentrate on the main term, i.e. on the sum
on the left hand side of  (\ref{9.4}).
We first remove $\lambda^{4\kappa}$ 
from the denominator in the main term in (\ref{9.4}).
To estimate this replacement error, we first notice that
the large regimes of $a$ or $b$  are harmless. Due to
$\delta(e(v_j)-a)$, the integrand is zero unless $a\in (0, 2d)$.
In the large $b$ regime,
each denominator can be estimated by $(|b| + 1)^{-1}$ since 
$|\e\lambda^{-2}(\nabla e)(v_j)\cdot \xi|\ll 1$ and
there are at least two denominators ($k\ge 1$), so the tail of the
$\rd b$ integration is small.

The integration  in $a$ is divided into two cases.
In the regime where
$\cI(a)\ge \lambda^{3\kappa/2}$, the error of the
replacement in one denominator is bounded by 
$$
     C\| F_j\|_\infty \int \frac{ \lambda^{4\kappa}\; 
 \delta(e(v)-a)}{  {\cal I} (a) [ {\cal I} (a)+\lambda^{4\kappa}] }\; \rd v
\leq O(\lambda^{5\kappa/2})\; ,
$$
by using a resolvent expansion. 
For the complement regime, 
$\cI(a)\leq \lambda^{3\kappa/2}$, we will use the trivial
estimate \eqref{trivv}.
Therefore we obtain
\begin{align}\label{9.31}
 \int &
   \frac{-2\pi iF_j(\xi, v_j) \; \delta(e(v_j)-a) }{  
b + \e\lambda^{-2}(\nabla e)(v_j)\cdot \xi -
       2 i [\cI (a)+\lambda^{\kappa/4}]}\;  \rd v_j  \nonumber \\
&
= \int
    \frac{-2\pi i F_j(\xi, v_j)\;
     \delta(e(v_j)-a)}{ b +  \e\lambda^{-2}(\nabla e)(v_j)\cdot \xi -
       2 i  {\cal I} (a)} \; \rd v_j
       + O(\lambda^{5\kappa/2})
 + O(1){\bf 1}( \cI(a)\leq \lambda^{3\kappa/2}) \; .
\end{align}

In the regime $\cI(a)\ge \lambda^{3\kappa/2}$,
similarly to the telescopic estimate leading to
(\ref{9.4}), we can remove the 
$\lambda^{\kappa/4}$ terms
from each denominator on the left hand side of (\ref{9.4})
 one by one, 
the error is controlled by $CK^2 \lambda^{-\delta d} 
\lambda^{5\kappa/2}\to0$ if $\delta$ is sufficiently small. 
For $\cI(a)\leq  \lambda^{3\kappa/2}$   we use  the
 trivial estimate (\ref{trivv}) for each denominator, and the fact
that
\begin{align}
    \int \rd a \; {\bf 1}(\cI(a)\leq \lambda^{3\kappa/2})& 
\int |\wh W_{\psi_0} (\e\xi, v_1)|\;  \delta( e(v_1)-a) \rd v_1 
    \nonumber\\
   & \leq C \|\wh \psi_0\|^2_\infty \int \cI(a)  \;
 {\bf 1}(\cI(a)\leq \lambda^{3\kappa/2})\rd a
    = C \lambda^{3\kappa/2}\|\wh \psi_0\|^2_\infty \; .
\label{eq:smalla}
\end{align}
Together with  the summation and the $\rd \xi$-integration,
this yields an error
 term of order $ K\lambda^{3\kappa/2}\lambda^{-\delta d}\to0$.
We therefore obtain
\begin{align}\label{fulll}
   \sum_{k<K}\Xi_k&  =  o(1)   \\
& +\sum_{k\leq K}  \int^* \rd \xi 
  \!\!  \iint_\bR \frac{\rd a  \rd b}{(2\pi)^2}   \; e^{it\lambda^2 b+ 2 t \eta}
\Bigg(\prod_{j=1}^{k+1} \int
   \frac{-2\pi i F_j(\xi, v_j) \; \delta(e(v_j)-a) }{ b + 
\e\lambda^{-2}
      (\nabla e)(v_j)\cdot \xi -
       2 i  \cI (a)} \rd v_j \Bigg) \; . \nonumber
\end{align}
Notice that we removed the constraint $\cI(a)\ge \lambda^{3\kappa/2}$
after having used it and we extended the $\rd a\rd b$ integration from
$D^*$ to $\bR^2$.  The corresponding errors are negligible
by arguments similar to the previous ones.

Next we perform  a change of variable
$b\to b(2 \cI (a))^{-1}$ so that
$$
\rd b  \; e^{it\lambda^2 b+ 2 t \eta}
\to  2 \cI (a)\;  \rd b  \; e^{i2\lambda^2 \cI(a) t b + 2 t \eta}
$$
Since $t \eta = \lambda^{3\kappa}\to0$, 
we shall drop the $e^{2t\eta}$ factor. Denote $H(v):=
 \frac{\nabla e (v)}{2\cI (a) }$.
Introduce the probability  measure $\rd\mu_a(v)$ on
the level surface $\Sigma_a=\{ e(v) = a\}$ by
$$
     \int h(v) \rd\mu_a(v): = \langle h \rangle_a = \frac{\pi}{\cI (a)}
     \int h(v) \delta(e(v)-a)\rd v\; .
$$
Then
$$
   \int \frac{-2\pi i F_j(\xi, v) 
\;\delta( e(v)-a) }{ b + \e\lambda^{-2} (\nabla e)(v)\cdot \xi -
       2 i  \cI(a)} \; \rd v \mapsto \;\;
    \int \frac{- i  F_j(\xi, v)}{ b +\e\lambda^{-2}H (v)
     \cdot \xi - i } \;\rd\mu_a(v)
$$
with the new variable $b$ on the right hand side.

From \eqref{2.1}, $H$ is bounded above, i.e.,
$    \sup_{v} |H(v)|\leq C$.
With this bound,
we  expand the fraction up to second order in
$\e\lambda^{-2}|\xi| \leq \lambda^{\kappa/2-\delta}\ll 1$
for $2\leq j\leq k$
\begin{align}\label{9.7}
&\int
   \frac{- i  }{ b +\e\lambda^{-2}H (v)
     \cdot \xi - i }\; \rd\mu_a(v) \nonumber \\
& = \frac{-i }{ b  -i  } \int
     \Bigg[ 1 - \frac{\e\lambda^{-2}
       H(v)\cdot \xi}{b- i }
     +\frac{ \e^2\lambda^{-4} [H (v)\cdot \xi]^2}{ (b-  i)^2}
     + O\Big( (\e\lambda^{-2} |\xi|)^3 \Big)\Bigg]  \rd\mu_a(v)
\end{align}
Since $\e=\lambda^{2+\kappa/2}$,
 $|\xi|\le \lambda^{-\delta}$ and $K=O(\lambda^{-\kappa-\delta})$,
the last error term, $O\Big( (\e\lambda^{-2} |\xi|)^3 \Big)$,
even after summation
in $k$, is negligible:
\be
       K (\e\lambda^{-2} |\xi|)^3 =o(1) \; .
\label{eq:higher}
\ee
By symmetry, $H(v)=-H(-v)$.
Therefore
the linear term in $\xi$  on the right hand side of (\ref{9.7})
 vanishes after the  $\rd\mu_a$ integration since $\rd\mu_a(v)=\rd\mu_a(-v)$.

For $j=1, k+1$ we will use the following simple estimate
\begin{align}\label{9.8}
\int
   \frac{- i F_j(\xi, v) }{ b +\e\lambda^{-2}H (v)
     \cdot \xi - i }\rd\mu_a(v)
& = \frac{-i }{ b  -i  } \Bigg[
   \int
   { F_j(\xi, v) }\rd\mu_a(v)+ o(1)\Bigg]\; .
\end{align}
We also
 define the matrix
$$
    D(a):
   = \frac{2 \cI (a)}{(2\pi)^2}\int \rd\mu_a(v)\;  H(v) \otimes H(v) 
    =  \frac{1}{2\pi\Phi(a)} \langle \; \sin (2\pi v)\otimes \sin (2\pi v)
   \rangle_a \;
$$
using  $\cI(e)= \pi\Phi(e)$. The quadratic form of $D(a)$ is denoted
by $(\xi, D(a)\xi)$, $\xi\in \bR^d$.

By applying \eqref{9.8} to $j=1, k+1$ and \eqref{9.7} to the rest
of the $\rd v_j$ integrals in \eqref{fulll},
we obtain 
\begin{align}\label{OW}
   \sum_{1\leq k<K} W_\lambda(t,k, \cO) 
=  & \frac{1}{(2\pi)^2}   \int^* \rd \xi
   \int_\bR \rd a \; 2\cI(a)
\int \ov{ \wh W_{\psi_0}}(\e\xi, v_1)
   \rd\mu_a(v_1)   \nonumber\\
& \times  \int \wh\cO(\xi, v) \rd\mu_a(v)
 \int_\bR \rd b  \; e^{2i\lambda^2\cI (a)tb}\\
&
\times\sum_{1\leq k< K}
\Big( \frac{-i }{ b-i }\Big)^{k+1} \Bigg[ 1
     + \frac{ (2\pi)^2\e^2\lambda^{-4} (\xi\cdot D(a)\xi)}{2 \cI(a)}
     \frac{ 1}{      (b-i )^2}\Bigg]^{k-1}\; ,\nonumber
\end{align}
modulo negligible errors. 
We shall also change the last exponent from $k-1$ to $k+1$ to simplify
 the computation.
Since $\e^2\lambda^{-4} \to 0$, this modification causes only negligible
errors.

Since $\e$ is small, we can replace
$\overline{\wh W_{\psi_0}}(\e\xi, v_1)$ by $\wh W_{\psi_0}(0, v_1)\in \bR$ 
with a negligible error, by 
recalling that $\wh\psi_0 \in C^1$ (Section \ref{sec:trun}),
in particular,
$$
\int \ov{\wh W_{\psi_0}}(\e\xi, v_1) 
   \rd\mu_a(v_1)  = 
g(a) + O(\e)\;, \qquad g(a):=
\big[ |\wh \psi_0|^2\big]_a\; .
$$
Let $A:= 2\lambda^2 \cI(a)\neq 0$ and
$B:= 2\pi \e\lambda^{-2}
\Big(\frac{ ( \xi, D(a)\xi)}{2\cI(a)}\Big)^{1/2} $.
Note that $B\ll 1$ uniformly in $a$.

Suppose that we can extend the summation in $k$  to infinity in 
(\ref{OW}). Then
$$
  \sum_{k=0}^\infty \Big( \frac{-i}{ b-i}\Big)^{k+1}
  \Bigg[ 1 +\frac{B^2}{(b-i)^2}\Bigg]^{k+1}
   = (-i)\frac{(b-i)^2 + B^2}{ (b-i)^3 +i (b-i)^2 + iB^2}
$$
and we compute
$$
(-i) \int_\bR \rd b  \; e^{itAb}\;
\frac{(b-i)^2 + B^2}{ (b-i)^3 +i (b-i)^2 + iB^2} \; .
$$
The roots of the denominator are $b_1, b_2, b_3$  and using that $B\ll 1$,
we obtain
$$
   b_1= iB^2 + O(B^4), \qquad b_2= i+iB + O(B^2), \qquad b_3= i-iB + O(B^2)\; .
$$
The roots $b_2, b_3$ give exponentially small contributions
since $e^{itAb_{2,3}} \leq e^{-tA} = \exp (-\lambda^{-\kappa}\Phi(a))$.
The main contribution comes from $b_1$, so
$$
  (-i) \int_\bR \rd b  \; e^{itAb}\;
\frac{(b-i)^2 + B^2}{ (b-i)^3 +i (b-i)^2 + iB^2}
  = 2\pi e^{-tAB^2} (1 + O(B)) + o(1)\; .
$$
We have
$$
    tAB^2 = (2\pi)^2\e^2\lambda^{-4-\kappa} T ( \xi, D(a)\xi)
   =(2\pi)^2 T ( \xi, D(a)\xi)  .
$$
Note that to get a nontrivial limit, the space scale $\e$ has to
be chosen as $\e=\lambda^{-2-\kappa/2}$ here.
Thus we obtain
\begin{align}\label{1}
\lim_{\lambda\to0} \sum_{1\leq k<K}& W_\lambda(t,k, \cO) = \\
 &=\int_{\bR^d} \rd\xi 
\int_\bR \frac{\rd a}{2\pi}
   \; 2\cI (a)
\Big( \int \cO(\xi, v) \rd\mu_a(v)\Big)
 g(a)  \exp\Big( - (2\pi)^2T (\xi, D(a)\xi) \Big) \nonumber  
\end{align}
Since $f(T, X, a)$ satisfies the heat equation \eqref{eq:heat}
with initial condition $f(0,X,a):= \delta(X) g(a)$, its
Fourier transform in $X$ is given by
$\wh f(T, \xi , a)= g(a)  \exp( - (2\pi)^2T ( \xi, D(a)\xi) )$.
Thus
\begin{align}
\label{12}
\lim_{\lambda\to0} \sum_{1\leq k<K} W_\lambda(t,k, \cO)
& =  (2\pi)^{-1} \int_{\bR^d} \rd\xi \int_\bR \rd a \; 2\cI(a)
\int \wh\cO(\xi, v) \rd\mu_a(v) \wh f(T, \xi, a)   \nonumber\\
& = \int_{\bR^d} \rd\xi \int \rd v \;
\wh \cO(\xi, v)  \wh f(T, \xi, e(v))   \; ,\nonumber
\end{align}
where we used the definiton of $\rd\mu_a$.
This proves Theorem \ref{main}.

\bigskip

It remains to prove the contribution from $k\ge K$ is negligible.
 By the residue theorem
\be\label{res}
\int_\bR \rd b  \; e^{itAb}\;
\Big( \frac{-i}{ b-i}\Big)^{k+1} \Bigg[ 1 +\frac{B^2}{(b-i)^2}\Bigg]^{k+1}
= 
(2\pi)  \frac{(At)^k}{k!}e^{-tA} \sum_{\ell=0}^{k+1}
   C_{k,\ell}
  \frac{1}{\ell!}\Big[-\frac{(B A t)^2}{k}\Big]^\ell
\ee
with
$$
   C_{k,\ell}: =  \frac{k! (k+1)! k^{\ell}}{(k+2\ell)!(k+1-\ell)!} .
$$
It is easy to check that
$$
   C_{k,0}=1, \qquad C_{k,\ell}\leq 1\; .
$$
If $k\ge K = O(\lambda^{-\kappa-\delta})$, then $At/k =O( \lambda^\delta)$
and $(BAt)^2/k\le \lambda^{\delta}$ (using that $\cI(a)$ and $D(a)$ 
are bounded),
 so \eqref{res} is smaller than
$$
      \frac{(At)^k}{k!}e^{-tA} \leq C^k\lambda^{k\delta} \; ,
$$
which is negligible even after summing up for all $k\ge K$.
 $\;\;\Box$

\appendix

\section{Estimates on Propagators}
\setcounter{equation}{0}

\subsection{Proof of Lemma \ref{le:opt}.}\label{sec:33}

Let $\| g \|_{C^1} = \| g \|_\infty+ \| \nabla g \|_\infty$ denote
 the $C^1$ norm
of a function on the torus $\Tor^d$. We start with the following Lemma.

\begin{lemma}\label{le:A1} For any $\alpha, \alpha'$ real and 
$\e \ge \e'> 0$, we have in
$d\ge 3$,
\be\label{A3}
\int \left [ \frac 1 {\alpha- e(p) + i \e}
- \frac 1 {\alpha'- e(p) + i \e'} \right ] g (p) \rd p
\le C\Big[ \e^{-1/2}|\alpha- \alpha'|
    + \e^{1/2} \Big] \| g \|_{C^1}\; .
\ee
In particular, with $g=1$ we have
\be
     | \Theta_\e (\alpha) -\Theta_{\e'} (\alpha')|  \le 
C\Big[ \e^{-1/2}|\alpha- \alpha'|
    + \e^{1/2} \Big] \; .
\label{eq:Thetaest}
\ee
The functions $\Phi$ (from \eqref{co-area}) and $\Theta$ are
 H\"older continuous with exponent $\frac{1}{2}$ for $d\ge3$:
\be\label{holder}
     |\Phi(e)-\Phi(e')|, \;\; |\Theta(e)-\Theta(e')| \leq C|e-e'|^{1/2} \; .
\ee
We also have
\be\label{A4}
\int \left [ \frac 1 {\alpha- \om(p)+ i \eta}
- \frac 1 {\alpha- e(p) + i \eta} \right ] g (p) \rd p
\le C \lambda^{1-4\kappa}|\log\lambda|\;  \| g \|_{C^1}
\ee
for any $\eta$ satisfying $\lambda^{2+4\kappa}\leq\eta\leq \lambda^{2+\kappa}$.
\end{lemma}

\noindent
{\it Proof.} 
We shall consider the case $d=3$ only, for $d > 3$ the proof is similar.
We first describe a local coordinate system.
Let $I_+^j = \{p_j: |p_j|\le 2\pi/3 \}$
and $I_-^j = \{p_j: |p_j|\ge \pi/3 \}$.
Let $\sigma \in \{(\pm, \pm, \pm)\}$. We can find
a smooth partition of unity $\chi_\sigma$ in the sense that
\be\label{unity}
\sum_\sigma \chi_\sigma(p) = 1
\ee
and $\chi_\sigma$ is supported in $I_\sigma= \prod_{j=1}^3  I^j_{\sigma_j}$.
Each $I_\sigma$ contains exactly one critical point of $e(p)$.

Clearly, $g(p) = \sum_\sigma \chi_\sigma(p) g(p)$ and we can prove the lemma
for each $\sigma$ fixed with $g$ replaced by $g \chi_\sigma$.
We now consider  the case $\sigma= (+, +, -)$ which corresponds to a 
hyperbolic critical point.
Define
\be\label{cov}
u_j= \sqrt 2 \sin  (\pi p_j), \; j=1, 2; \quad u_3 =  \sqrt 2 \cos (\pi p_3)
\ee
We have
\be
\rd p_1 \rd p_2 \rd p_3= J(u) \rd u_1 \rd u_2 \rd u_3
\label{JA}
\ee
where $J(u)$ is a smooth, non-vanishing function on $I_\sigma$.
 In terms of $u$, the dispersion
relation in $I_\sigma$ can be written in the canonical form
\be
e(p) = u_1^2+ u_2^2 - u_3^2 + 2
\label{canform}
\ee
Thus $u=0$ is a hyperbolic critical point.
 For all $\sigma$ fixed, we can perform
a similar change of variables. There are two elliptic points and six
hyperbolic points. The following argument for
$\sigma= (+, +, -)$ applies to all hyperbolic points. For the elliptic points,
after the change of variables into the canonical form,  the same
result 
was proved in Lemma 3.10 of \cite{EY}.

Returning to the case $\sigma= (+, +, -)$, we need to estimate
\be\label{A1}
\int \left [ \frac 1 {\alpha- (u_1^2+ u_2^2 - u_3^2) + i \e}
- \frac 1 {\alpha'- (u_1^2+ u_2^2 - u_3^2) + i \e'} \right ] f (u) \rd u
\ee
where we have replaced $\alpha-2$ by $\alpha$ and $\alpha'-2$ by $\alpha'$.
Here $f(u)= g(p(u))\chi_\sigma(p(u)) J(u)$, where the regular function $p(u)$
is given by the inverse of the change of variables
formula \eqref{cov}.

Define $u_1= Q^{1/2} \cos \phi, u_2= Q^{1/2} \sin \phi$,
\begin{align}
f^\ast (Q, u_3): = & \int_0^{2 \pi} f(Q^{1/2} \cos \phi, Q^{1/2} \sin \phi, u_3)\rd \phi \\
Y(z, u_3) : = & \int_0^2 \frac {f^\ast (Q, u_3)} {z-Q} \rd Q
\end{align}
Let $z = \alpha+u_3^2+ i \e$ and $z'= \alpha'+u_3^2+ i \e'$. For any fixed $u_3$,
we can follow the proof of (3.69) in the appendix of \cite{EY} to  have
\be\label{A2}
 |Y(z, u_3)-Y(z', u_3)|\le C \|f^\ast \|_\infty |\log z - \log z'|
+ C |z-z'| \e^{-1/2} \| f \|_{C_1} \; .
\ee
We point out that there is a typo on the right side
of equation (A.7) of \cite{EY}. The correct bound
should be const.$|\mbox{Im } \xi|^{-1/2} \| f \|_{C_1}$, which leads
to \eqref{A2}.

Since
$$
|z-z'|
\le |\alpha-\alpha'| + \e\;, 
$$
the last term on the right hand side of \eqref{A2}, after integration in $u_3$,
is of the form stated in \eqref{A3}. For the first term, we use
$$
\int_0^{\sqrt{2}}
 \rd u_3 \big| \log (\alpha+ u_3^2 + i \e)- \log (\alpha'+ u_3^2 + i \e')\big|
= \int_0^2 \frac {\rd Q} {Q^{1/2}} 
\big| \log (\alpha+ Q + i \e)- \log (\alpha'+ Q + i \e')\big|\; .
$$
The last integral is very similar to the equation following (A.6) in \cite{EY}.
We can follow the proof in \cite{EY} after (A.6) to estimate this integral
by $C|z-z'|\e^{-1/2}$.
We thus bound both terms in \eqref{A2} and prove \eqref{A3}.
Equation \eqref{holder} follows from optimizing $\e$ in
\eqref{A3}.

To prove \eqref{A4}, we rewrite it as
\be\label{A5}
\begin{split}
&  \int \left [ \frac 1 {\alpha- \om(p)+ i \eta}
- \frac 1 {\alpha- e(p) + i \eta} \right ] g (p) \rd p \\
= &  \int \frac {\lambda^2 (\Theta(e(p))- \Theta(\alpha))}
 {(\alpha- e(p)- \lambda^2 \Theta(e(p))+ i \eta)
(\alpha- e(p) - \lambda^2 \Theta(\alpha)+ i \eta) } \; g (p) \rd p \\
& +\int \left [ \frac 1 {\alpha- e(p)- \lambda^2 \Theta(\alpha)+ i \eta}
- \frac 1 {\alpha- e(p) + i \eta} \right ] g (p) \rd p \; ,
\end{split}
\ee
recalling that $\om(p)=e(p)+ \lambda^2 \Theta(e(p))$.
From the H\"{o}lder continuity
estimate \eqref{holder}, we can bound the first term on the right hand side by
\be
C\| g\|_\infty\int \Bigg| \frac {|e(p)-\alpha|^{1/2}}
 {(\alpha- \om(p) + i \eta)}
\frac {\lambda^2 }{(\alpha- e(p)
- \lambda^2 \Theta(\alpha)+ i \eta) } \, \Bigg|  \rd p\; .
\label{eq:ghol}
\ee
Since
$$
|e(p)-\alpha|^{1/2} \le |\om(p)-\alpha |^{1/2}
+O(\lambda )\; ,
$$
the integral in  \eqref{eq:ghol} is bounded by
$$
(\lambda^2 \eta^{-1/2}+\lambda^3 \eta^{-1}) \int
 \frac {\rd p} {|\alpha- e(p)
- \lambda^2 \Theta(\alpha) + i \eta|} \; .
$$
By the co-area formula (\ref{coa}), the last integral is bounded by
\begin{equation}\label{A9}
\int
 \frac {\rd p} {|\alpha- e(p)
- \lambda^2 \Theta(\alpha) + i \eta|}
=    \int_{0}^{2d} \frac{\rd s}{|\a - s -\lambda^2 \Theta(\alpha) +i\eta|}
\;  \Phi(s)
\le C  |\log \eta|\;.
\end{equation}
Here we have used that
$\Phi$ is bounded according to
\eqref{eq:Phi}. This bounds the first term in \eqref{A5}.
The second term is bounded $O(\lambda)$ from  \eqref{holder}.
This proves Lemma~\ref{le:A1}. $\Box$

\bigskip

{\it Proof of Lemma \ref{le:opt}}. 
 The proof for \eqref{eq:logest}
and \eqref{eq:2aint} is similar to the argument
from \eqref{A5} to \eqref{A9} and we shall not repeat it here. We now prove
the more accurate estimate \eqref{eq:ladderint}; it is
 is similar  to 
the proof of (2.9) in \cite{ESYII} (Appendix B.1).

From the Schwarz inequality and a change of variables, it suffices to prove
only the first estimate of \eqref{eq:ladderint}.
We can assume that $|\a| \leq 4d$,
since otherwise there is no singularity at
all and the Lemma trivially holds.

Recall $\omega(p):= e(p) +\lambda^2 \theta(p), \theta(u)=\Theta(e(u))$
and $\Theta(e)=  {\cal R} (e)- i{\cal I}(e)$ with ${\cal I}(e) \ge 0$.
We have
$$
\frac{ \lambda^{2}  }{|\alpha- \ov \om(u)-i\eta|^{2}}
=
  \frac{ \lambda^{2}  }{\lambda^2  {\cal I} (e(u))+ \eta} \; \mbox{Im} \;
  \frac 1 {\alpha- e(u) - \lambda^2  {\cal R} (e(u))
-i(\lambda^2  {\cal I} (e(u))+ \eta)} \; .
$$
From the resolvent identity and with the notations $e= e(u)$,
$\wt \alpha = \alpha-\lambda^2  {\cal R} (\alpha)$,  this is equal to
 $(I)+ (II) + (III)$, where
\begin{align}
 (I) & :=  \frac{ \lambda^{2}  }{\lambda^2  {\cal I}(\wt\alpha)
+ \eta} \; \mbox{Im} \;  \frac 1 {\wt\alpha- e
-i(\lambda^2  {\cal I}(\wt\alpha) + \eta)}
\label{2res} \\
 (II) &:= - \frac{ \lambda^{2}  }{\lambda^2  {\cal I}(\wt\alpha) + \eta} \;
\frac{\lambda^2( {\cal I}(e) -{\cal I}(\wt\a) )}{\lambda^2 {\cal I}(e) +\eta}
\; \mbox{Im} \frac 1 {\alpha- e - \lambda^2\Theta(e) -i\eta}
\nonumber\\
 (III) &:=- \frac{ \lambda^{2}  }{\lambda^2  {\cal I}(\wt\alpha)+ \eta} \;
   \mbox{Im} \Bigg[  \frac 1 {\wt \alpha- e
-i(\lambda^2  {\cal I}(\alpha) + \eta)} \;
    \frac{\lambda^2 (\Theta(\a)-\Theta(e))}{\a -e -\lambda^2\Theta(e) -i\eta}
\Bigg] \; .
\nonumber
\end{align}
We will estimate $\int \rd u [ (I)+ (II)+ (III)]$. 
The main term will be the first one.
In this term we first use the definition \eqref{theta} and
the estimate \eqref{eq:Thetaest} with
$\e:= \lambda^2  {\cal I}(\wt\alpha) + \eta =O(\lambda^2)$
 to obtain 
$$
\int (I) \; \rd u \leq 
 \frac{ \lambda^{2}  }{\lambda^2  {\cal I}(\wt\alpha) + \eta}
 \Big[  {\cal I}(\wt\alpha)
+ O(\lambda)\Big]
\leq 1 + O( \lambda^3\eta^{-1})\; .
$$
In the second term of (\ref{2res}) we use (\ref{holder})
$$
  |(II)|
   \leq   C\lambda^2 \Big( \frac{\lambda^2}{\eta}\Big)^2
  \frac{  |\wt\a -e|^{1/2}}{|\wt\a -e
 + \lambda^2({\cal R}(\wt\a)-{\cal R}(e))|^2
   + \eta^2} \; ,
$$
where we also 
used  that $\Big[ \lambda^2( {\cal R}(\wt\a) - {\cal R}(\a))\Big]^2
=O(\lambda^6)\ll \eta^2$ based upon  (\ref{holder}).

To perform the $\rd u$ integration (recall $e=e(u)$),
 we distinguish two regimes
depending on whether $|\wt\alpha - e(u)|$ is bigger or smaller than 
$K_0\lambda^4$
for a sufficiently large fixed $K_0$. When $|\wt\alpha - e(u)|\ge 
K_0\lambda^4$,
then $\lambda^2|{\cal R}(\wt\a)-{\cal R}(e(u))| <
 \frac{1}{2}|\wt\a -e (u)|$.  Hence
$$
  |(II)|\leq  C\lambda^2 
\Big( \frac{\lambda^2}{\eta}\Big)^2
 \frac{\eta^{-1/2} }{  |\wt\a -e(u)| +\eta} \; ,
$$
and the corresponding integral is of
 order $O(\lambda^6\eta^{-5/2}|\log\eta|)$
using the co-area formula
and the boundedness of $\Phi$. When $|\wt\alpha - e(u)|\le K_0\lambda^4$, then
we can trivially estimate
$|(II)| \leq C(\lambda^2\eta^{-1})^4$ and the volume  factor is given by
$$
    \int_0^{2d} {\bf 1}( |\wt\a -s|\le K_0\lambda^4) \Phi(s)
 \rd s = O(\lambda^4) \; .
$$
Therefore the integral is of order $O(  (\lambda^3\eta^{-1})^4)$.

Finally, the last term  in (\ref{2res}) is estimated as
$$
  |(III)| \leq C\lambda^2 \Big( \frac{\lambda^2}{\eta}\Big) 
 \frac{1}{ |\wt\a -e|+\eta}
   \;\frac{ |\a-e|^{1/2}} {| \a -e + \lambda^2 {\cal R}(e)| + \eta} \;
$$
In the regime where $|\a - e(u)| \ge K_0\lambda^2$ (with some large $K_0$)
 we obtain
$$
   |(III)| \leq C\lambda^2 \Big( \frac{\lambda^2}{\eta}\Big)
   \;\frac{ |\a-e|^{1/2}} {(| \a -e| + \eta)^2} \leq C\lambda^2 \eta^{-1/2}
   \Big( \frac{\lambda^2}{\eta}\Big)  \frac{1}{ |\a -e|+\eta}
$$
and after integration we collect $O( \lambda^4\eta^{-3/2}|\log\eta|)$.
In the regime where $|\a - e(u)| \le K_0\lambda^2$ we have 
$|(III)|\leq O(\lambda^5\eta^{-3})$
and the volume factor is $O(\lambda^2)$. Therefore the 
integral is $O(\lambda^7\eta^{-3})$.
Collecting the error terms, we arrive at the proof of Lemma 
\ref{le:opt}. $\;\;\Box$

\subsection{Proof of Lemma \ref{lemma:opt1}}

We can assume that $f$ is a real function.
We can also assume that $|\alpha - \beta|\leq \lambda$, otherwise the
two singularities are separated by $O(\lambda)$ and at least one
of the denominator can be estimated by $O(\lambda^{-1})$
and the other one integrated out by (\ref{eq:logest}) to give
$O(\lambda|\log\lambda|)$.

We use the partition of unity $1=\sum_\sigma \chi_\sigma$ introduced
in (\ref{unity}) and let $f_\sigma:=f\chi_\sigma$.
Fix a $\sigma$ around a hyperbolic critical point;
the proof for the elliptic point is similar but easier.
We choose the case $\sigma=(+,+,-)$ as in the proof of Lemma \ref{le:A1}.
Recall the new coordinate system on $I_\sigma$ from \eqref{cov},
where $|u_j|\leq \sqrt{2} - c_0$ with some positive $c_0$.
 In terms of $u$, the dispersion
relation in $I_\sigma$ can be written in the canonical form
\eqref{canform}.
Introduce the notations
$$
G_j(u):= 2\pi \sqrt {2-u_j^2}  \; u_j, \; j=1, 2; \qquad
G_3(u):= -2\pi \sqrt {2-u_3^2} \; u_3\; ,
$$
then the vector-field $G$ is regular on the integration domain.
Note that
$$
    (\nabla e)(p(u)) = G(u)\; .
$$
Let
\be\label{ftilde}
\wt f(u) = f_\sigma(p(u)) J(u)\; ,
\ee
where $J(u)$ is the Jacobian \eqref{JA}.
We expand in $r$ as $e(u\pm r)= e(u) \pm \nabla e(u)\cdot r +O(r^2)$
and we also use the H\"older
continuity of $\Theta$ given in (\ref{holder}):
$$
    \Theta(e(p(u)\pm r)) = \Theta(\gamma) + 
O(|\gamma - e(p(u))|^{1/2}) + O(|r|^{1/2}) \; .
$$
Thus we can write the integral \eqref{eq:optical} around
 the selected critical point as
\begin{align}\label{9.1}
& \int  \rd u    { \wt f(u) }  \\ &
  \times    \frac{\lambda }{\wt\a - (u_1^2+ u_2^2 - u_3^2) - G(u)\cdot r
    + O(r^2+ \lambda^2 |r|^{1/2}) - \lambda^2\overline{\Theta}(\gamma)
    +O(\lambda^2|\gamma-e|^{1/2})- i\eta }  \nonumber \\
&
    \times  \frac{\lambda }{ \wt\beta-(u_1^2+ u_2^2 - u_3^2) + G(u)\cdot r
    + O(r^2+ \lambda^2 |r|^{1/2}) - \lambda^2 {\Theta}(\gamma)
    +O(\lambda^2|\gamma-e|^{1/2})+ i\eta }  \nonumber
\end{align}
where $e=e(p(u))$, $\wt\alpha:= \alpha-2$, $\wt\beta:=\beta-2$.

The error terms can be removed from the denominators by an argument similar
to the proof of Lemma \ref{le:opt}
in Appendix \ref{sec:33}. Using the resolvent identity for the
first denominator in (\ref{9.1})
$$
\frac{1}{ A    + B- i\eta }
=\frac{1}{A- i\eta }
-\frac{ B }{ ( A    + B
    - i\eta)(A- i\eta )}\;, 
$$
with $A:=\wt\a - (u_1^2+ u_2^2 - u_3^2) - G(u)\cdot r
- \lambda^2\overline{\Theta}(\gamma)$,  $B:=O(r^2+\lambda^2 |r|^{1/2})
    +O(\lambda^2|\gamma-e|^{1/2})$, and $|\gamma-e|^{1/2}\leq |\alpha -e|^{1/2}
    + O(\lambda^{1/2})\leq |A|^{1/2} + O(\lambda^{1/2})$,
     we can estimate the error term by
$$
   \Big| \frac{ B }{ ( A    + B
    - i\eta)(A- i\eta )}\Big|\leq
    O \big ( \, \eta^{-2} \big [ \, \lambda^{5/2}
    + r^2 + \lambda^2 |r|^{1/2} \, \big ] \, \big )+O(\lambda^2\eta^{-3/2})=O(\lambda^{-3/2-8\kappa})\;.
    $$
Then we can integrate out the denominator with $\wt\beta$ to collect an extra $|\log\lambda|$
factor. This term, together with the $\lambda^2$ in the numerators
of (\ref{9.1}), is included in the error term in Lemma \ref{lemma:opt1}.

Now we compute the main term
\begin{align}
M:=\int  \rd u    { \wt f(u) } &
    \Bigg[ \frac{\lambda }{ \wt\a - (u_1^2+ u_2^2 - u_3^2) - G(u)\cdot r
    - \lambda^2\overline{\Theta}(\gamma)
    - i\eta } \\ \nonumber
&
    \times  \frac{\lambda }{ \wt\beta -(u_1^2+ u_2^2 - u_3^2) + G(u)\cdot r
     - \lambda^2 {\Theta}(\gamma) + i\eta } \Bigg] \; .
\end{align}

We introduce spherical coordinates in the $(u_1, u_2)$ plane.
 Let $Q:= u_1^2 + u_2^2$
and
$$
u_1= Q^{1/2} \cos \phi, \qquad u_2= Q^{1/2} \sin \phi\; .
$$
We shall use $Q, \phi, u_3 = y$ as coordinates instead of $u_1, u_2, u_3$
then
$$
   \rd u_1 \rd u_2\rd u_3 = \frac{1}{2}\rd Q\rd y \rd \phi \; .
$$
Thus
\begin{align}\label{M}
M= \frac{1}{2}\int  \rd y \rd Q \rd \phi   { \wt f(Q, y, \phi) } &
    \Bigg[ \frac{\lambda }{ \wt\a + y^2 - Q  - G(Q, \phi, y)\cdot r
     - \lambda^2\overline{\Theta}(\gamma ) - i\eta } \\ \nonumber
&
    \times  \frac{\lambda }{ \wt\beta + y^2 - Q  + G(Q, \phi, y)\cdot r
     - \lambda^2 {\Theta}(\gamma ) + i\eta } \Bigg]\; .
\end{align}
The integration domains are $Q\in [0,4]$, $y\in [-\sqrt{2}, \sqrt{2}]$,
$\phi\in [0,2\pi]$, but since $\wt f$ is compactly supported on
this domain, we can assume that $Q\in [0, \infty)$ and $y\in \bR$.
Set $\rho:= \wt\gamma+ y^2$, $\wt\gamma=\gamma-2$, 
and distinguish three cases (subdomains of integration).

\medskip

{\it Case 1: $|\rho |< 3\lambda$.}
After a Schwarz inequality, estimating $\wt f$ by its maximum and
a change of variable
 $Q\to \wt Q:=Q  \pm G(Q, \phi, y)\cdot r$
we  estimate
$$
 \int  \rd y \rd \wt Q \rd \phi  \; 
\frac{ \lambda^2\cdot1( |\wt\gamma+y^2|\leq 3\lambda)}
    { |\wt\a + y^2 - \wt Q
     - \lambda^2\overline{\Theta}(\gamma ) - i\eta |^2}
     \leq C\lambda^2\eta^{-1} \int \rd y \; {\bf 1}
( |\wt\gamma+y^2|\leq 3\lambda)
     \leq O(\lambda^{1/2-4\kappa})\; . 
$$

\bigskip
{\it Case 2: $\rho < -3\lambda$.} The contribution of this
regime, after a Schwarz inequality
and the estimate ${\bf 1}(\wt\gamma + y^2 \leq -3\lambda)\leq
{\bf 1}(\wt\alpha + y^2 \leq -2\lambda)$, is bounded by
\begin{align}
     \lambda^2 \int  \rd y \rd \phi \int_{-\lambda^2}^{5} & \rd \wt Q \;
     \frac{ {\bf 1}(\wt\alpha + y^2 \leq -2\lambda)}    { |\wt\a + y^2 - \wt Q
     - \lambda^2\overline{\Theta}(\gamma ) - i\eta |^2}
     \leq C\lambda^2 \int  \frac{ \rd y  \rd \phi \rd \wt Q }
     { |- \wt Q - i\eta |\; |\wt\a + y^2
      - i\eta |} \\ \nonumber
      & \leq O(\lambda^2 \eta^{-1/2}|\log\lambda|^2)\leq O(\lambda^{1/2})\; .
\end{align}
The integration domain of $\wt Q$ comes from the definition of $\wt Q$,
$Q\ge 0$  and $r \ll \lambda^2$.
\bigskip

{\it Case 3: $\rho \ge 3\lambda$.} 
We replace $G(Q, \phi, y)$ by $G(\wt\gamma + y^2, \phi, y)$ in \eqref{M}.
Since $G$
is regular on the integration domain, the error of this replacement
in $M$ is estimated by $O(\lambda^{1/2-8\kappa})$ by a similar argument
as we removed the error terms in (\ref{9.1}).
Therefore the contribution of the regime  $\rho \ge 3\lambda$
to \eqref{M} is (up
to negligible errors)
\begin{align*}
M_3:=& \frac{1}{2}\int_{\wt\gamma+y^2 \ge 3\lambda}
\rd y\rd\phi \int \rd Q \; \wt f(Q, y, \phi) \\ \nonumber
&    \times \Bigg[ \frac{\lambda }{ \wt\alpha +y^2 - Q  
 + G(\wt \gamma+ y^2, \phi, y)\cdot r
     - \lambda^2\overline{\Theta}(\gamma) - i\eta } \\ \nonumber
&
    \quad\times  \frac{\lambda }{ \wt\beta + y^2 - Q +
 G(\wt \gamma+ y^2, \phi, y)\cdot r
     - \lambda^2 {\Theta}(\gamma) + i\eta } \Bigg] 
\end{align*}
\begin{align}\label{9.3a}
= & \lambda^2  \int   \frac{ 1(\wt\gamma+y^2 \ge 3\lambda)  \rd y  \rd \phi}
  { \wt\a-\wt\beta  +  2 G(\wt\gamma+ y^2, \phi, y)\cdot r
    -  2 i\lambda^2 \cI (\gamma)   - 2 i\eta } \\ \nonumber
& \times\frac{1}{2} \; \int_0^\infty \rd Q   \; \wt f(Q, y, \phi)
 \Bigg[ \frac{1}{\wt \beta - Q + R
     - \lambda^2 {\Theta}(\gamma) + i\eta } -
 \frac{1 }{ \wt\alpha - Q  + R
     - \lambda^2\overline{\Theta}(\gamma) - i\eta }
 \Bigg] \nonumber
\end{align}
with $R:=y^2  + G(\wt\gamma+ y^2, \phi, y)\cdot r$
and recalling that $\Theta= \cR-i\cI$. 
 
\begin{lemma}\label{lemma:Y}
Let $F$ be a $C^1$-function on $\bR$ with compact support and let
$$
    Y(z):= \int_0^\infty \frac{F(Q)}{z-Q} \; \rd Q
$$
for any $z=\alpha + i\Lambda$ with $\Lambda>0$. Then
$$
      |Y(z)-Y(z')|\leq C |z-z'| |\log \Lambda|
$$
where $z'=\alpha' + i\Lambda'$ and $\Lambda\ge \Lambda'>0$.
\end{lemma}
The proof is analogous to that of Lemma 3.10 in \cite{EY}
and will not be repeated here.

\bigskip

We change $\wt\beta$ to $\wt\alpha$ in the first denominator in the big
bracket in (\ref{9.3a}),
the error is estimated by 
$\lambda^2\eta^{-1}|\wt\alpha-\wt\beta||\log\eta| \leq O(\lambda^{1/2})$
using Lemma \ref{lemma:Y}.
Then we compute
$$
\frac{1}{2}\int_0^\infty \rd Q  \; \wt f(Q, y, \phi) \Bigg[
 \frac{1}{ \wt\alpha - Q + R     - \lambda^2 {\Theta}(\gamma) + i\eta }
- \frac{1 }{ \wt\alpha - Q  + R
     - \lambda^2\overline{\Theta}(\gamma) - i\eta }
\Bigg]
$$
$$
    = i \, \mbox{Im}\int_0^\infty \rd Q  \;\frac{ \wt f(Q, y, \phi)}{ \wt\alpha - Q  + R
     - \lambda^2\Theta(\gamma) + i\eta } \; .
$$

Note that $\wt\alpha + R \ge \wt\gamma + y^2 - \lambda - O(r) \ge \lambda$
since $\varrho\ge 3\lambda$,
so we can use the
estimate (for $\e >\e'>0$)
$$
      \mbox{Im} \int_{-\e}^\infty \frac{g(x)}{x+i\e'}\;\rd x 
= -\pi g(0)  + O(\e'/\e) +
      O(\e'|\log\e'|)\; 
$$
where the constants depend on $\| g\|_{C^1}$.
We obtain
\begin{align*}
     \mbox{Im}\int_0^\infty \frac{ \wt f(Q, y, \phi)\;\rd Q}{
 \wt\alpha - Q  + R
     - \lambda^2\Theta(\gamma) + i\eta }  & =-\pi \wt f\Big( \,
     \wt\alpha +R -\lambda^2 \cR (\gamma), y, \phi\, \Big) +O(\eta/\lambda)
\\
&     = -\pi \wt f(\wt\gamma +y^2, y, \phi) + O(\lambda) \; ,
\end{align*}
where  $\cR$ is the real part of $\Theta$ and in the last estimate we used
 the smoothness of $\wt f$ and $|\wt\alpha-\wt\gamma| =O(\lambda)$.
 The result of these calculations is summarized as
 $$
      M_3 =-\lambda^2i \pi\int \frac{ {\bf 1}(\wt\gamma+y^2 \ge 3\lambda)
      \wt f(\wt\gamma +y^2, y, \phi)\rd y  \rd \phi}
  { \wt\a-\wt\beta  +  2 G(\wt\gamma+ y^2, \phi, y)\cdot r
    -  2 i\lambda^2 \cI (\gamma)   - 2 i\eta }
    + O(\lambda^{1/2-8\kappa}|\log\lambda|)  
$$
as the contribution for Case 3.

Using (\ref{ftilde}) and changing back the variables
and adding the errors from Cases 1 and 2,
we obtain
$$
      M =-\lambda^2 i\pi\int \frac{  f_\sigma(\wt\gamma +y^2, y, \phi)
      J( \wt\gamma +y^2, y,\phi)
      \rd y  \rd \phi}
  { \wt\a-\wt\beta  +  2 G(\wt\gamma+ y^2, \phi, y)\cdot r
    -  2 i\lambda^2 \cI (\gamma)   - 2 i\eta }
    + O(\lambda^{1/2-8\kappa}|\log\lambda|)
$$
$$
      =-2\lambda^2 i\pi\int
       \frac{  f_\sigma(p) \, \delta( e(p)-\gamma) \; \rd p}
  { \a-\beta  +  2 (\nabla e)(p)\cdot r
    -  2 i[\lambda^2 \cI (\gamma)   +\eta] }
    + O(\lambda^{1/2-9\kappa}) \; . \qquad \Box
$$

\subsection{Proof of Lemma \ref{lemma:newrow}.}\label{secnewroww}

In this section  we
prefer to avoid the $2\pi$ factors in the arguments
of the trigonometric functions. 
Therefore we change variables $p\to \frac{p}{2\pi}$,
we {\it redefine} the dispersion relation
\be
   \wt e(p) : = e(\frac{p}{2\pi})= \sum_{i=1}^d (1-\cos (p^{(i)})) \;, 
 \label{red}
\ee
the integration domain, $\wt \Tor^d:= [-\pi, \pi]^d$, and
the triple norm, $\tri u\tri\wt{} : = \tri \frac{u}{2\pi} \tri$.
The task of proving Lemma \ref{lemma:newrow} with the redefined
 data is  equivalent to the original formulation up to changing 
the  universal constant. For simplicity, we then remove all tildes from
the notation. Hence in this section
the integration domain is $p\in [-\pi, \pi]^d$ and 
$e(p)= \sum_{i=1}^d (1-\cos (p^{(i)}))$. As before, we work in $d=3$
dimensions.

\medskip

To prove the estimates in Lemma \ref{lemma:newrow}, we 
first replace $\om(p)$ with $e(p)$ in the
propagators  at the expense of an
extra factor $\lambda^2\eta^{-1}= \eta^{-\kappa/(2+\kappa)}$ using 
 a straightforward resolvent expansion:
\be
  \frac{1}{|\a - \om(p) +i\eta|} \leq \frac{1}{|\a - e(p) +i\eta|}+
   \frac{c\lambda^2}{|\a - e(p) +i\eta||\a - \om(p) +i\eta|}
  \leq \frac{1+c\lambda^2\eta^{-1}}{|\a - e(p) +i\eta|} \; .
\label{resex}
\ee
Therefore  Lemma \ref{lemma:newrow} will immediately follow from
\begin{lemma}\label{lemma:main}
\be
  I:=\int \frac{\rd p}{|\a - e(p) +i\eta|}\frac{1}{|\beta - e(p+q) +i\eta|}
  \leq  \frac{c\eta^{-3/4}|\log\eta|^3}
   {\tri q \tri }
\label{nopont}
\ee
\be
  I(r):=\int \frac{\rd p}{|\a - e(p) +i\eta|}\frac{1}{|\beta - e(p+q) +i\eta|}
  \frac{1}{\tri p-r\tri }\leq  \frac{c\eta^{-7/8}|\log\eta|^3}
   {\tri q \tri }
\label{withp}
\ee
uniformly in $r,\alpha, \beta$,
and
\be
  \wt I:= \int_{-4d}^{4d} d\alpha
  \int \frac{\rd p}{|\a - e(p) +i\eta|}\frac{1}{|\a - e(p+q) +i\eta|}
  \leq  \frac{c\eta^{-1/2}|\log\eta|^3}
   {\tri q \tri } \; .
\label{bnopont}
\ee
\end{lemma}

We remark that these estimates are far from being optimal, e.g.
\eqref{nopont} and \eqref{withp}
are expected to hold 
with a prefactor $\eta^{-1/2}$
instead of  $\eta^{-3/4}$ and $\eta^{-7/8}$.

\medskip

Before the proof of Lemma \ref{lemma:main}, we establish the
following 
\begin{proposition}\label{lemma:1pon}
\be
    \int \frac{\rd p}{|\a - e(p) +i\eta|}\frac{1}{\tri p-r\tri }
  \leq  c|\log\eta|^3 \;,
\label{1dee}
\ee
\be
    \int \frac{\rd p}{|\a - \om(p) +i\eta|}\frac{1}{\tri p-r\tri }
  \leq  c\eta^{-\kappa/2}|\log\eta|^3 \;.
\label{logpo}
\ee
\end{proposition}

{\it Proof of Proposition \ref{lemma:1pon}.} 
The second inequality follows immediately from the first one by
(\ref{resex}). For the proof of the first one,
 we consider
the integral restricted on the set where
 $\tri p-r\tri =|p-r|+\eta$, i.e.
the closest critical point to $p-r$ is 0.
The other regimes are treated analogously.
Since $|p-r|\ge |p_1-r_1|$, we have
$$
\int \frac{\rd p}{|\a - e(p) +i\eta|}\frac{1}{|p-r|+\eta}
\leq \int \frac{\rd p_1}{|p_1-r_1|+\eta}
  \int \frac{\rd p_2 \rd p_3}{|\a - \cos p_1 - \cos p_2 - \cos p_3 +i\eta|} \;.
$$
The inner integral is computed for any fixed $p_1$ by
using 
\be
   \sup_c \int \frac{\rd p_2 \rd p_3}{|c- \cos p_2 - \cos p_3 +i\eta|}
   \leq c|\log\eta|^2\; .
\label{cos}
\ee
Then (\ref{1dee}) follows after the $\rd p_1$ integration.

To see (\ref{cos}), 
let $f(p):=\cos p_2 + \cos p_3$. On the set $\{ (p_2, p_3)\; : \;
|\nabla f(p)|\leq \eta\}$,
we use the trivial $\eta^{-1}$ bound on the integrand. The volume 
of this set is $O(\eta^2)$, so this contribution is negligible.
On the complement set, $|\nabla f|\ge \eta$, we use the
 coarea formula,
$$
    \int \frac{{\bf 1} (|\nabla f|\ge \eta)
\rd p_2 \rd p_3}{|c- f(p) +i\eta|}
    = \int_{-2}^2 \frac{\rd a}{|c-a+i\eta|} \int_{S(a)}
    \frac{{\bf 1} (|\nabla f|\ge \eta)\rd \ell}{|\nabla f(p)|}
$$
with  $S(a)=\{ (p_2,p_3) \; : \;
f(p)=a\}$
 and $\rd \ell$ is the arclength on
$S(a)$ that is an analytic curve with
finite length. Since $|\nabla f(p)|\ge \tri p \tri+\eta$, the
integral on the set $S(a)$ is bounded by $|\log\eta|$
since in the worst case
it is a regularized $1/|x|$ singularity on a 
one dimensional piecewise smooth curve. $\Box$

\bigskip

{\it Proof of estimate  (\ref{withp}) in Lemma \ref{lemma:main}.} 
We first show how  the estimate (\ref{withp}) follows
from (\ref{nopont}). We distinguish two cases.
If $\tri q\tri \le 100\eta^{1/8}$,
then we estimate the $\beta$ denominator trivially by $\eta^{-1}$ and
use Proposition \ref{lemma:1pon}. The total estimate is
$$
     c\eta^{-1}|\log\eta|^3\leq  \frac{c\eta^{-7/8}|\log\eta|^3}
   {\tri q \tri  }\;.
$$

Now we assume that $\tri q\tri \ge 100\eta^{1/8}$.
 Let $S:= \{ p \; : \; \tri p-r\tri \le \frac{1}{100}\eta^{1/8}\}$.
On the set $S^c$ we will estimate the $\tri p-r\tri ^{-1}$
term by $c\eta^{-1/8}$ and 
we use  (\ref{nopont}).

Finally, we consider the integral on the set $S$. From
$$
  \inf_x\Big( |\sin x|+|\sin(x+y)|\Big)
 \ge \frac{1}{10}\min\{ |y-k\pi| \; : \; k\in\bZ\}\; ,
$$
we see that
$$
   |\nabla e(r)| + |\nabla e(r+q)|\ge \frac{1}{10}\tri q\tri \; .
$$
Hence we can assume that
 $|\nabla e(r)|\ge \frac{1}{20}\tri q\tri$
(the other case being identical). Since $\sup_u\| e''(u)\|\leq 1$
for the usual matrix norm of the Hessian, we see that, for
 all $p\in S$,
\be
  |\nabla e(p)|\ge \frac{1}{20}\tri q\tri -  \frac{1}{100}\eta^{1/8}
  \ge c\tri q\tri \; .
\label{lowdd}
\ee

We again estimate the $\beta$ denominator trivially by $\eta^{-1}$
in $I(r)$ and use  the coarea formula to obtain
$$
  \eta^{-1} \int
\frac{{\bf 1}(p\in S)\rd p}{|\a - e(p) +i\eta|}\frac{1}{\tri p-r\tri }
 = \eta^{-1} \int_0^6 \frac{\rd a}{|\a - a +i\eta|}
   \int_{\Sigma(a)}\frac{{\bf 1}( \tri 
p-r\tri \leq \frac{1}{100}\eta^{1/8})\rd\sigma }
{\tri p-r\tri \; |\nabla e(p)|}
$$
where $\Sigma_a$ is the level surface $\{ p\; : \; e(p)=a\}$,
consisting of finitely many smooth pieces and $\rd\sigma$
is the surface measure. Using (\ref{lowdd}), we continue
 with
$$
   \eta^{-1} \int\frac{{\bf 1}(p\in S)\rd p}{|\a - e(p) +i\eta|}
\frac{1}{\tri p-r\tri }
 \leq  \frac{c\eta^{-1}}{\tri q\tri}
 \int_0^6 \frac{\rd a}{|\a - a +i\eta|}
   \int_{\Sigma(a)}\frac{{\bf 1}( \tri 
p-r\tri 
\leq \frac{1}{100}\eta^{1/8})\rd\sigma }
{\tri p-r\tri }
$$
$$
 \leq  \frac{c\eta^{-7/8}|\log\eta|}{\tri q\tri} 
\int_0^6 \frac{\rd a}{|\a - a +i\eta|}
   \leq \frac{c\eta^{-7/8}|\log\eta|^2}{\tri q\tri} \; .
$$
Here we used that the regularized $1/|x|$ singularity
on two dimensional regular surfaces $\Sigma$ can be estimated as
$$
    \int_\Sigma \frac{{\bf 1}(|x|\leq \e)\rd\sigma(x)}
   {|x|+\eta}\sim \int_{\bR^2}\frac{{\bf 1}(|x|\leq \e)}{|x|+\eta}
 \leq \e|\log\eta| \; .
$$
This completes the proof of  (\ref{withp}). $\;\;\Box$

\bigskip

 {\it Proof of estimate  (\ref{nopont}) 
 in Lemma \ref{lemma:main}.}
We fix $q$ throughout the proof
and we assume that $\tri q\tri =|q|+\eta$; the proof for the neighborhood
of other critical points are analogous.
We can assume that $|q_2|\ge \frac{1}{3} \tri q \tri$
by permuting the indices
and we can also assume $q_2\in [-\pi/2, \pi/2]$, by a shift.

We use the coarea formula
\be
  I:=\int \frac{\rd p}{|\a - e(p) +i\eta|}\frac{1}{|\beta - e(p+q) +i\eta|}
  = \int_{0}^6\int_{0}^6\frac{\rd a \rd b}{ |\a-a+i\eta| \; |\beta- b+i\eta|}
  \int_{S(a,b)} \frac{\rd\ell}{|F(p)|}\; ,
\label{coare}
\ee
where we define the vector
\be
   F(p):= \nabla e(p)\wedge \nabla e(p+q) =
   \begin{pmatrix} \sin p_2 \sin (p_3+q_3) - \sin p_3 \sin (p_2+q_2)\cr
\sin p_3 \sin (p_1+q_1) - \sin p_1 \sin (p_3+q_3)\cr
\sin p_1 \sin (p_2+q_2) - \sin p_2 \sin (p_1+q_1) \; 
\end{pmatrix} \; ,
\ee
the set
$$
   S(a, b):= \{ p\in \Tor^3\; : \; e(p)=a, \; e(p+q)=b\}
$$
and let $\rd\ell$ denote the arclength measure. Since $e(p)$
is analytic, the set $S(a,b)$ consists of
piecewise analytic curves with the exception
of finitely many $a,b$. The lengths of the curves are bounded.

Let $f(p_2)$ be the first component of $F(p)$ as a function of $p_2$
with fixed $p_3$,
\be
  f(p_2):=\sin p_2 \sin (p_3+q_3) - \sin p_3 \sin (p_2+q_2)\; ,
\label{deff}
\ee
and compute
$$
f'(p_2):=\cos p_2 \sin (p_3+q_3) - \sin p_3 \cos (p_2+q_2)\; .
$$
For $q_2\in [-\pi/2, \pi/2]$ a simple estimate shows that
$$
   |f(p_2)|^2 + |f'(p_2)|^2 \ge
\Big(|\sin (p_3+q_3)|+|\sin p_3| \Big)^2\sin^2 \frac{q_2}{2}\; .
$$
Therefore
\be
   |f(p_2)| + |f'(p_2)|\ge c \Big(|\sin (p_3+q_3)|+ |\sin p_3 |\Big) |q_2|
\label{flow}
\ee
uniformly in $p_2$.
We also note that
$$
  |f''(p_2)| = |\sin (p_3+q_3)|+ |\sin p_3 | \; .
$$
We now need the following one dimensional lemma:

\begin{lemma} \label{lemma:1dim}
Let $f\in C^2[-\pi, \pi]$ satisfying
$$
   \inf_p (|f(p)| + |f'(p)|) \ge \delta, \qquad  \| f''\|_\infty \leq M \; .
$$
Then for $\e\leq 1/2$
$$
  \mbox{Vol} \Big\{ p \; : \; |f(p)|\leq \e \delta \Big\}
  \leq 8\e \min\{ 8\pi M \delta^{-1}, 1\}\; .
$$
\end{lemma}

{\it Proof.} We can clearly assume that $\e M \leq \delta/32$,
otherwise the statement is trivial.
Suppose now that for some $p_0$ we have $|f(p_0)|\leq \e\delta$.
Then $|f'(p_0)|\ge \delta/2$. For definiteness,  assume
that $f'(p_0)\ge \delta /2$. Clearly
$f'(p)\ge \delta/4$ for all $|p-p_0|\leq \frac{1}{4}\delta M^{-1}$
since $f'(p)\ge f'(p_0)- \| f''\|_\infty |p-p_0|$.
For any $p$ with $8\e < |p-p_0|\leq  \frac{1}{4}\delta M^{-1}$
we thus have $|f(p)|> \e\delta$ and this set is non-empty
since $\e M \leq \delta/32$.

Define the intervals
$I_1:=\big\{ p\in [-\pi, \pi] \;: \;
 |p-p_0|\leq \frac{1}{4}\delta M^{-1}\big\}$ 
and $I_2: = \big\{ p\in [-\pi, \pi] \;: \;
 |p-p_0|\leq 8\e\}$, then clearly  $I_2\subset I_1$
and  $|f(p)|>\e\delta$
for any $p\in I_1\setminus I_2$.
Thus $I_2$ occupies at most
$$
  \frac{8\e}{|I_1|} 
\leq \min \{ 32 M\e\delta^{-1} , 4\e/\pi\}
$$
proportion of  $I_1$.
Repeating this argument
 for each interval where $|f(p)|\le \delta\e$,
we obtain the lemma. $\Box$

\bigskip

Using this lemma for $f$ from \eqref{deff}, we obtain 
\be
   \mbox{Vol}\Big\{ p_2\; : \;
   |F_1(p)| \leq c\e \Big(|\sin (p_3+q_3)|+ |\sin p_3 |\Big) |q_2|
  \Big\} \leq \frac{c\e}{|q_2|}
\label{F1}
\ee
for any $\e\leq\frac{1}{2}$ and uniformly in all other variables.
It is trivial to see that
$$
   \mbox{Vol}\Big\{ p_3\; : \;
   |\sin (p_3+q_3)|+ |\sin p_3 |\leq \mu\Big\} \leq c\mu
$$
for any $\mu>0$, uniformly in $q_3$.
Let
$$
   M_j: = \Big\{ p_3\; : \;
  2^{-j}\leq |\sin (p_3+q_3)|+ |\sin p_3 |\leq 2^{-j+1}\Big\} 
$$   
then
$$ 
   \mbox{Vol}(M_j) \leq c\cdot 2^{-j}\; .
$$
We consider these sets for $j$'s such that  $2^{-j}\ge 2\eta^{3/4}$ and
set
$$
  M^*: =\Big\{ p_3\; : \;
   |\sin (p_3+q_3)|+ |\sin p_3 |\leq 2\eta^{3/4}\Big\} 
$$   
for the rest.

Let
$$
M:=\Big\{ (p_2, p_3)\; : \;
   |F_1(p)| \leq c\e \mu |q_2|
  \Big\} 
$$
then
$$
   \mbox{Vol}(M) \leq \mbox{Vol}(M^*)+
\sum_j \mbox{Vol}(M\cap M_j)
$$
$$
  \leq c\eta^{3/4}+\sum_j\Big\{ (p_2, p_3)\; : \; p_3\in M_j\; , \;
   |F_1(p)| \leq c\e \mu \cdot 2^j(|\sin (p_3+q_3)|+ |\sin p_3 |) |q_2| 
  \Big\}
$$
$$
  \leq c\eta^{3/4}+\sum_j \mbox{Vol}(M_j)\frac{\e\mu \cdot 2^j}{|q_2|} \leq
  c\eta^{3/4}+\frac{c\e\mu |\log\eta|}{|q_2|}
$$
from (\ref{F1}) with $\e$ modified to $\e\mu \cdot 2^j$
 and using the uniformity of these estimates.
At the end we will choose $\e\mu=\eta^{3/4}$, so the
constraint $2^{-j}\ge 2\eta^{3/4}$ will
guarantee that the condition $\e\leq 1/2$ 
in Lemma \ref{lemma:1dim} holds for the modified $\e$.

We split the $I$ integral into two parts:
$$
  I =\int \frac{[{\bf 1}(M)+{\bf 1}(M^c)]\rd p}
{|\a - e(p) +i\eta|}\frac{1}{|\beta - e(p+q) +i\eta|}\; .
$$
On $M$ we use the trivial $\eta^{-1}$
bound for the second   denominator and integrate out the other one
in the $p_1$ variable (independent of $M$) to bound
its contribution to $I$ by
\be
   C\eta^{-3/2}\Big(\eta^{3/4}+|\log\eta|\cdot \frac{c\e\mu}{|q_2|}\Big)\; ,
\label{errr}
\ee
since
\be
   \sup_c \int_{-\pi}^\pi \frac{\rd p_1}{| c-\cos p_1 +i\eta|} 
\leq c\eta^{-1/2}\; .
\label{supc}
\ee

On $M^c$ we use \eqref{coare} and the estimate $|F|^{-1}\le 
(c\e \mu |q_2|)^{-1}$ to get
\be
  \frac{c |\log\eta|^2 }{\e \mu |q_2|} \; .
\label{errrr}
\ee
The bounds in (\ref{errr}) and (\ref{errrr}) are optimized for
$\e\mu=\eta^{3/4}$ and give (\ref{nopont}). 
Finally the proof of (\ref{bnopont}) is almost identical, except
that instead of (\ref{supc}) we integrate out $\alpha$
to collect only $|\log\eta|$ instead of $\eta^{-1/2}$.
$\Box$

\end{document}